\documentclass{aa}  

\usepackage{graphicx}
\usepackage{txfonts}
\usepackage{subcaption}

\begin{document}

   \title{Probing 3D and NLTE models using APOGEE observations of globular cluster stars}

   \author{T. Masseron \inst{1,2}
          \and
          Y. Osorio \inst{1,2}
          \and 
         D.A.Garc\'ia-Hern\'andez \inst{1,2}         
         \and 
         C. Allende Prieto\inst{1,2}
           \and
         O. Zamora\inst{1,2}   
                  \and 
         Sz. M\'esz\'aros\inst{3,4}
 }
\institute{Instituto de Astrof\'isica de Canarias, E-38205 La Laguna, Tenerife, Spain
        \and 
Departamento de Astrof\'isica, Universidad de La Laguna, E-38206 La Laguna, Tenerife, Spain\\
              \email{tmasseron@iac.es}
        \and
ELTE E\"otv\"os Lor\'and University, Gothard 
Astrophysical Observatory, 9700 Szombathely, Szent Imre H. st. 112, Hungary
\and
MTA-ELTE Exoplanet Research Group
        \and
    Premium Postdoctoral Fellow of the Hungarian Academy of Sciences
    }

   \date{Received ; accepted}

 
  \abstract
   {Hydrodynamical (or 3D) and non-local thermodynamic equilibrium (NLTE) effects are known to affect abundance analyses. However, there are very few observational abundance tests of 3D and NLTE models.}
   {We developed a new way of testing the abundance predictions of 3D and NLTE models, taking advantage of large spectroscopic survey data.}
   {We use a line-by-line analysis of the Apache Point Observatory Galactic Evolution Experiment (APOGEE) spectra (H band) with the Brussels Automatic Code for Characterizing High accUracy Spectra (BACCHUS).  We compute line-by-line abundances of Mg, Si, Ca, and Fe for a large number of globular cluster K giants in the APOGEE survey. We compare this line-by-line analysis against NLTE and 3D predictions.}
   {While the 1D--NLTE models provide corrections in the right direction, there are quantitative discrepancies between different models. We observe a better  agreement with the data for the models including reliable  collisional cross-sections. The agreement between data and models is not always satisfactory when the 3D spectra are computed in LTE. However, we note that for a fair comparison, 3D corrections should be computed with self-consistently derived stellar parameters, and not on 1D models with identical stellar parameters. Finally, we focus on 3D and NLTE effects on Fe lines in the H band, where we observe a systematic difference in abundance relative to the value from the optical. Our results suggest that the metallicities obtained from the H band are more accurate in metal-poor giants.}
   {Current 1D--NLTE models provide reliable abundance corrections, but only when the atom data and collisional cross-sections are accurate and complete. Therefore, we call for more atomic data for NLTE calculations. In contrast, we show that 3D corrections in LTE conditions are often not accurate enough, thus confirming that 3D abundance corrections are only valid when NLTE is taken into account. Consequently, more extended self-consistent 3D--NLTE computations need to be made. The method we have developed for testing 3D and NLTE models could be extended to other lines and elements, and is particularly suited for large spectroscopic surveys.}

   \keywords{
               }

   \maketitle
%

\section{Introduction}
 Accurate  chemical analysis of cool stellar spectra (T$\rm_{eff}\lesssim 8000K$) rely on a deep understanding of radiative transfer physics. As extensively explored over many astronomical applications in the review by \citet{Nissen2018},  currently the two main challenges for abundance derivation involve the three-dimensional (3D) modelling of hydrodynamics of stellar atmospheres  and the modelling of line formation in conditions of non-local thermodynamic equilibrium  (NLTE)  \citep[see][for the physics principles]{Nordlund2009,Bergemann2014,Bergemann20143DNLTE}.\\
Among the large number of publications on NLTE effects \citep[each  generally focused on one particular element and on a limited number of lines; see review by][]{Mashonkina2014_review}, some elements have been particularly studied.

Magnesium is the archetypal species for NLTE calculations, notably because the Mg triplet in the solar optical spectrum clearly shows the NLTE signature in the core of the lines. Its model atom is also fairly well established now \citep[][]{Osorio2015_Mgatom} and also has  good collisional cross-sections. Therefore, there are a large number of NLTE computations for this element in late-type stars \citep[e.g.][]{Merle2011,Bergemann2015_Mg,Osorio2016_Mg,Zhang2017,Aleexeva2018_Mg}. An important difference between the atom model \cite{Osorio2015_Mgatom} and other atom models presented in later works lies in the adoption of collisional data involving high-lying levels of Mg I (from which many IR lines are formed). Both \cite{Zhang2017} and \cite{Aleexeva2018_Mg} ignored hydrogen collisions involving high-lying levels of Mg I while \cite{Osorio2015_Mgatom} uses the formula from \cite{1984JPhB...17.4485K}. Electron collisional ionisation for high-lying levels of Mg I were also updated in \cite{Osorio2015_Mgatom}, where instead of the classical formula \citep{1962amp..conf..375S} the method from \cite{Vrinceanu:2005em} was  adopted. 

Calcium is fundamental for stellar spectroscopy, notably because of the triplet in the near-IR that represents an inevitable feature for metallicity estimate and more particularly for the Gaia mission. Therefore, a large number of NLTE studies have been dedicated to this element \citep[e.g.][]{Merle2011,Mashonkina2017_Ca}.  It now also has   a quite accurate model atom and good collisional data \citep{Osorio2019_Ca}. Similar improvements to the Mg atom are present in the \cite{Osorio2019_Ca} model atom.  

Like Mg and Ca, silicon is an $\alpha$-element, and thus its measurement reflects the chemical signature of core-collapse supernovae nucleosynthesis. Estimates for NLTE of Si have been published for various stellar spectra \citep[e.g.][]{Bergemann2013_Si,Mashonkina2016_Si,Zhang2016}. 

Iron is fundamental in stellar spectroscopy;  it is used as the main metallicity indicator due to the numerous lines present in the spectra, which are also used to infer the main stellar parameters (T$\rm_{eff} and \log g$). Therefore, its NLTE effects have been extensively studied in literature \citep{Bergemann2012_Fe,Mashonkina2011_Fe,Mashonkina2016_CaFe}. However, despite its extremely deep implication in all stellar studies, its model atom and collisional cross-sections are currently both theoretically and experimentally incomplete, and sometimes empirical estimates are necessary \citep{Ezzeddine2018,Caffau2019,Korotin2020}. 

Despite the large amount of atomic data simultaneously required, it is now possible to perform NLTE synthesis on a large scale \citep[e.g.][]{Kovalev2019,Osorio2019_inprep}. However, the main limitations and uncertainties regarding NLTE calculations depend on the fundamental knowledge of the collisional data and the structure of the atom. \\
In contrast, hydrodynamical simulations of stellar atmospheres are still very expensive with regard to   computational resources, which means  that  currently the largest grids \citep[][]{Ludwig2009,Magic2013} provided by the three main codes (MuRAM \citealt{Vogler2005}, CO$^5$BOLD \citealt{Freytag2012}, and Stagger \citealt{Galsgaard1996}) only include a few tens of models. From these limited number of models, 3D corrections on individual elements are still being estimated (e.g. \citealt{Spite2012} for Ca, \citealt{Cerniauskas2018} for Mg,  and \citealt{Collet2006} for several elements in a single star). 

Naturally, the ultimate goal regarding abundance determination is to achieve consistent 3D and NLTE calculations. The first developments have recently been made, but fully consistent 3D and NLTE analysis\footnote{Nearly fully consistent since the computation of 3D atmosphere structure assumes LTE.} are  limited to a few elements in one star or one element in a few stars \citep{Asplund2003,Sbordone2010,Lind2013,Steffen2015,Klevas2016,Amarsi2016,Nordlander2017,Amarsi2019}. Those works illustrate that 3D and NLTE computed simultaneously in a self-consistent manner clearly provides  results that are different to the simple sum of independently computed 3D and NLTE effects. A more computationally efficient alternative consists in using a horizontal average of the 3D atmosphere and then computing the NLTE corrections \citep{Bergemann2012_Fe3D,Mashonkina2013}, but this is still a limited approximation \citep[e.g.][]{Lind2017}. 

 This illustrates  that different approaches exist for computing 3D and NLTE effects on line formation. There are various prescriptions between the available models and codes, as shown for example by \citet{Beeck2012} by comparing 3D codes. Therefore, it is important that  models are probed against observations. The standard way to test NLTE effects consists in synthesising individual line profiles and comparing them to very high-resolution spectra \citep[e.g.][] {AllendePrieto2004,Mashonkina2011_Fe,Lind2017}. Another alternative consists in comparing the abundances obtained between neutral and ionised lines of the same element \citep[e.g.][]{Korn2003}. On the contrary, it is not so straightforward to test 3D effects against line profiles. Only the solar spectrum  \citep{Asplund2000a,Asplund2000b,Caffau2008,Pereira2009,Lind2017} and Procyon's spectrum \citep{AllendePrieto2002}  are  used for direct comparison against 3D line profiles. \\
In summary, both 3D and NLTE line formation predictions have been mostly tested only against a few available extremely high-resolution spectra with  high signal-to-noise ratios. Moreover, because absolute elemental abundances need to be fixed in the comparison, the solar spectrum turned out to be nearly the only reliable data to be used for  testing. 

\begin{figure}
\centering
        \begin{subfigure}[b]{\columnwidth}
        \includegraphics[angle=-90,width=0.495\columnwidth]{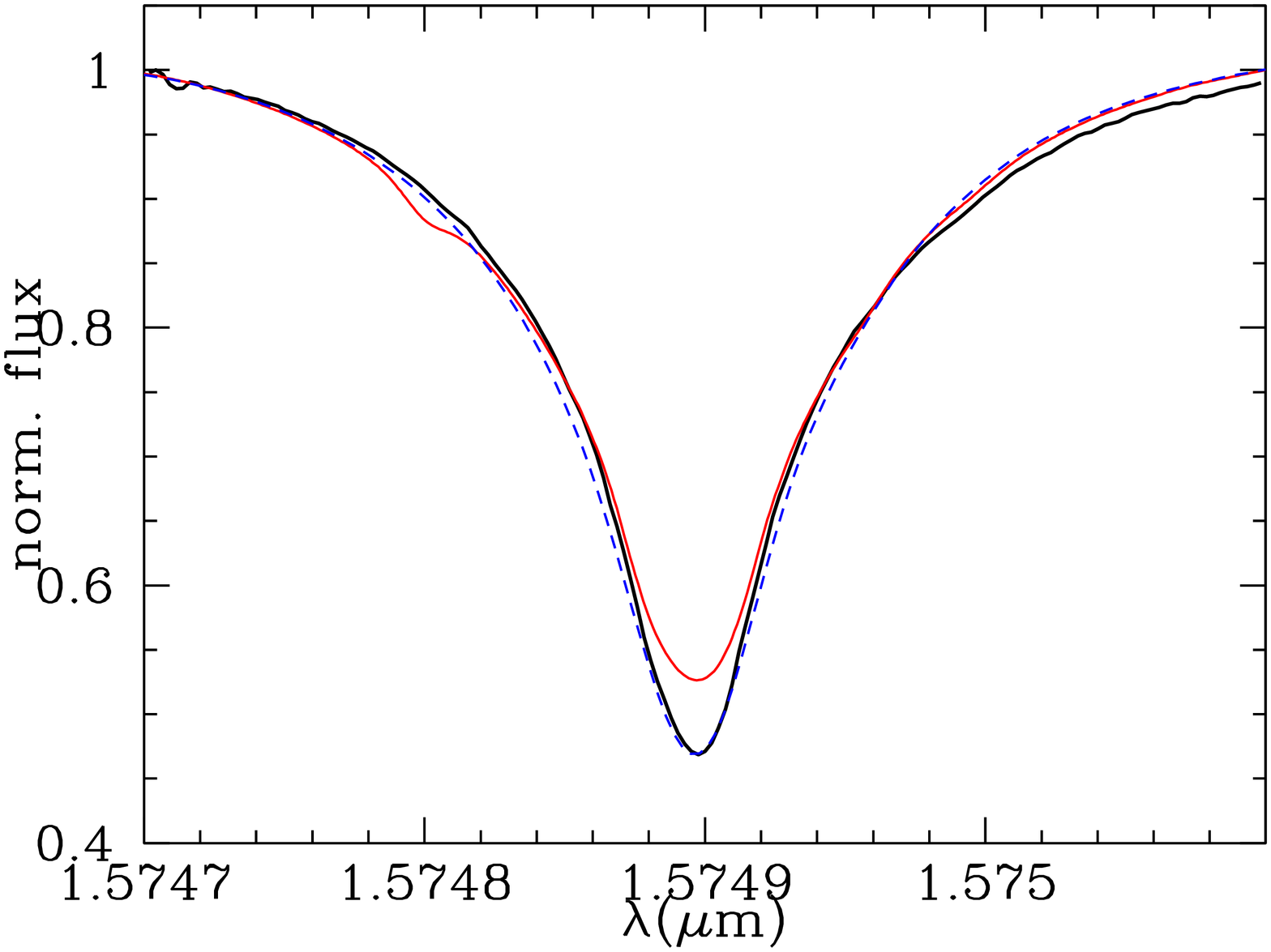}
        \includegraphics[angle=-90,width=0.495\columnwidth]{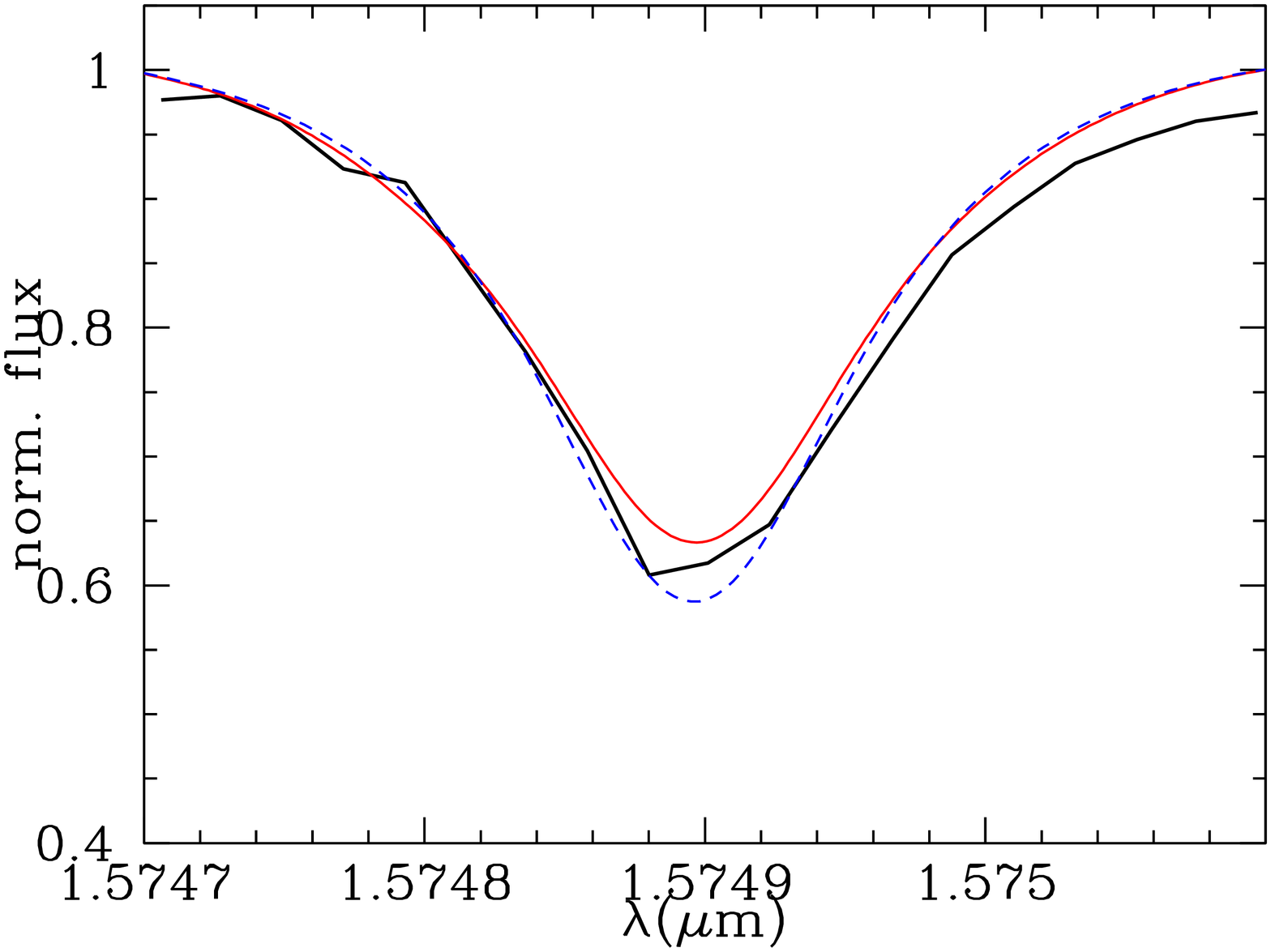}
    \end{subfigure}
        \begin{subfigure}[b]{\columnwidth}
        \includegraphics[angle=-90,width=0.495\columnwidth]{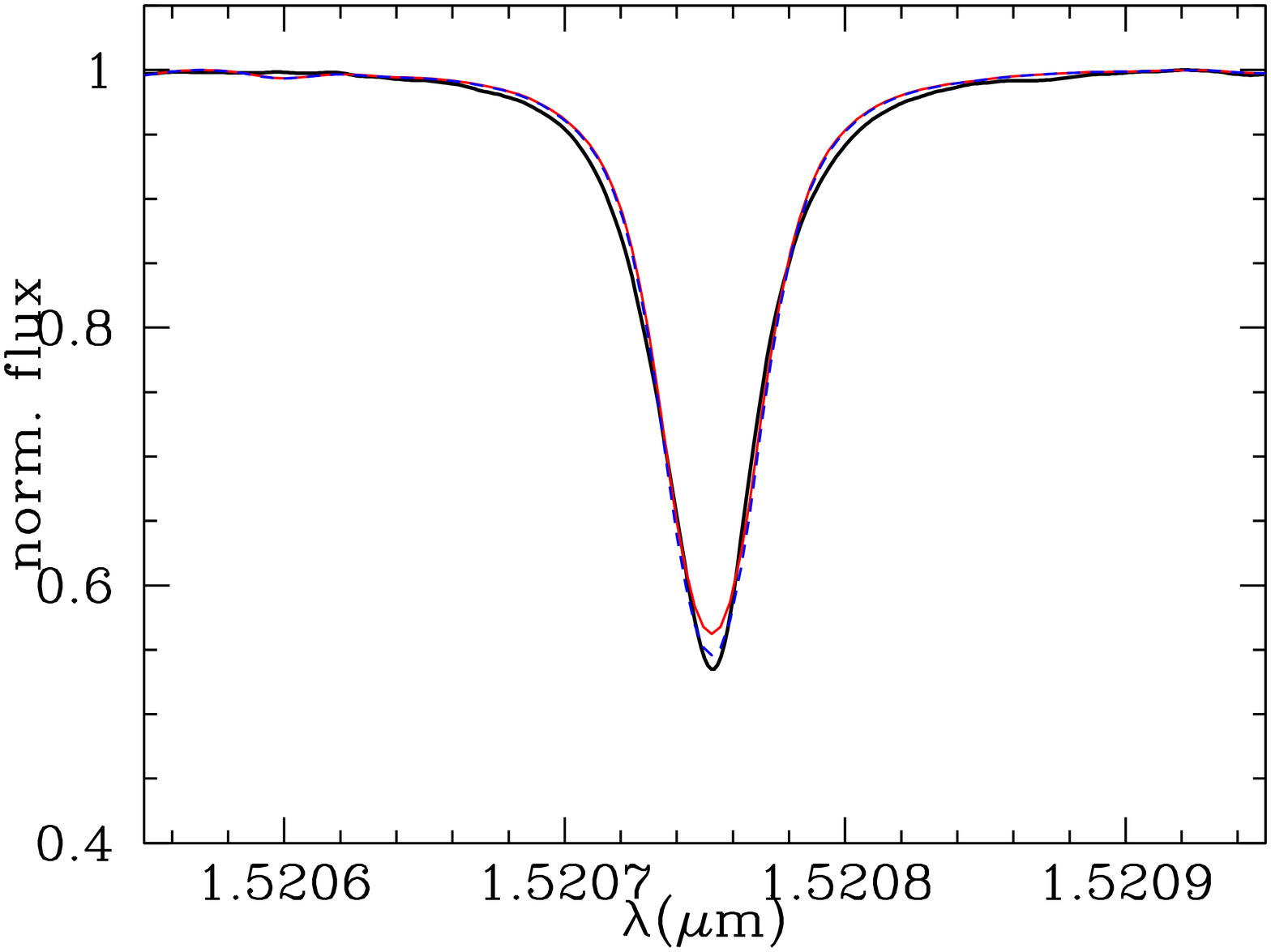}
        \includegraphics[angle=-90,width=0.495\columnwidth]{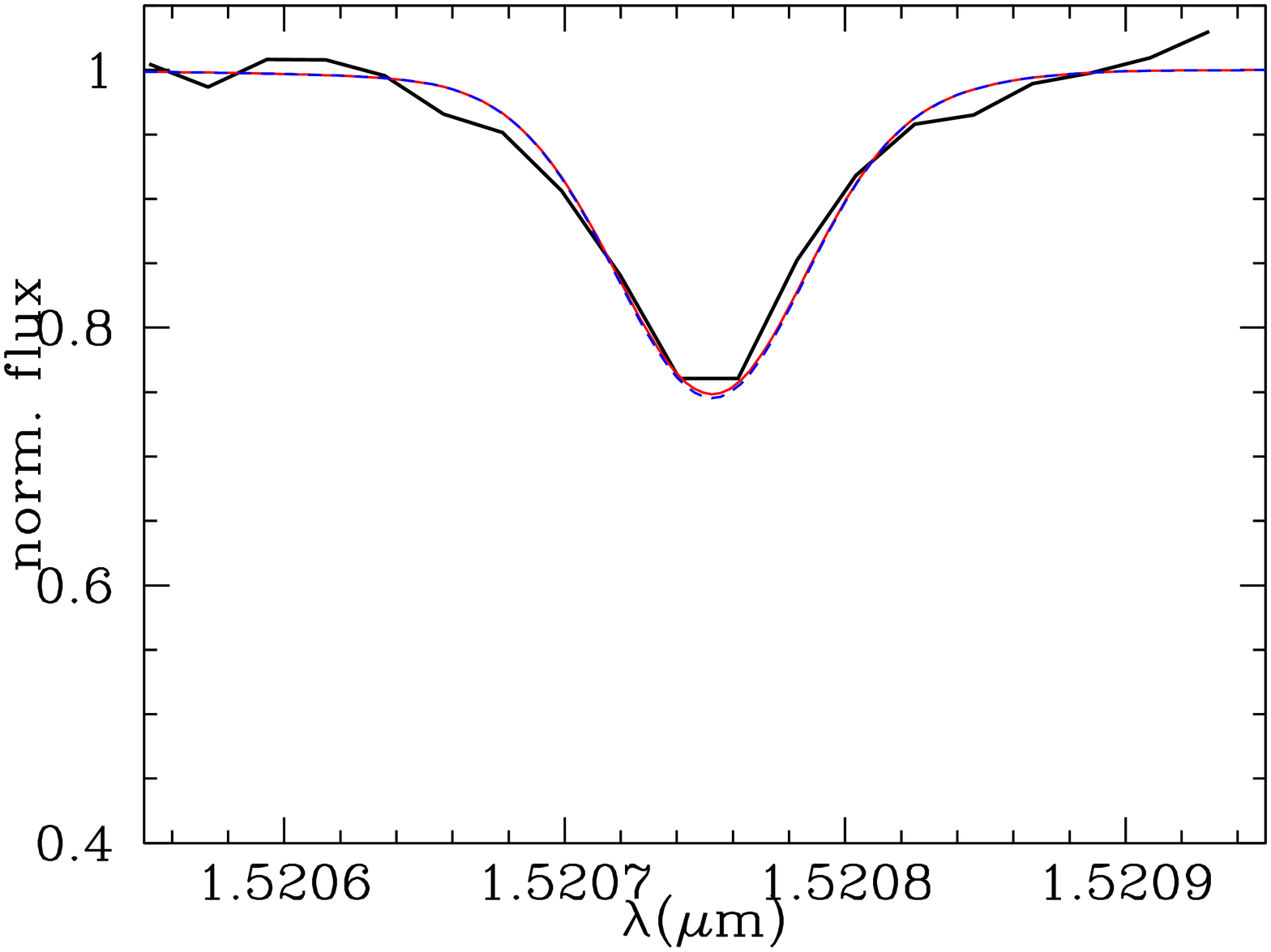}
    \end{subfigure}
\caption{\label{fig:Sun_lines} Illustration of an observed Mg line (upper panels) and an Fe line (lower panels) and their respective 1D--LTE (red) and 1D--NLTE (blue) models in the solar spectrum \citep{Livingston1991}. The left panels show high-resolution (R$>$ 100000) and very high signal-to-noise ratio (S/N$>$800) observations, while the right panels show the same spectrum at lower resolution (R$\approx$ 22000) and lower signal-to-noise ratio (S/N = 100). Although the high-resolution spectra clearly highlight some modelling issues around the core of the lines, the effect is less clearly visible in the lower resolution spectrum. }
\end{figure}

Meanwhile, several large spectroscopic surveys now offer a large amount of stellar spectra (several hundred thousand)  (APOGEE, \citealt{Majewski2017}; Gaia-ESO, \citealt{Gilmore2012,Randich2013}; GALAH, \citealt{Desilva2015}). These surveys have enough resolution power to resolve individual lines, and are therefore strongly affected by systematic errors due to NLTE and 3D effects that impact their 1D--LTE abundance determination.  For example, a systematic discrepancy in the derived metallicity of globular clusters between the IR APOGEE data and optical literature has been found  across the various releases \citep{Meszaros2013,Holtzman2015}, and has been confirmed by independent analysis of the same data \citep{Meszaros2015,Masseron2019}. This suggests that a 3D or NLTE effect can impact the Fe lines. Moreover, \citet{Hawkins2016} suspected that NLTE effects on some of the Ti and Al lines are affecting the abundance measurements in the APOGEE spectra. Unfortunately, despite these possible signatures of NLTE and/or 3D effects, the resolutions of the surveys spectrographs do not usually allow a detailed line profile examination (see Fig.~\ref{fig:Sun_lines}). Therefore, abundance results from these surveys more generally rely on 3D and NLTE models to correct the abundances \citep[e.g.][]{Kovalev2019}, and do not question them.

Still, despite the relatively limited resolution power of large spectroscopic surveys, we propose here an empirical method to test 3D and NLTE effects based on a line-by-line differential approach. Although a few other studies have attempted to compare line abundances to detect lines biased by substantial NLTE or 3D effects, they   relied on a small number of spectra \citep[e.g. $\approx$140 in][]{Roederer2014}.  Here we take  advantage of the large number of observations of the APOGEE survey of globular clusters (GCs) stars to quantify and discuss the occurrence of 3D and NLTE effects in stellar spectra of cool metal-poor stars.

\begin{figure*}[!ht]
\centering
        \begin{subfigure}[b]{\textwidth}
\centering
        \includegraphics[angle=-90,width=0.5\textwidth]{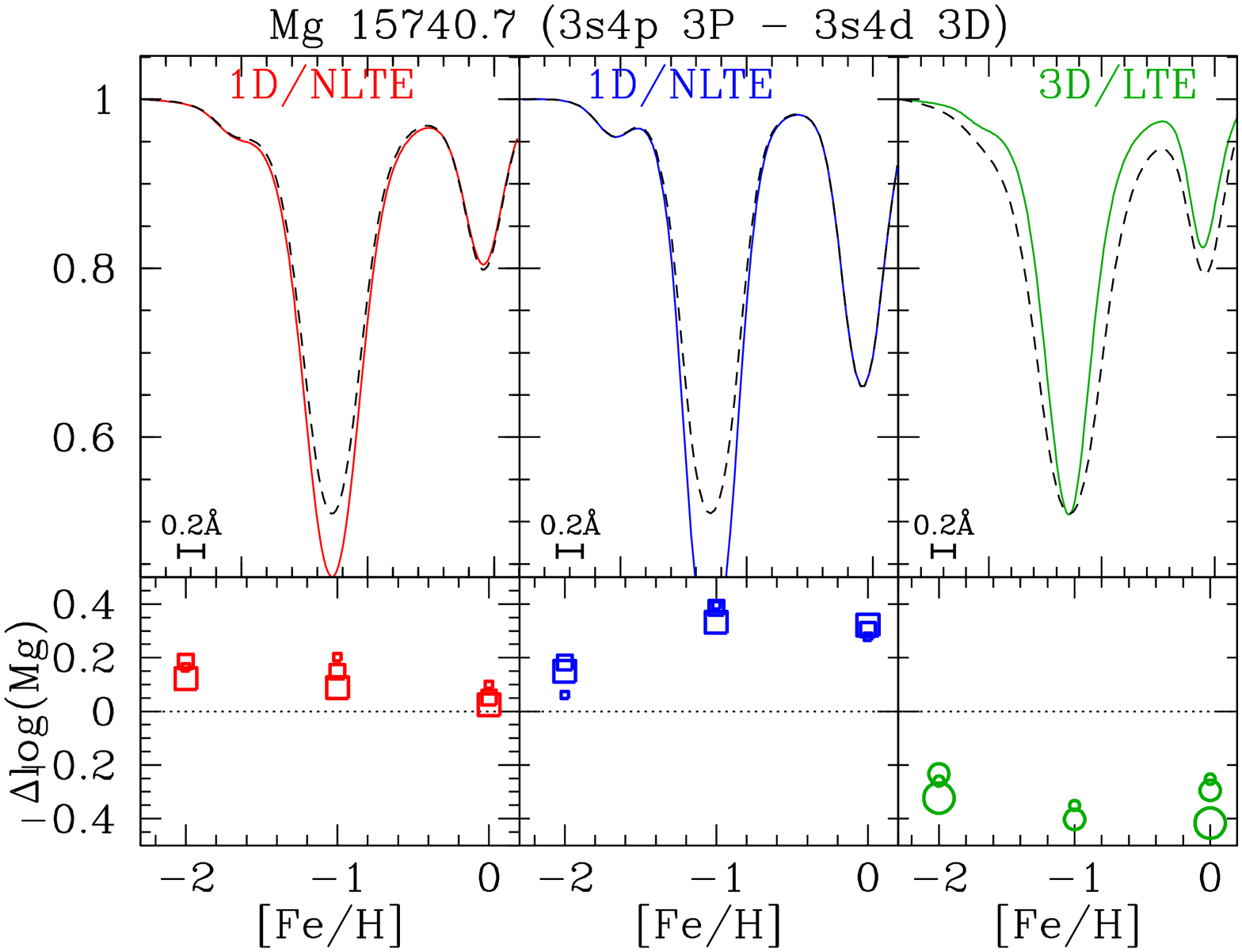}
    \end{subfigure}
        \begin{subfigure}[b]{\textwidth}
        \includegraphics[angle=-90,width=0.5\textwidth]{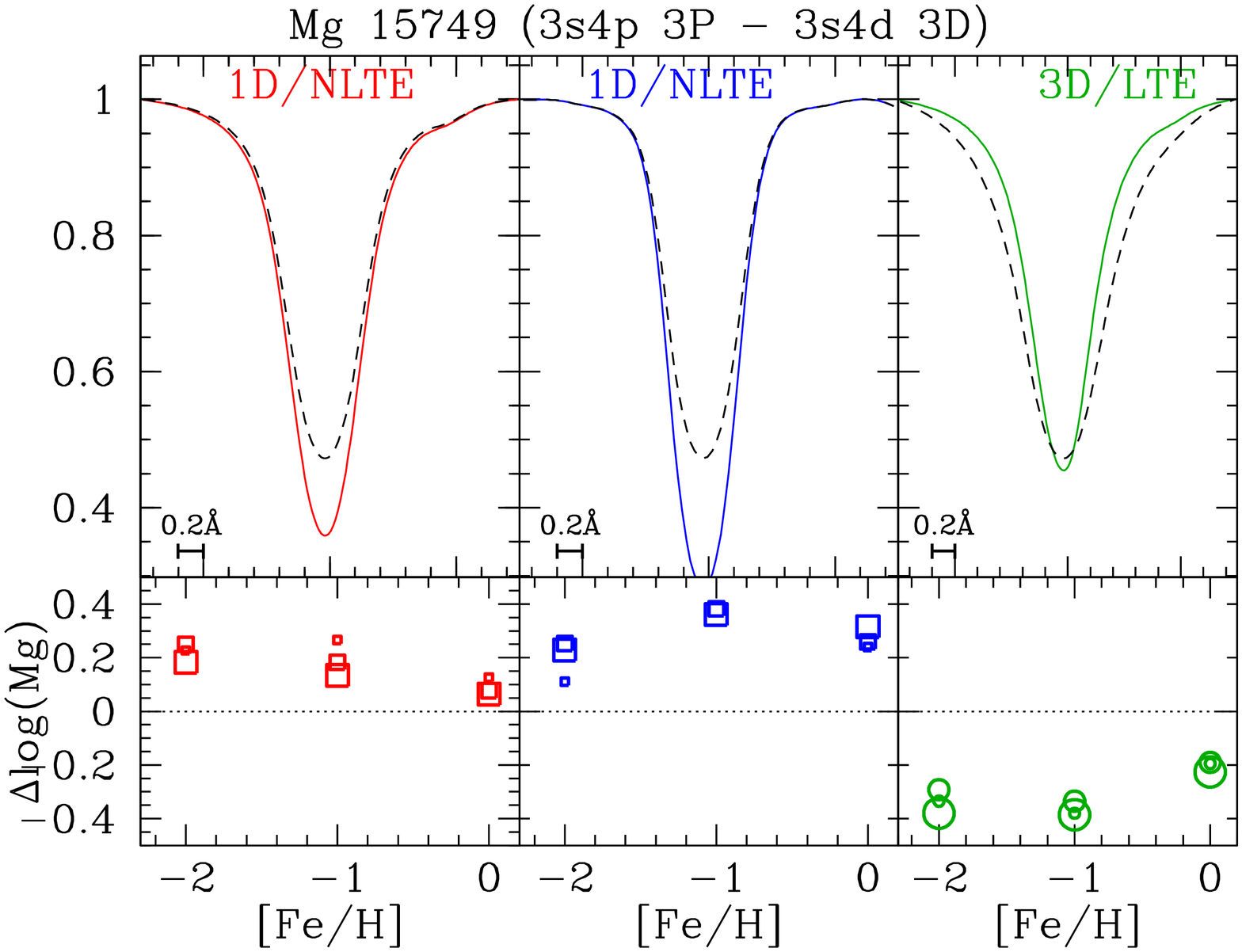}
        \includegraphics[angle=-90,width=0.5\textwidth]{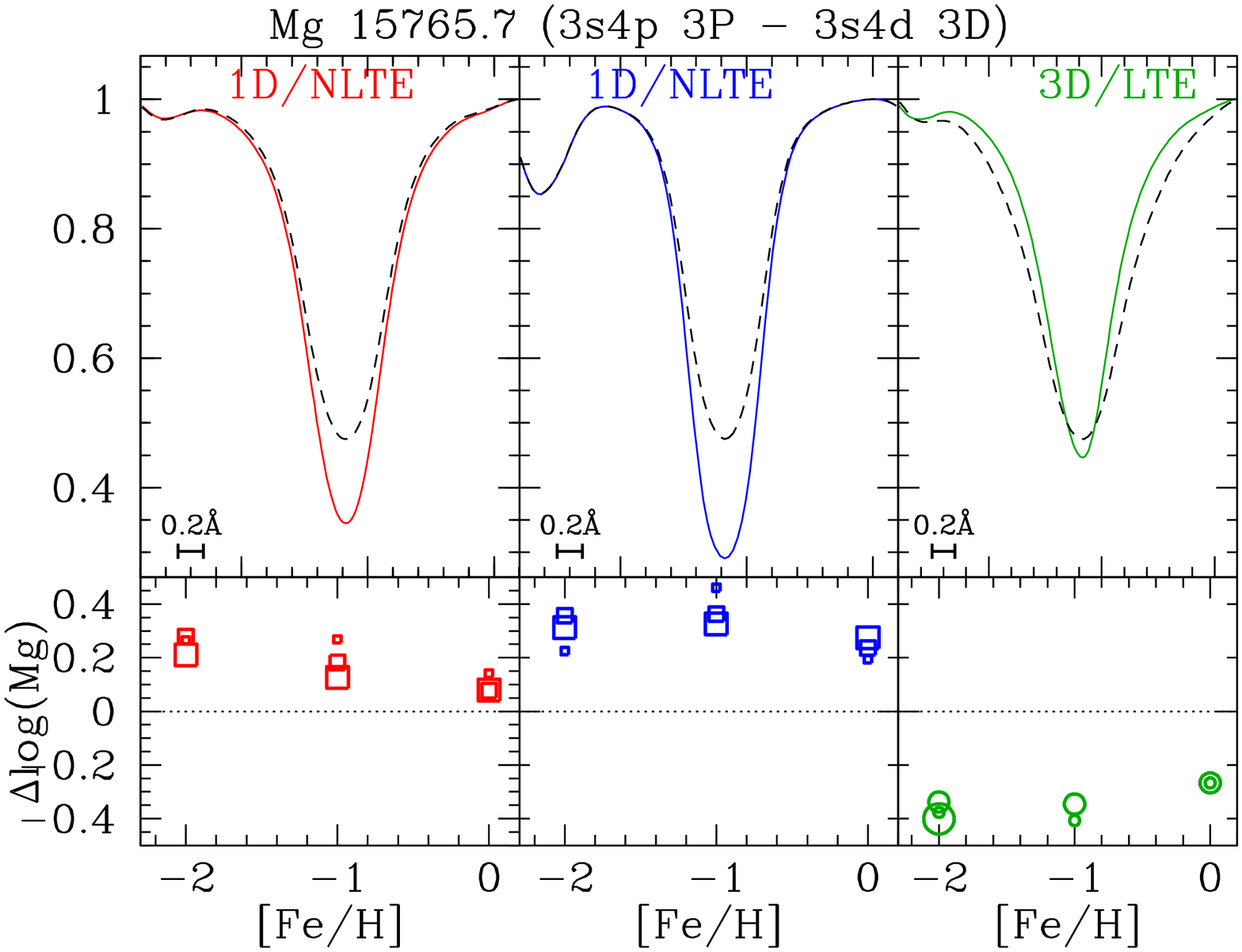}
    \end{subfigure}
        \begin{subfigure}[b]{\textwidth}
        \includegraphics[angle=-90,width=0.5\textwidth]{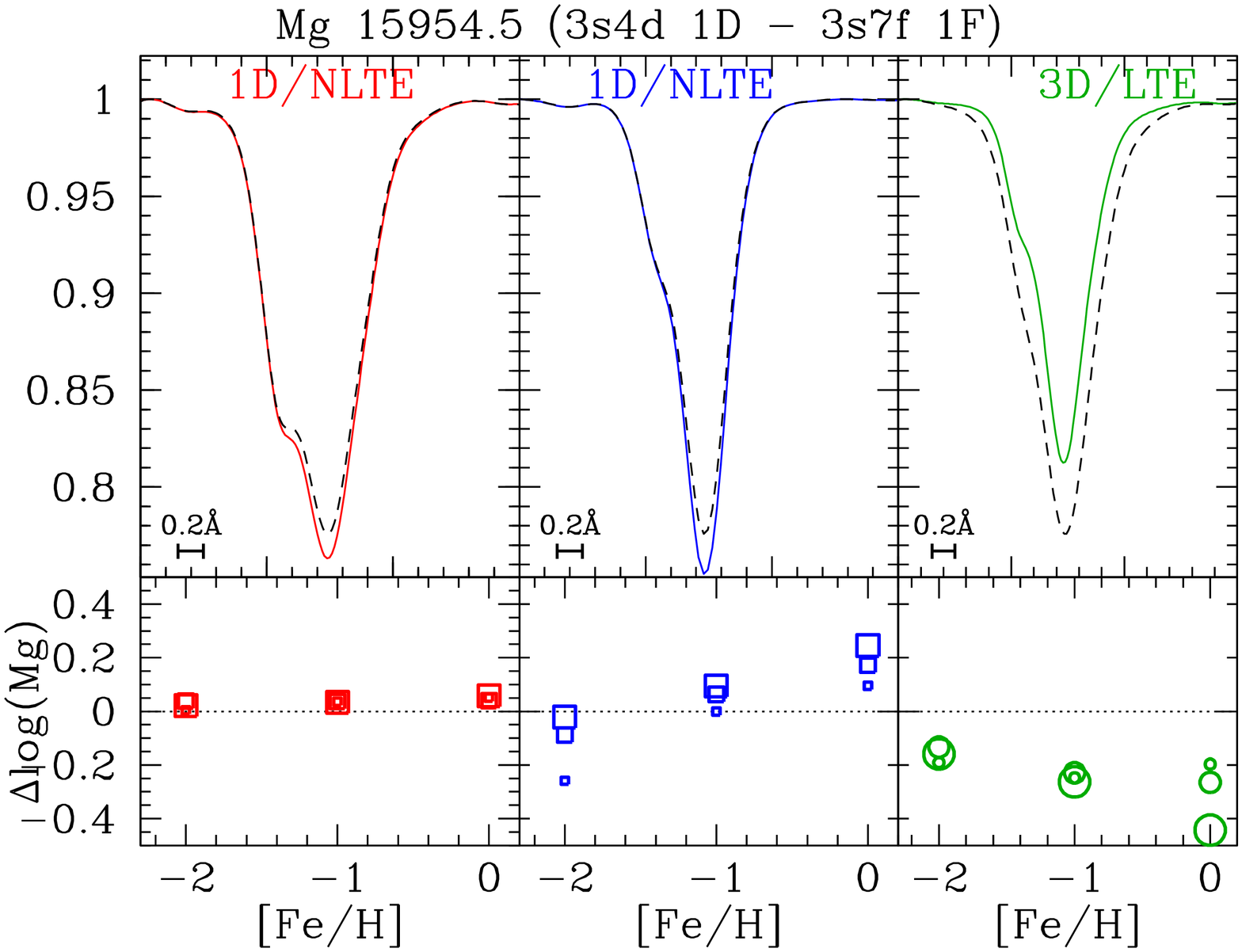}
        \includegraphics[angle=-90,width=0.5\textwidth]{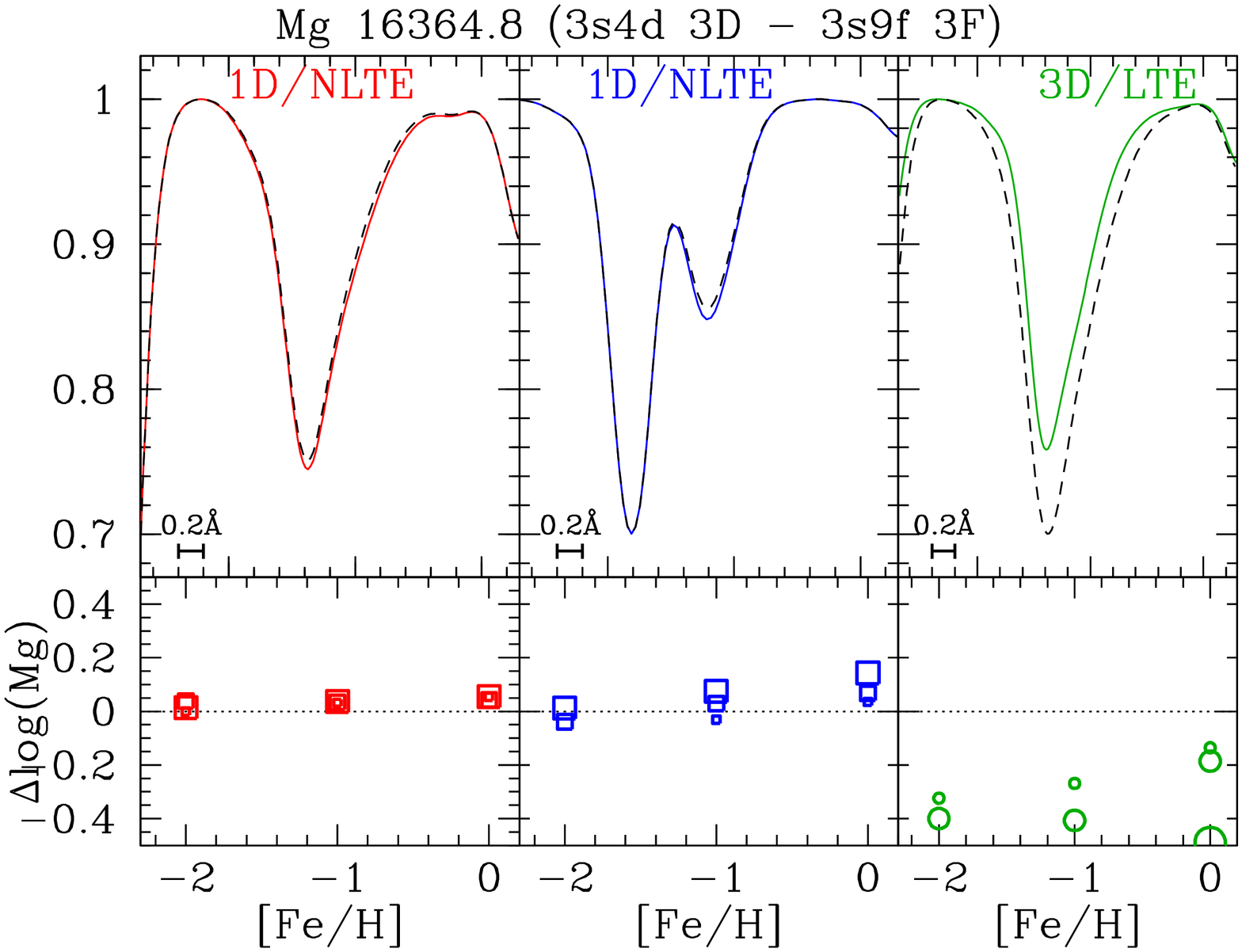}
    \end{subfigure}
\caption{\label{fig:Mg_lines} Synthesis of the Mg lines used in this work. In each sub-panel the upper part  shows the 1D--LTE synthesis (dashed black line), the 1D--NLTE synthesis from this work (red) and \citet{Kovalev2018} (blue), and the 3D--LTE synthesis (green)  for a T$\rm{eff}$=4500K, $\rm \log$g=2.5, [M/H]=$-$1.0 model. The lower parts display the corresponding differences in abundances of the same line between the 1D--LTE model and the NLTE or 3D models for various temperatures and metallicities (large symbols T$\rm{eff}$=4000K, $\rm \log$g=1.5; medium-size symbols  T$\rm{eff}$=4500K, $\rm \log$g=2.5; small symbols  T$\rm{eff}$=5000K, $\rm \log$g=2.5).}
\end{figure*}

\begin{figure*}[!ht]
\centering
        \begin{subfigure}[b]{\textwidth}
        \includegraphics[angle=-90,width=0.5\textwidth]{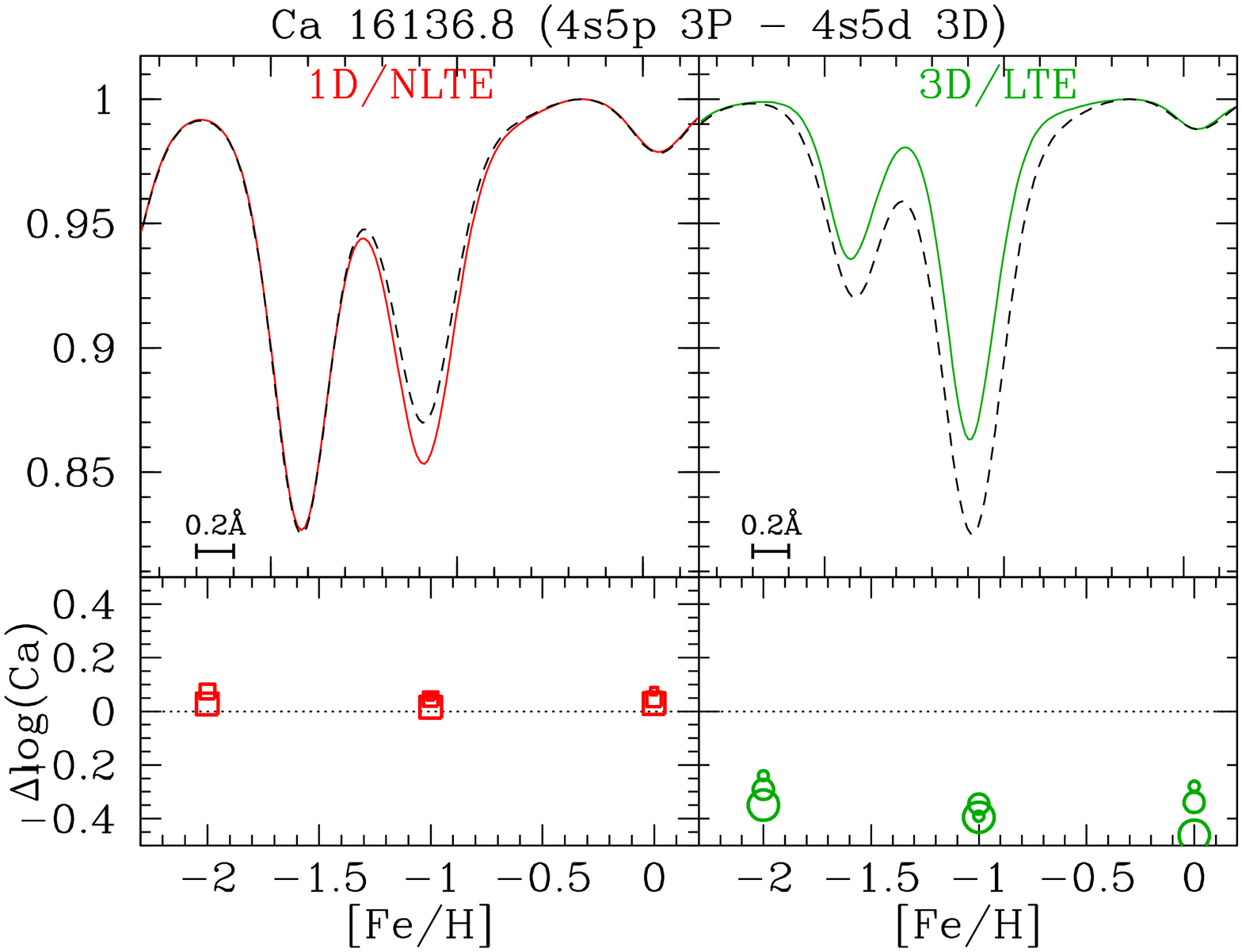}
        \includegraphics[angle=-90,width=0.5\textwidth]{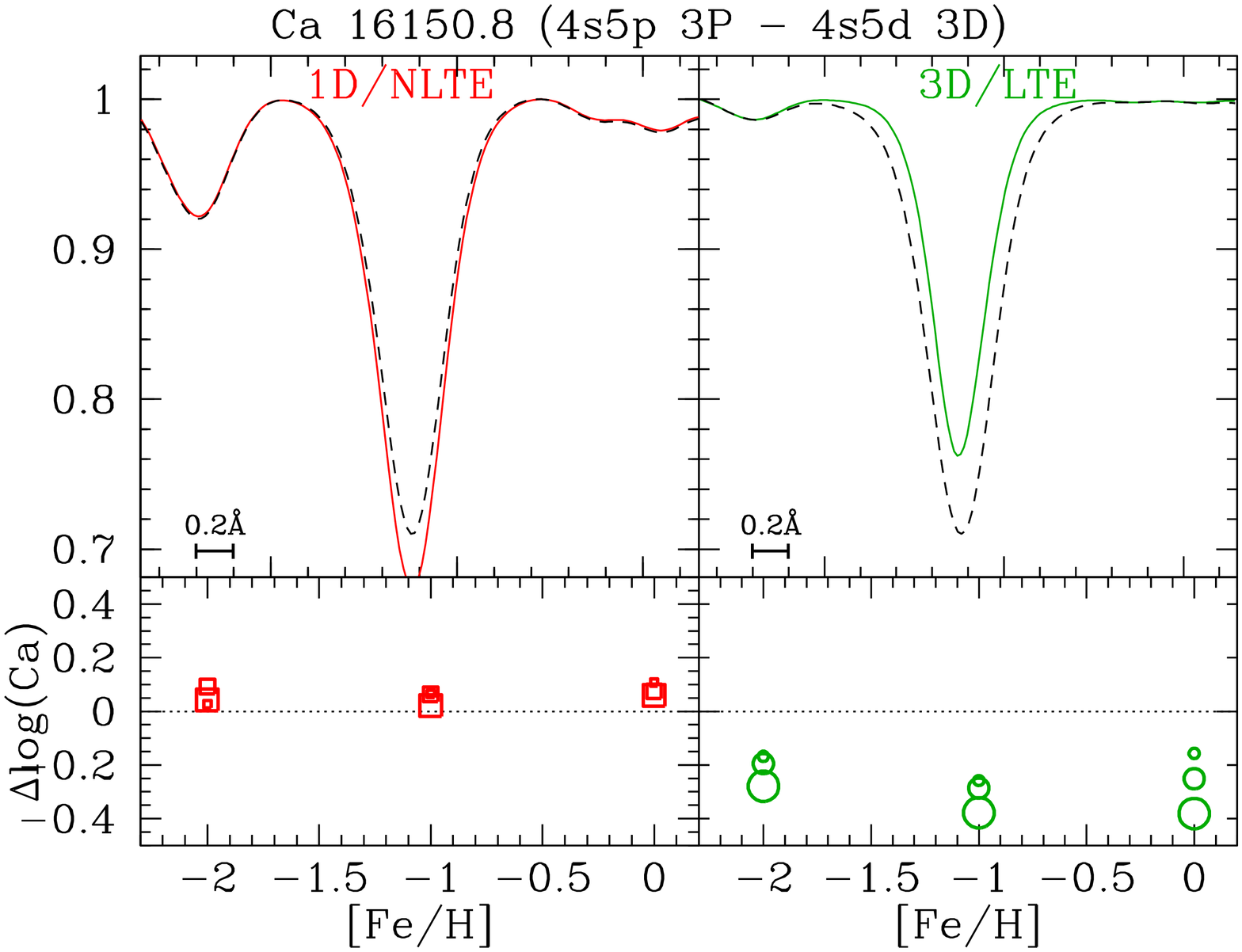}
    \end{subfigure}
        \begin{subfigure}[b]{\textwidth}
        \includegraphics[angle=-90,width=0.5\textwidth]{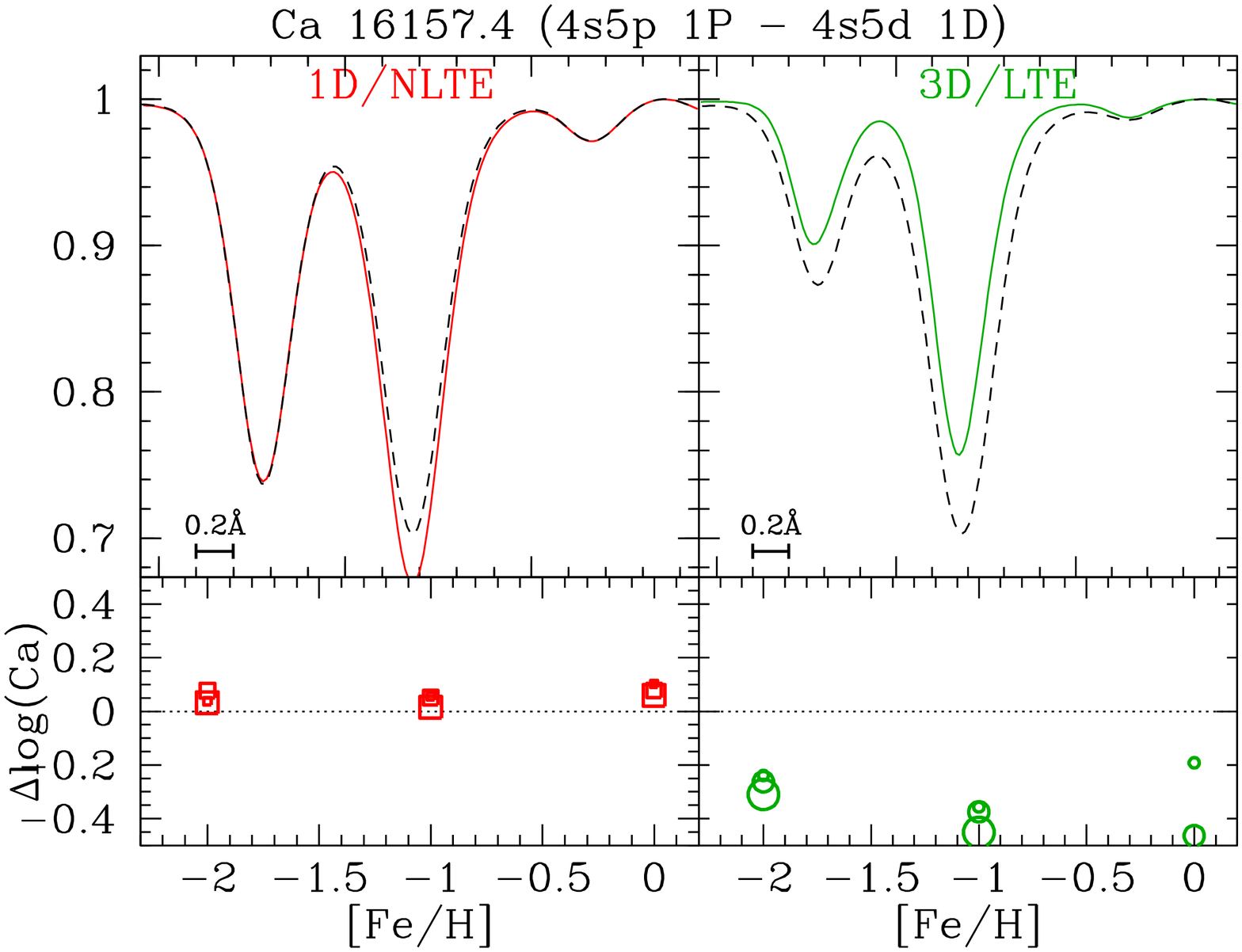}
        \includegraphics[angle=-90,width=0.5\textwidth]{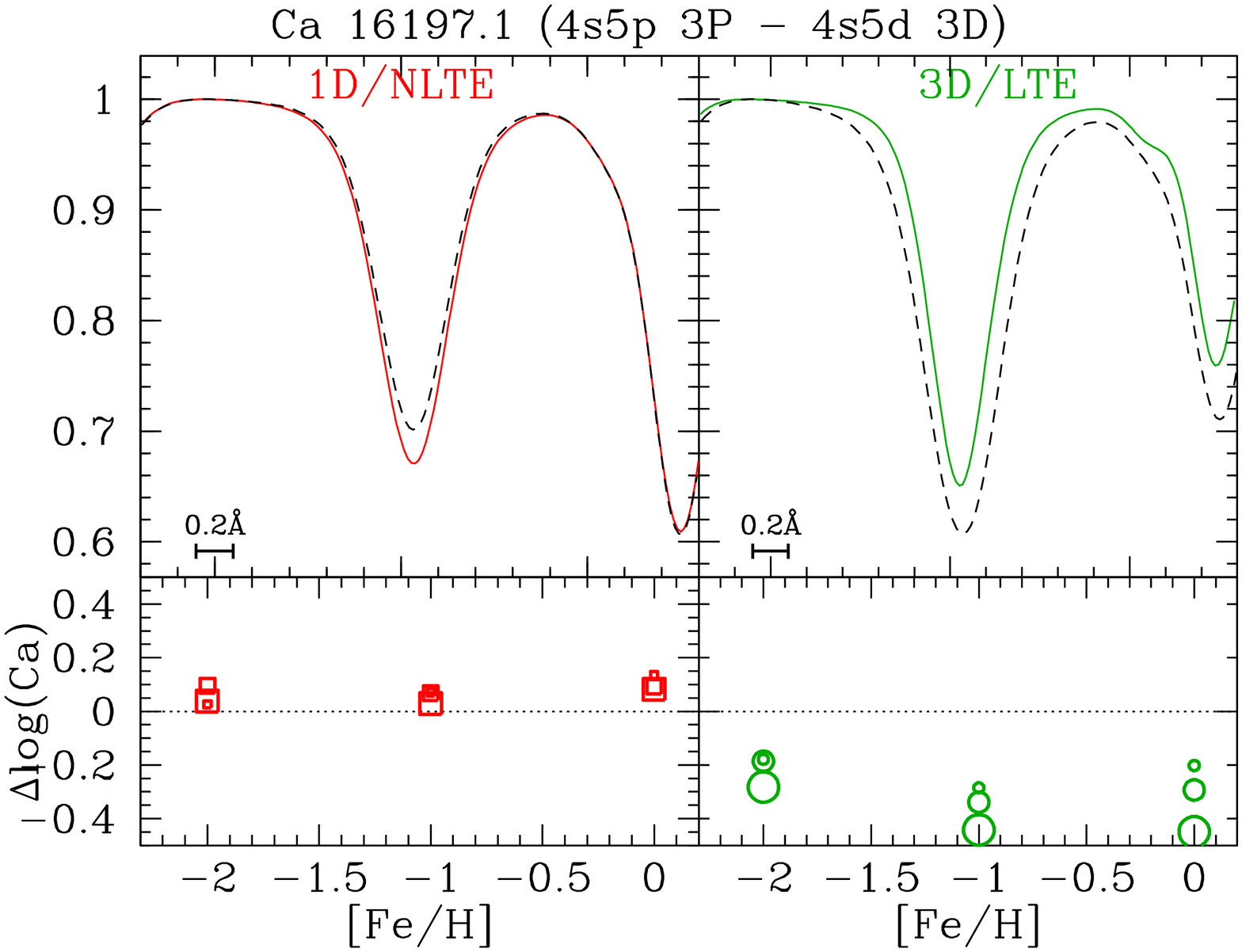}
    \end{subfigure}
\caption{\label{fig:Ca_lines} Synthesis of Ca lines used in this work. In each sub-panel the upper part  shows the 1D--LTE synthesis (dashed black line), the 1D--NLTE synthesis from this work (red), and the 3D--LTE synthesis (green)  for a T$\rm{eff}$=4500K, $\rm \log$g=2.5, [M/H]=$-$1.0 model. The lower parts display the corresponding differences in abundances of the same line between the 1D--LTE model and the NLTE or 3D models for various temperatures and metallicities (large symbols T$\rm{eff}$=4000K, $\rm \log$g=1.5; medium symbols  T$\rm{eff}$=4500K, $\rm \log$g=2.5; small symbols T$\rm{eff}$=5000K, $\rm \log$g=2.5).}
\end{figure*}

\begin{figure*}[!ht]
\centering
        \begin{subfigure}[b]{\textwidth}
        \includegraphics[angle=-90,width=0.33\textwidth]{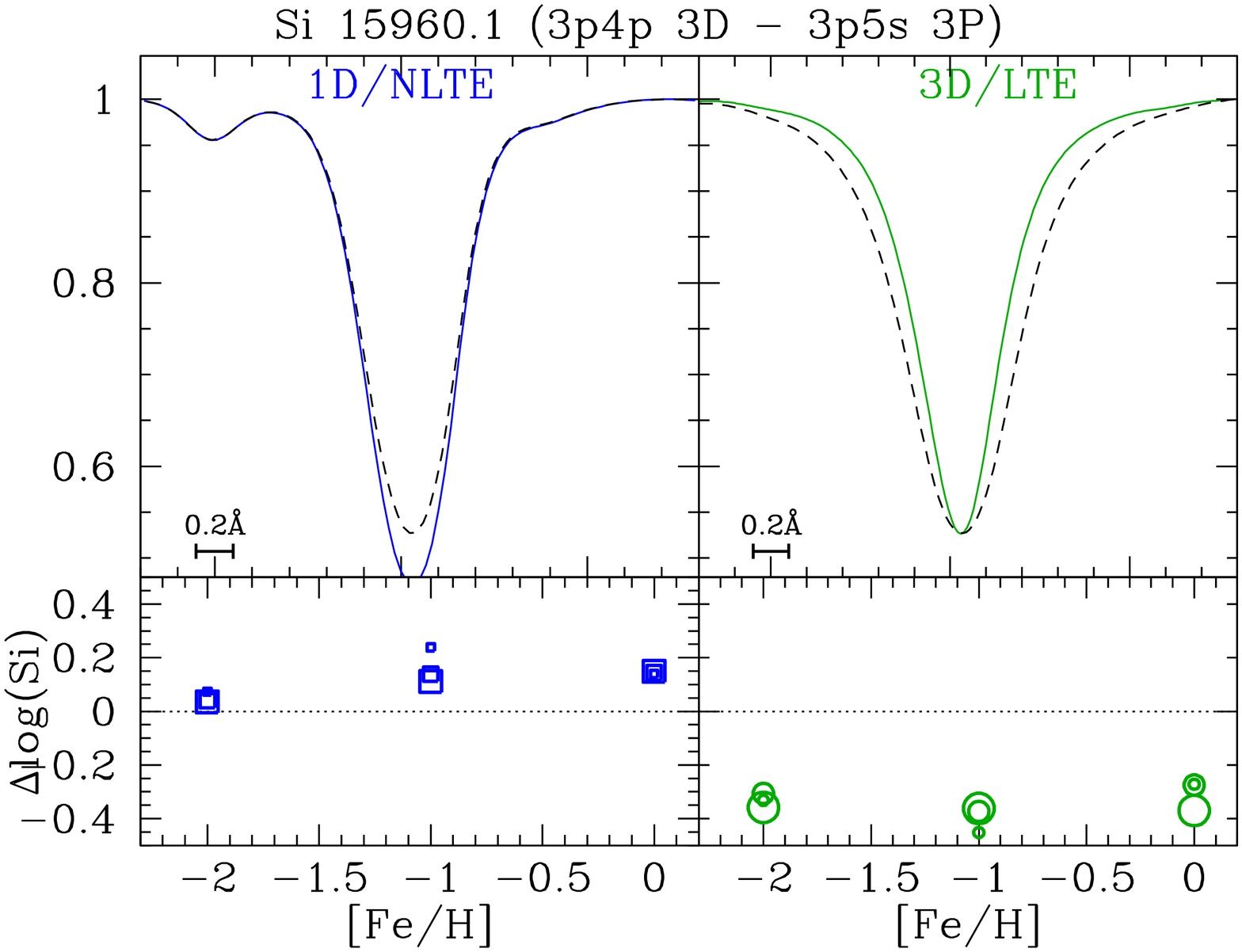}
        \includegraphics[angle=-90,width=0.33\textwidth]{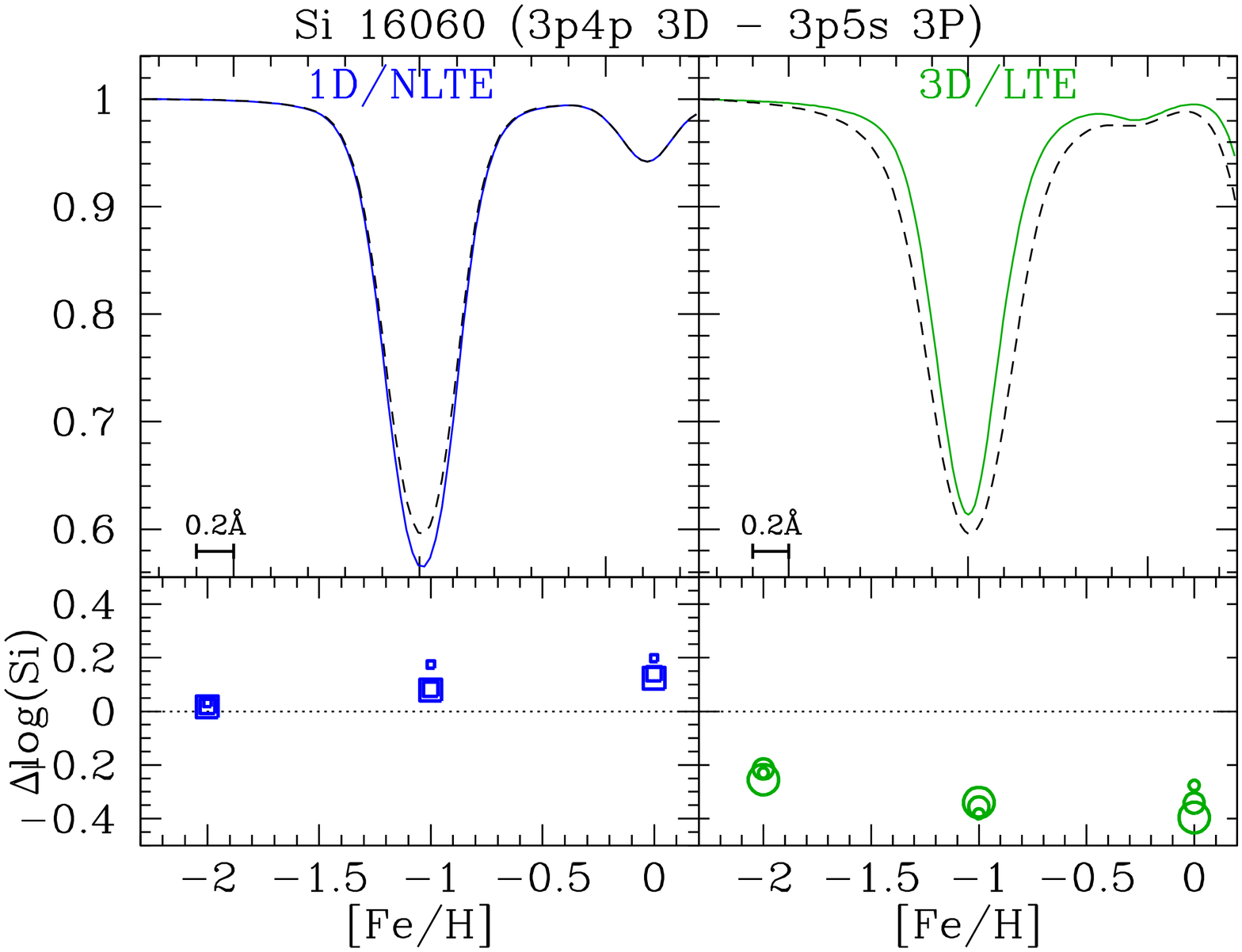}
        \includegraphics[angle=-90,width=0.33\textwidth]{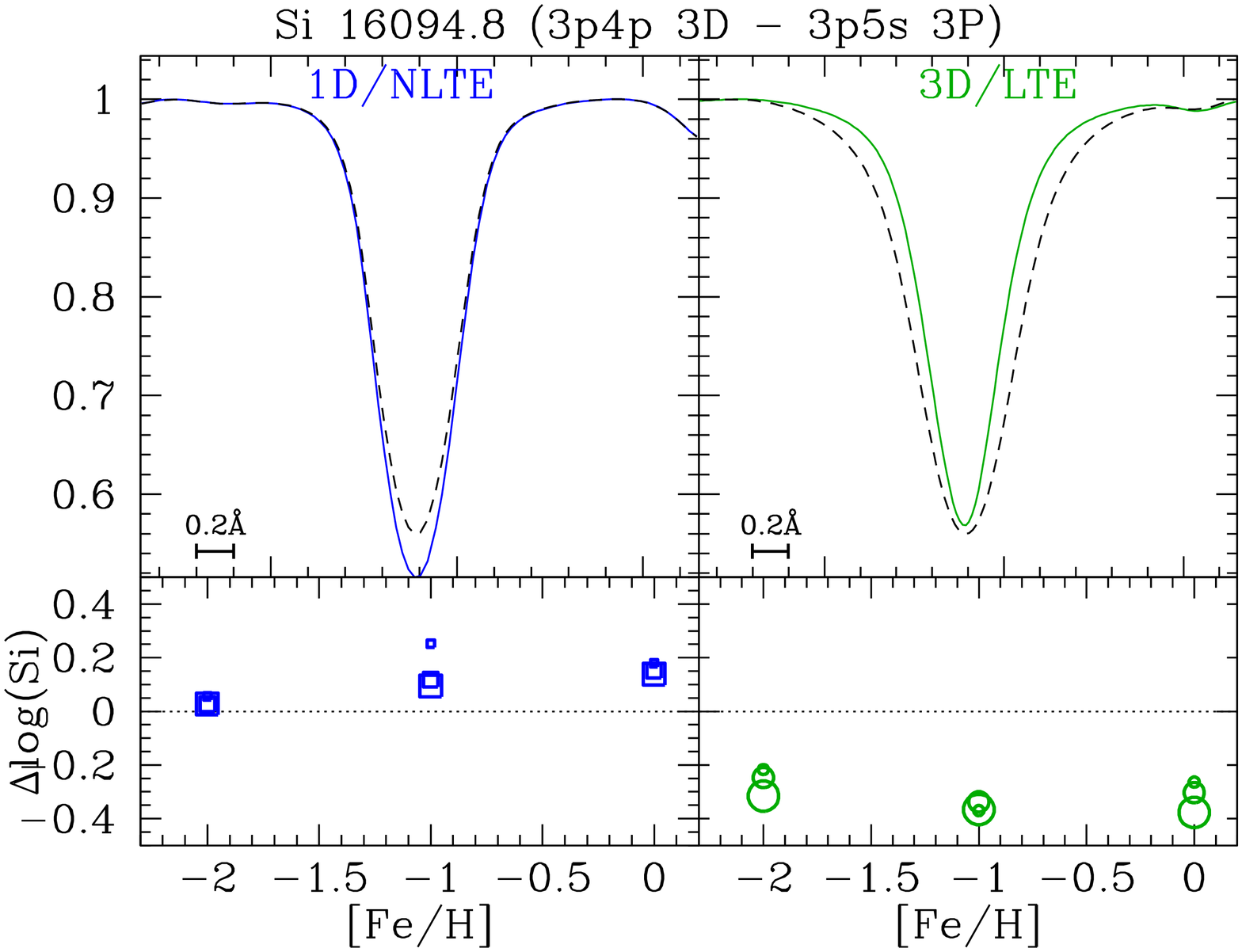}
    \end{subfigure}
        \begin{subfigure}[b]{\textwidth}
        \includegraphics[angle=-90,width=0.33\textwidth]{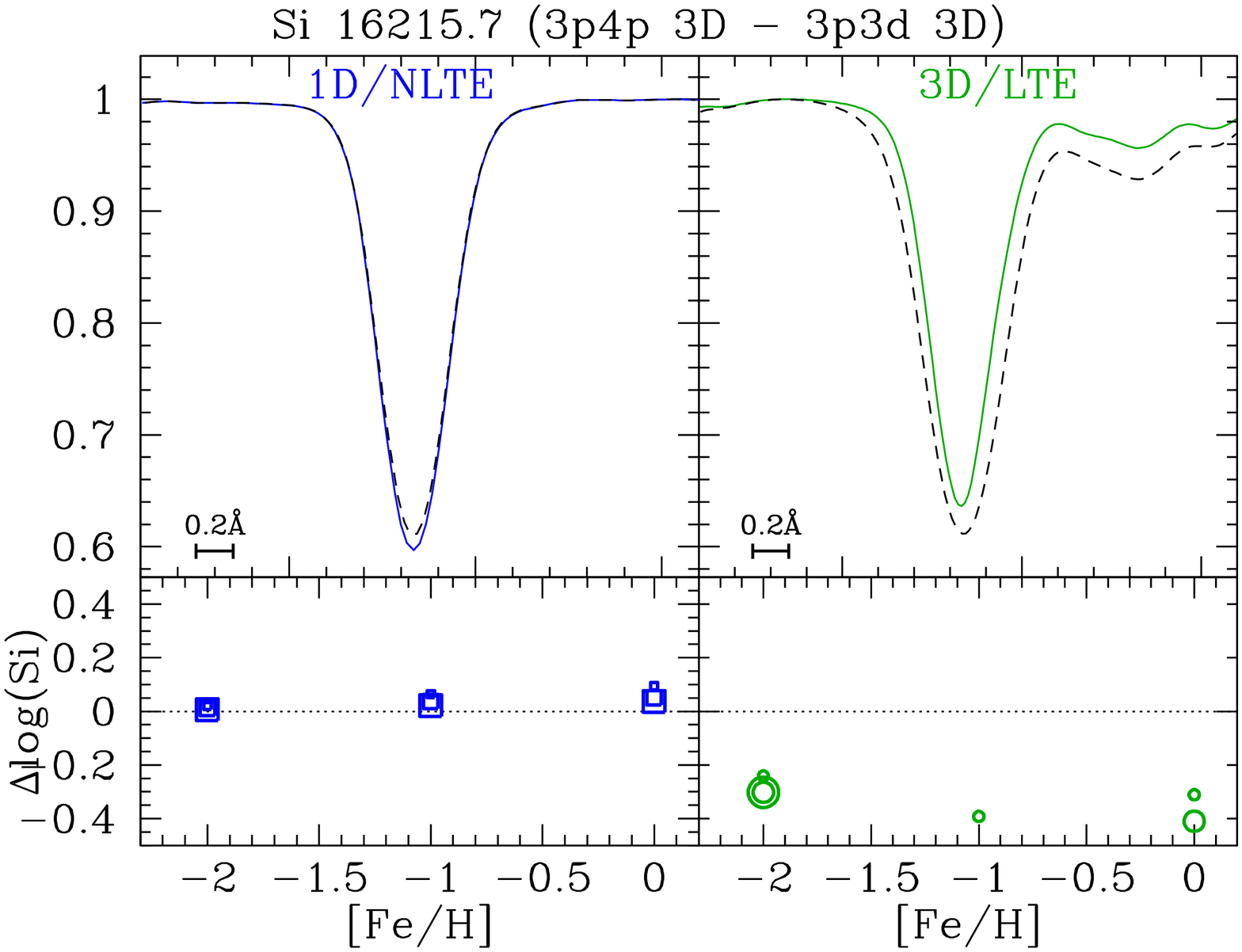}
        \includegraphics[angle=-90,width=0.33\textwidth]{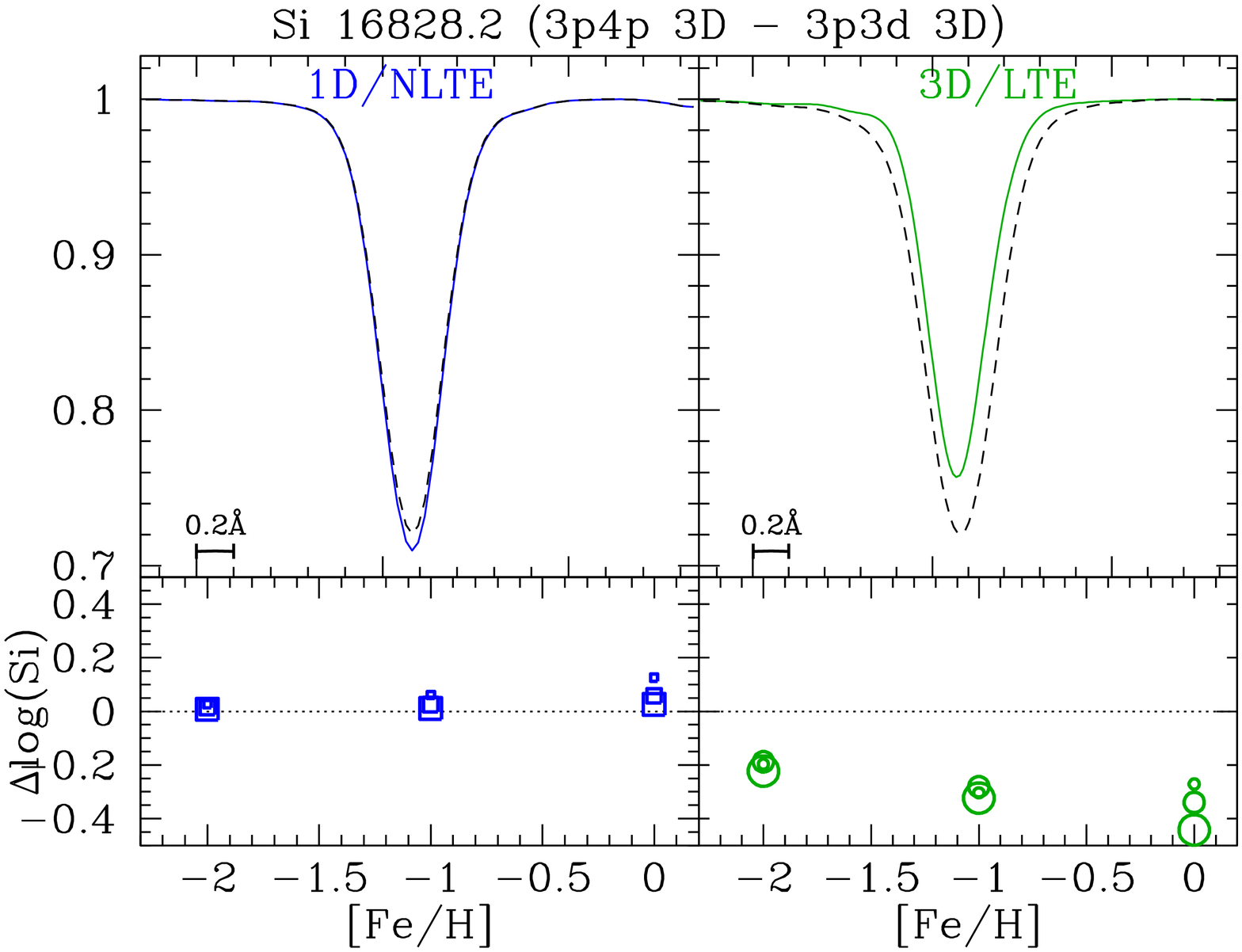}
        \includegraphics[angle=-90,width=0.33\textwidth]{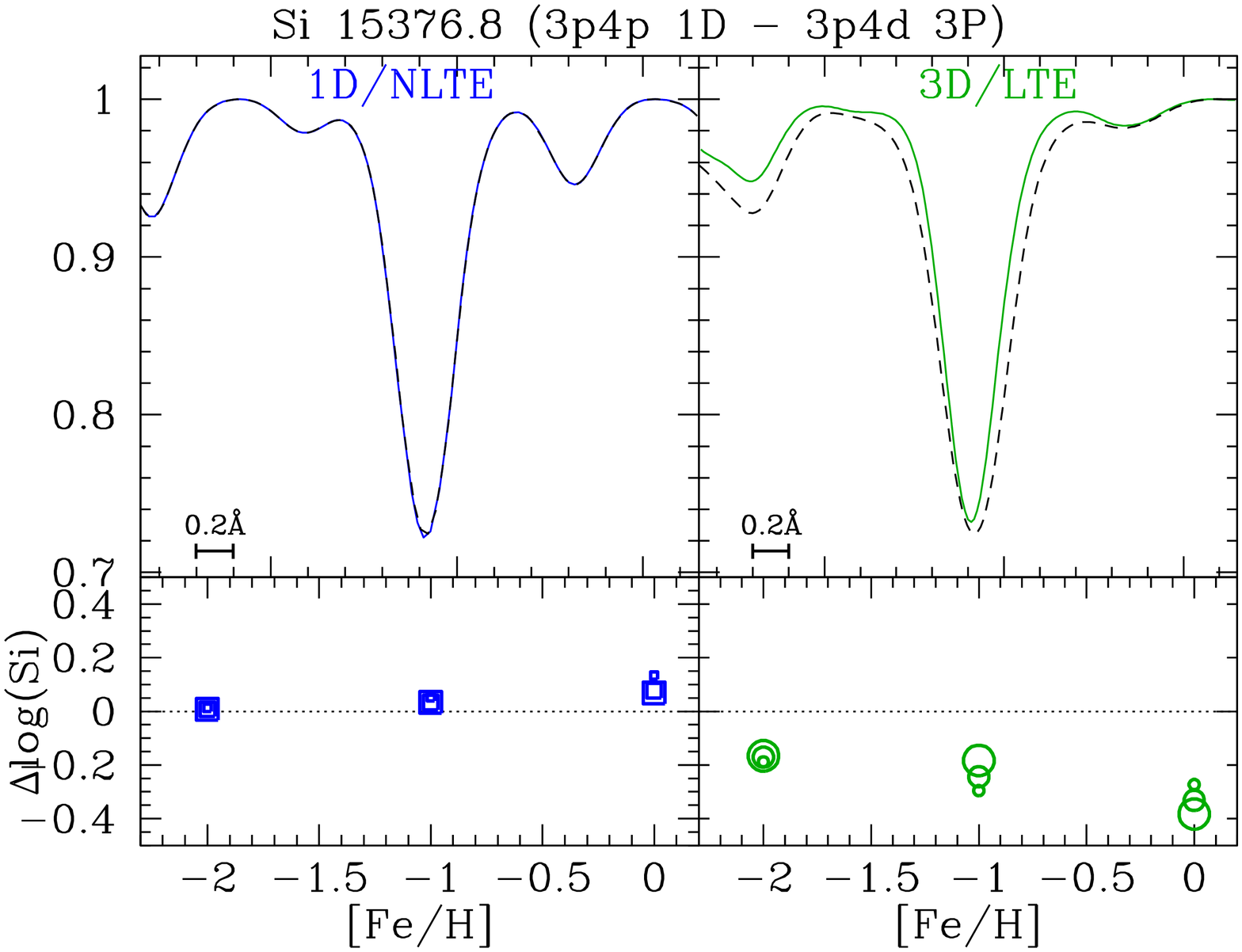}
    \end{subfigure}
\caption{\label{fig:Si_lines} Synthesis of Si lines used in this work. In each sub-panel the upper part  shows the 1D--LTE synthesis (dashed black line), the 1D--NLTE synthesis from \citet{Kovalev2018} (blue), and the 3D--LTE synthesis (green)  for a T$\rm{eff}$=4500K, $\rm \log$g=2.5, [M/H]=$-$1.0 model. The lower parts display the corresponding differences in abundances of the same line between the 1D--LTE model and the NLTE or 3D models for various temperatures and metallicities (large symbols T$\rm{eff}$=4000K, $\rm \log$g=1.5; medium-size symbols  T$\rm{eff}$=4500K, $\rm \log$g=2.5; small symbols T$\rm{eff}$=5000K, $\rm \log$g=2.5).}
\end{figure*}
 \begin{figure*}[!ht]
\centering
        \begin{subfigure}[b]{\textwidth}
        \includegraphics[angle=-90,width=0.33\textwidth]{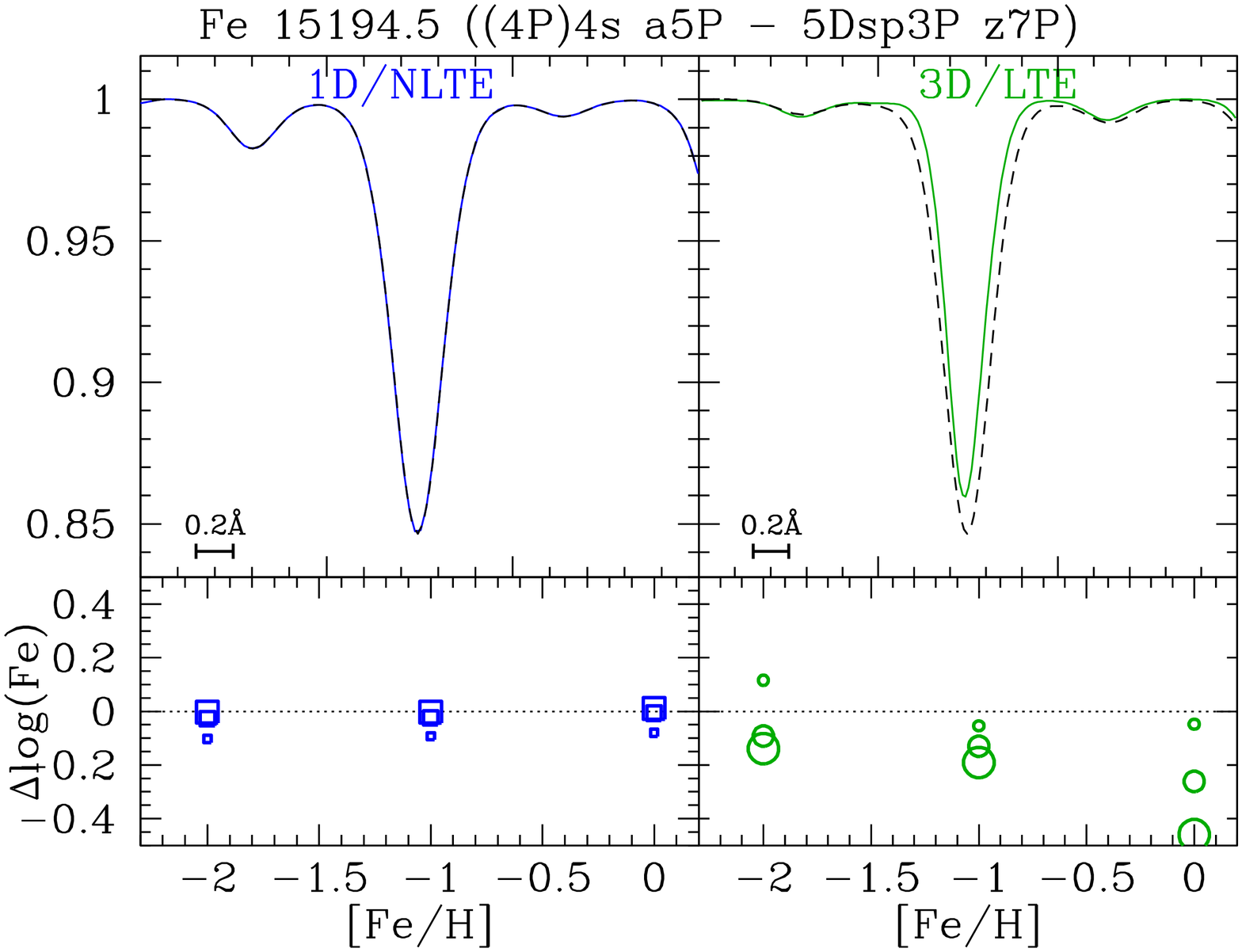}
        \includegraphics[angle=-90,width=0.33\textwidth]{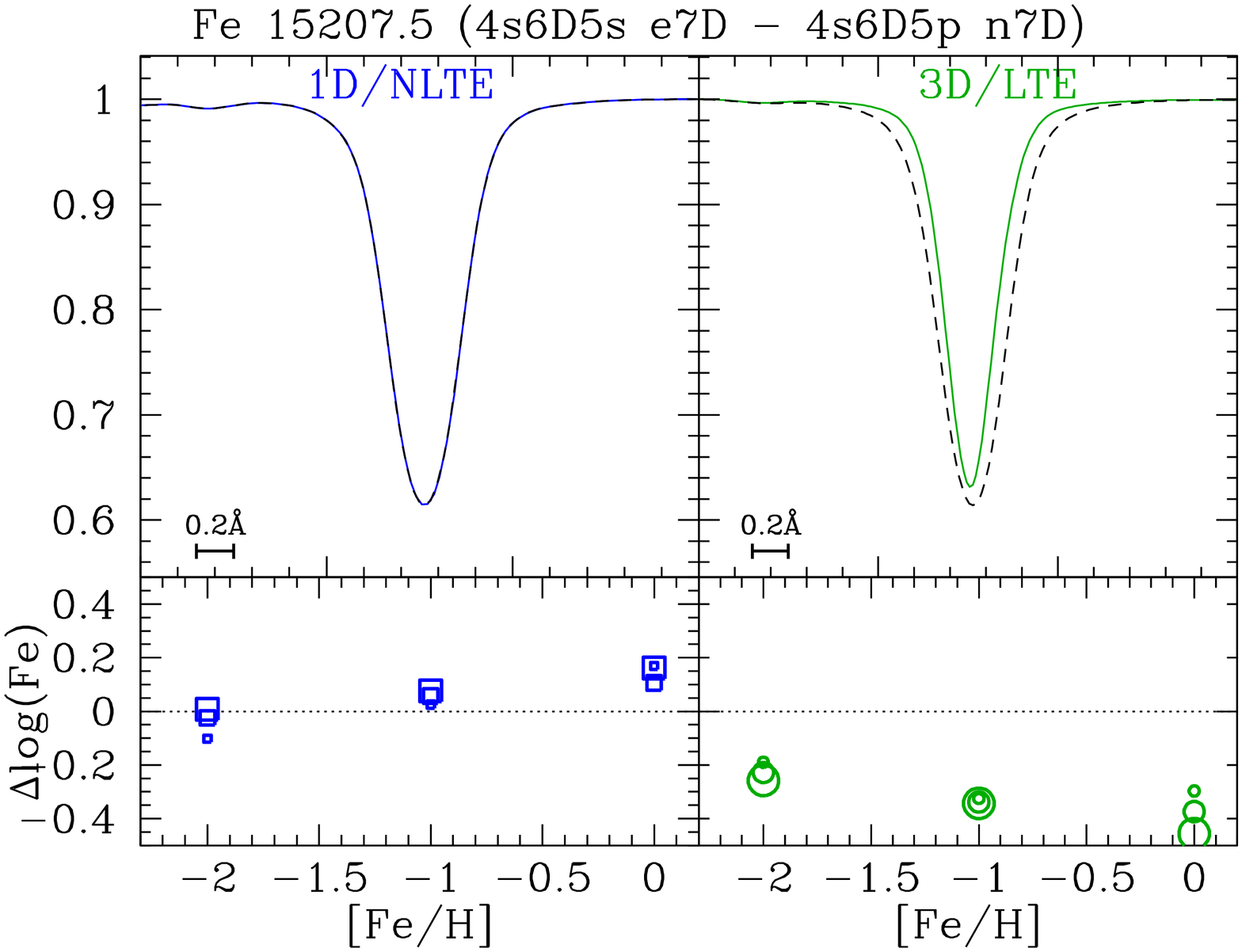}
        \includegraphics[angle=-90,width=0.33\textwidth]{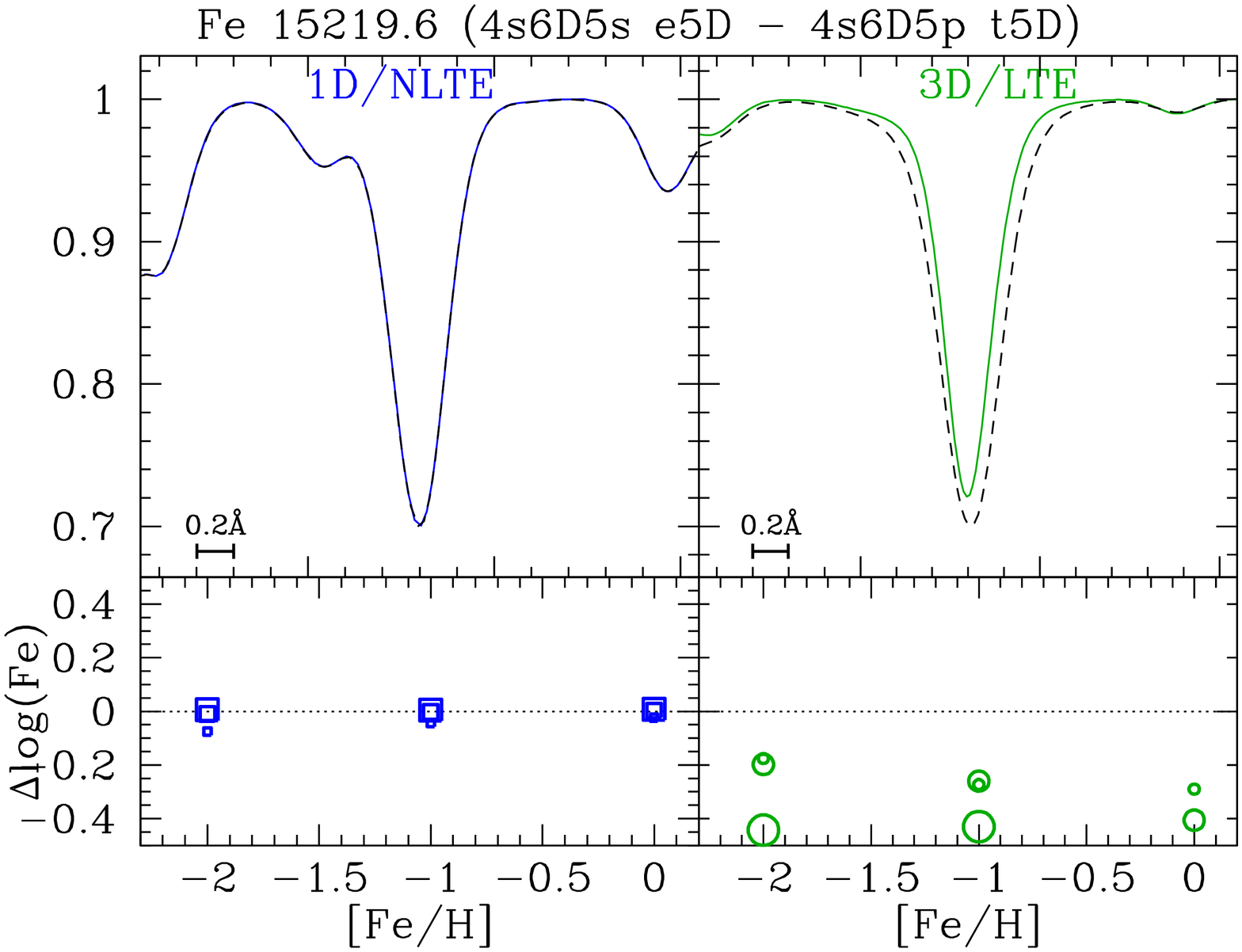}
    \end{subfigure}
        \begin{subfigure}[b]{\textwidth}
        \includegraphics[angle=-90,width=0.33\textwidth]{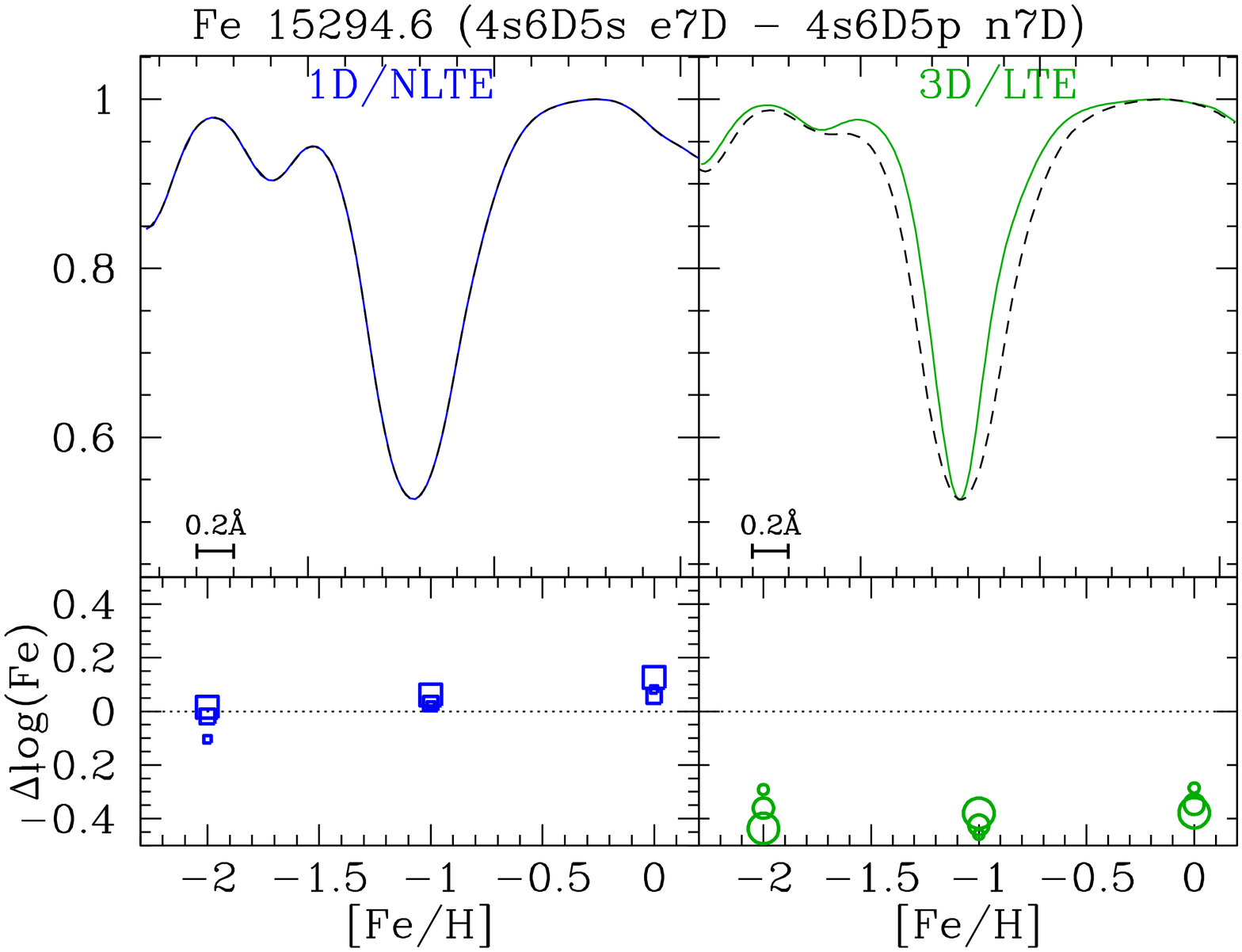}
        \includegraphics[angle=-90,width=0.33\textwidth]{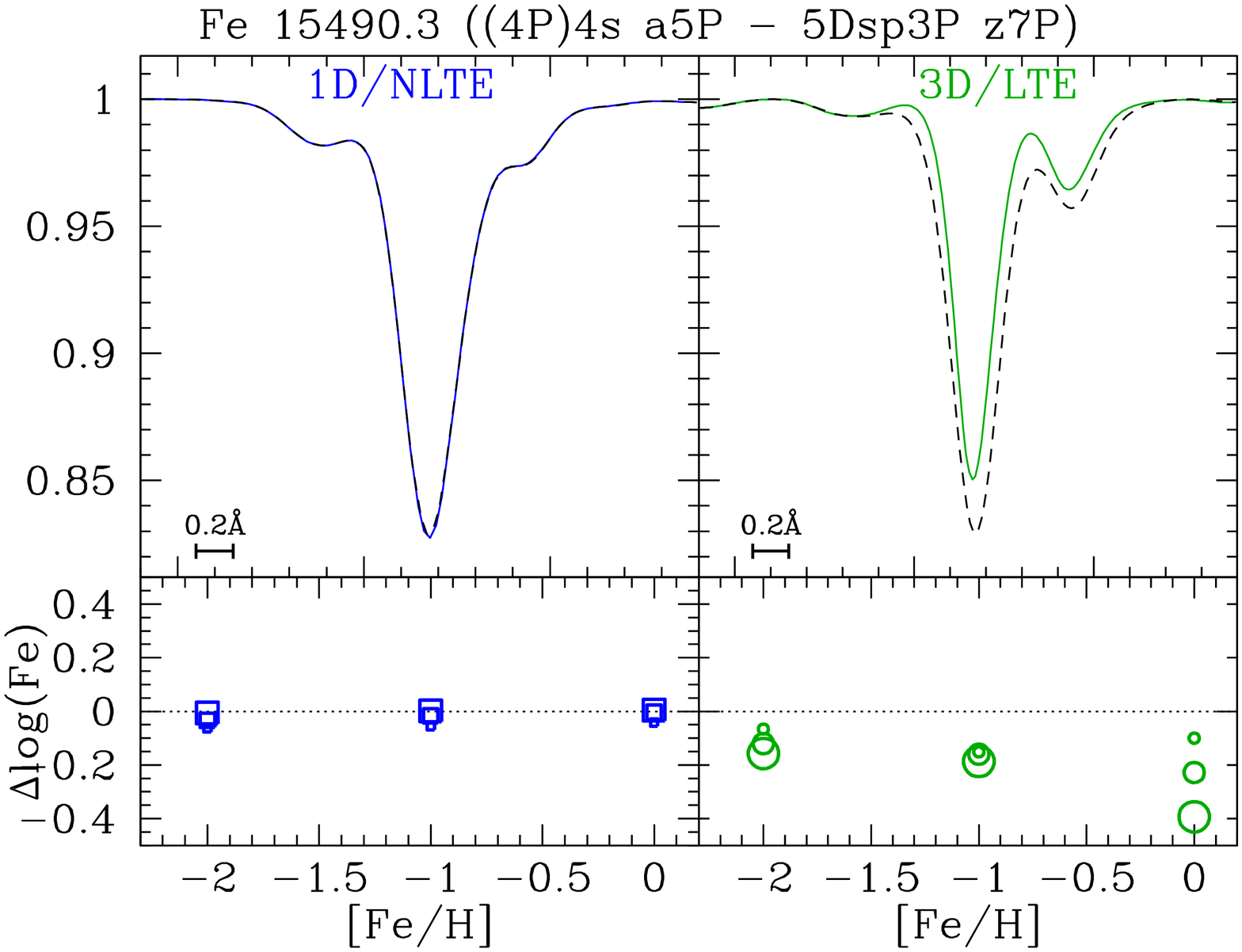}
        \includegraphics[angle=-90,width=0.33\textwidth]{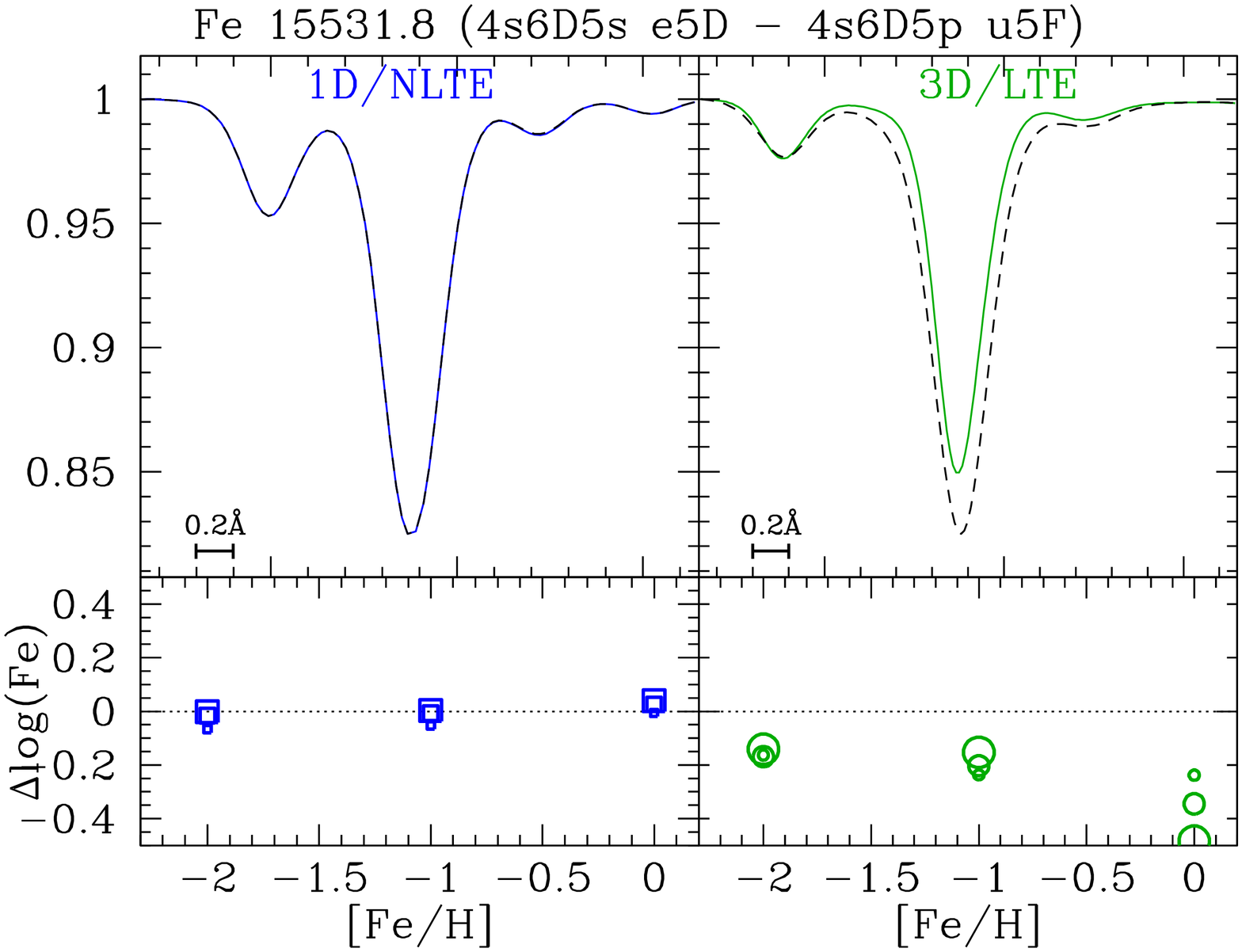}
    \end{subfigure}
        \begin{subfigure}[b]{\textwidth}
        \includegraphics[angle=-90,width=0.33\textwidth]{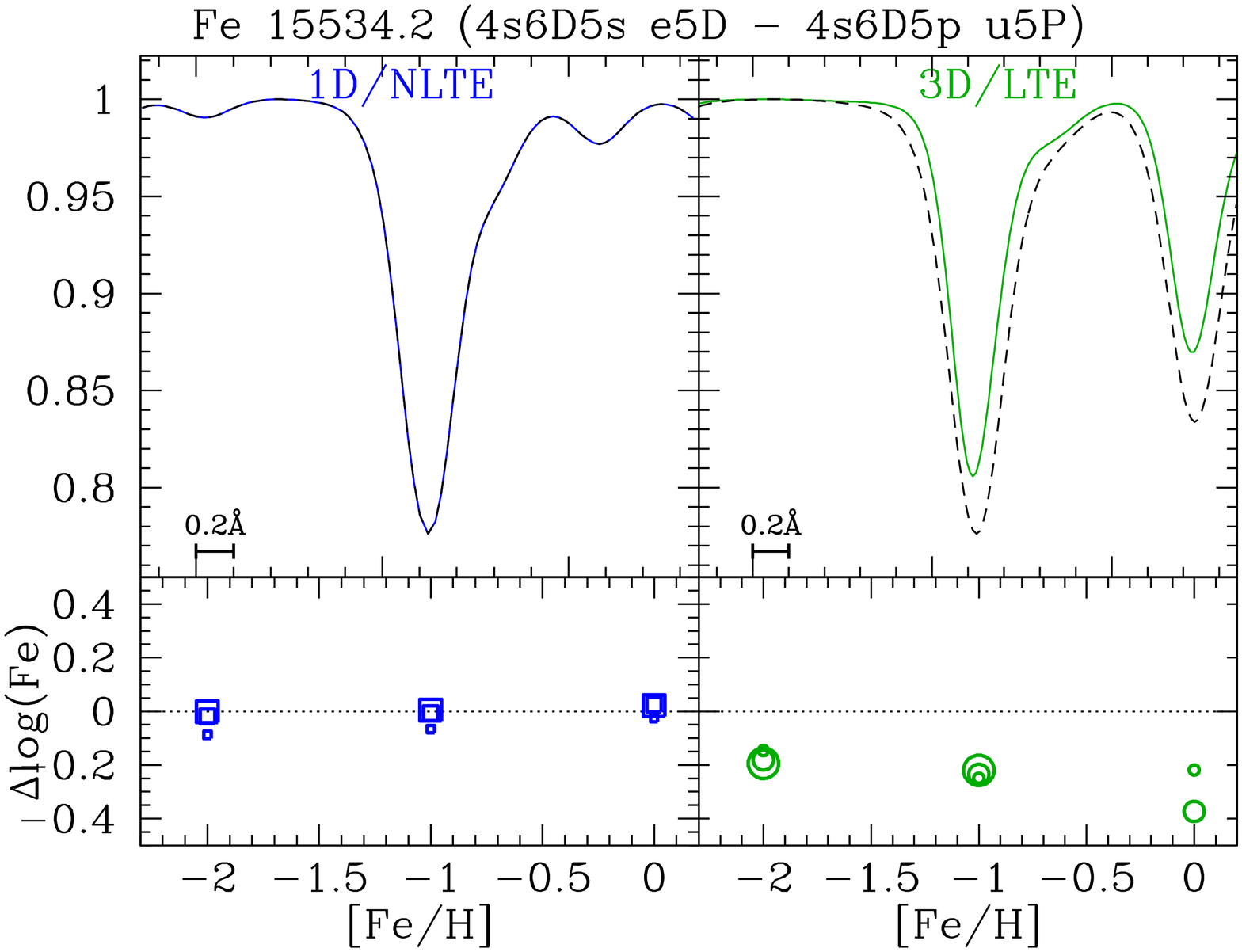}
        \includegraphics[angle=-90,width=0.33\textwidth]{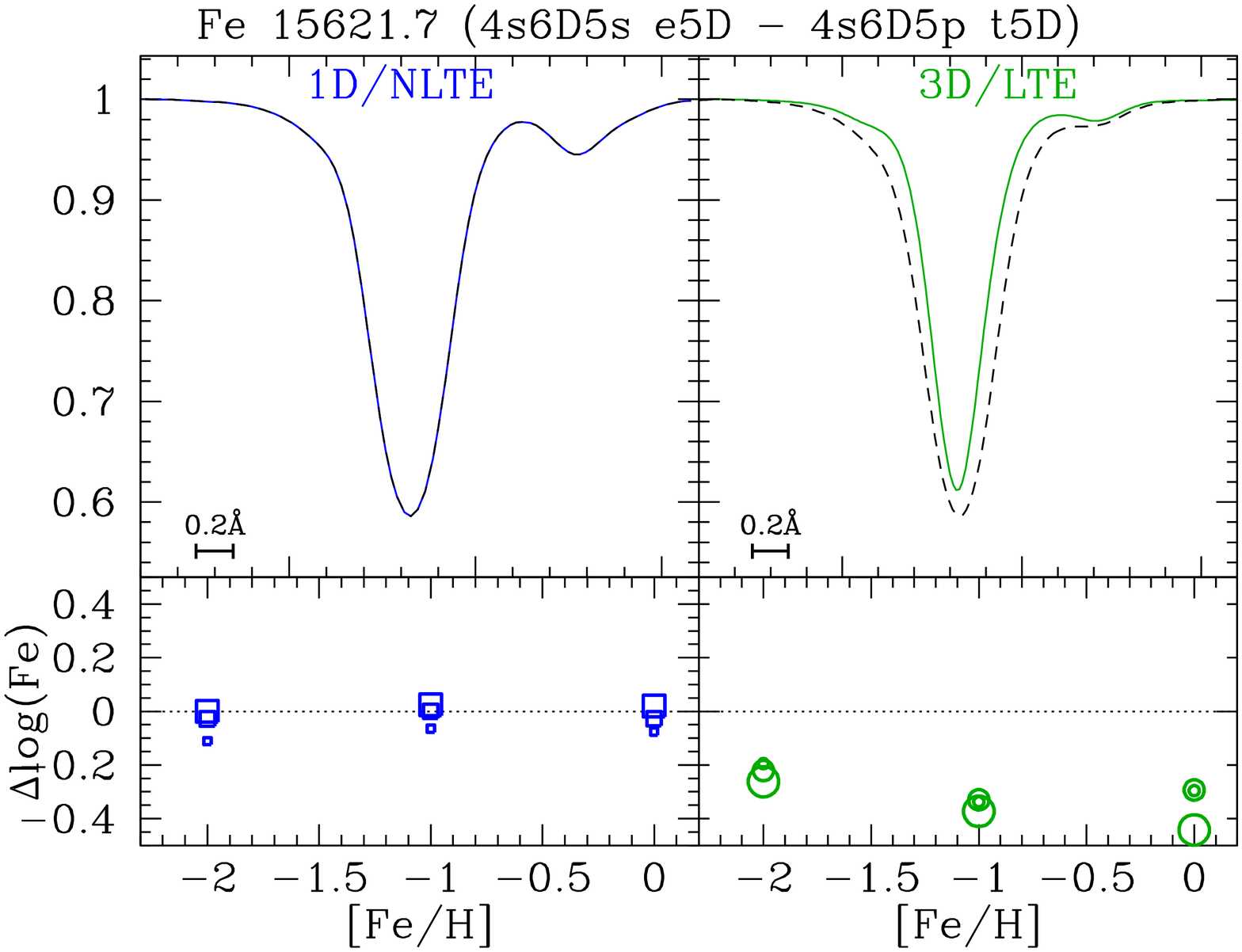}
        \includegraphics[angle=-90,width=0.33\textwidth]{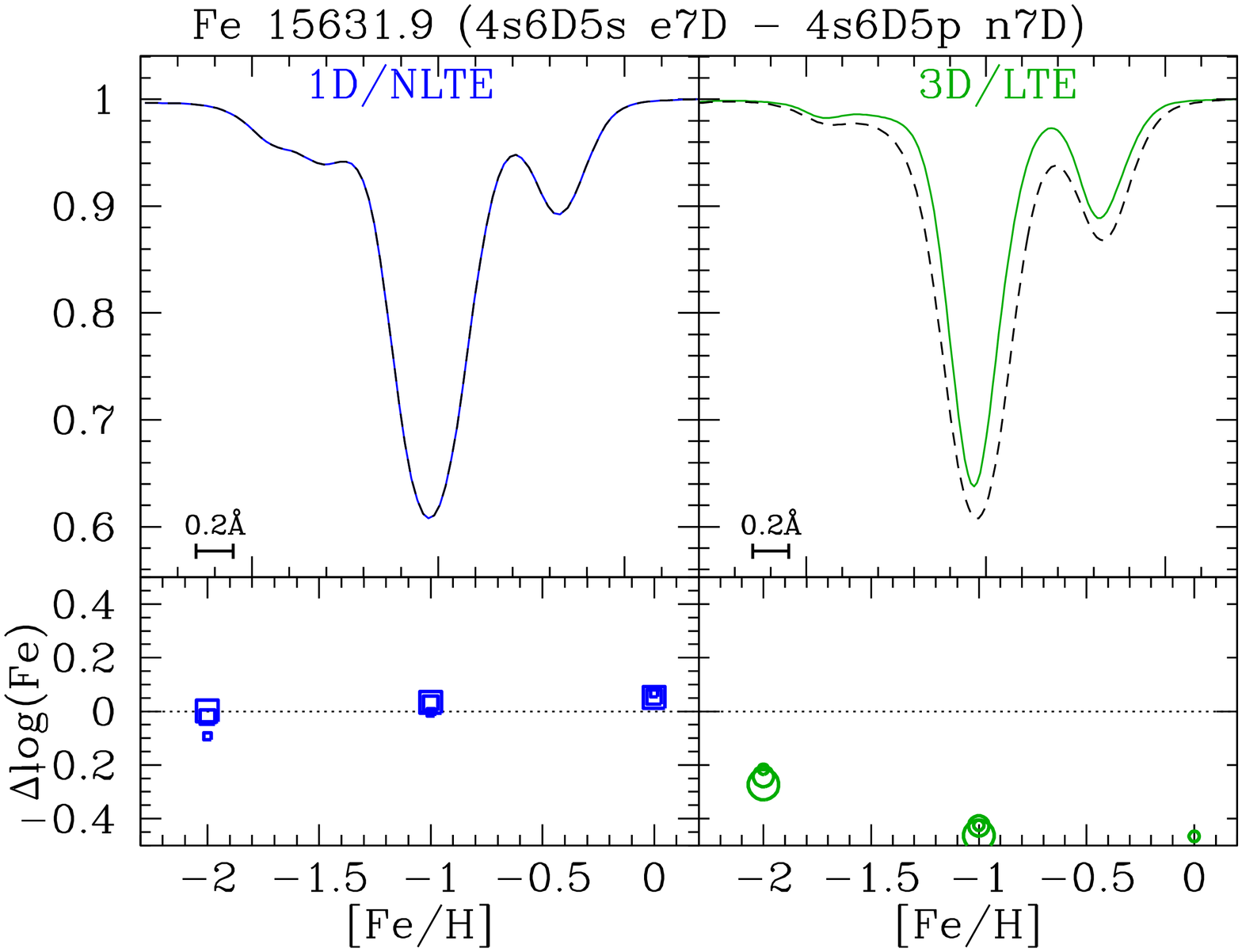}
    \end{subfigure}
        \begin{subfigure}[b]{\textwidth}
        \includegraphics[angle=-90,width=0.33\textwidth]{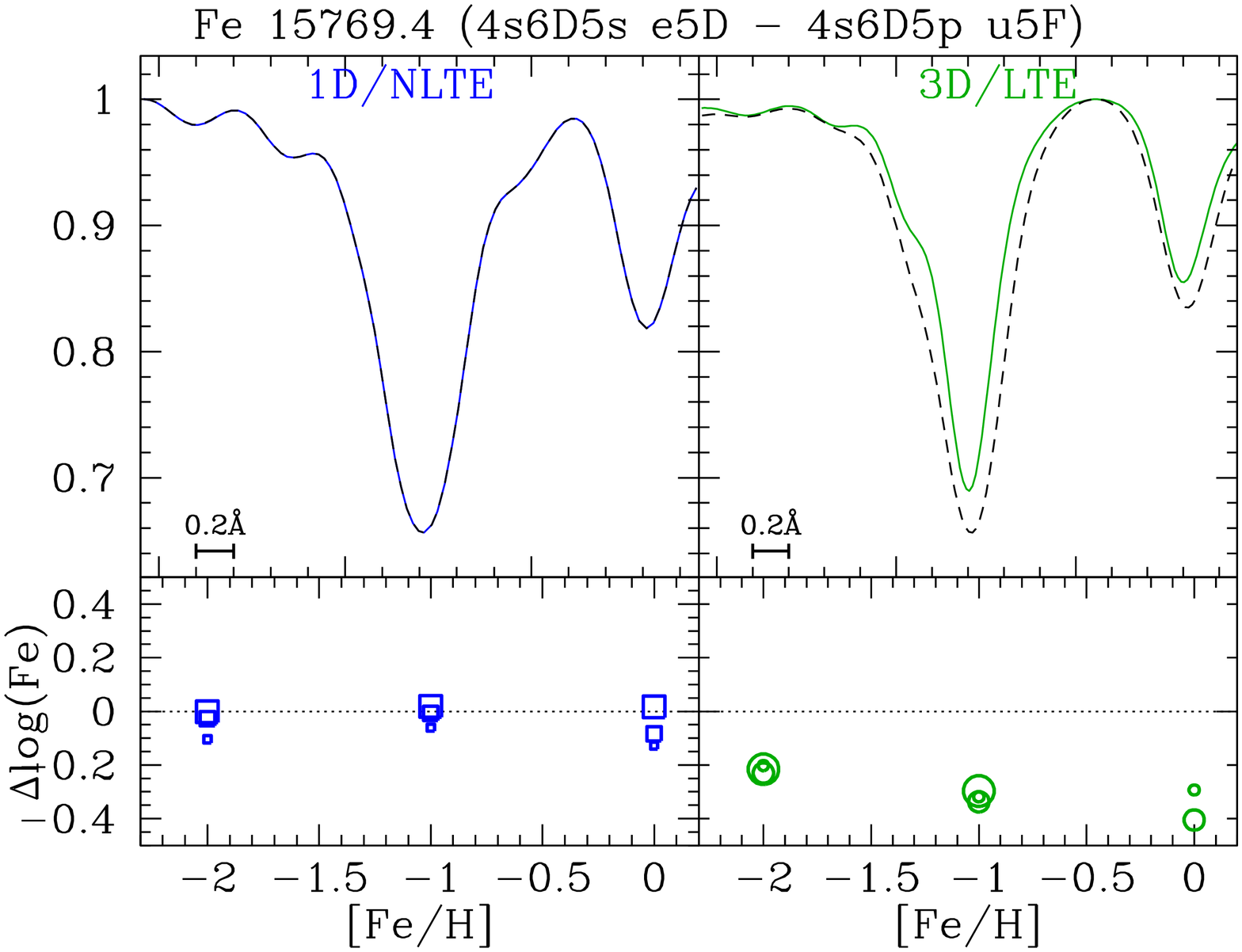}
        \includegraphics[angle=-90,width=0.33\textwidth]{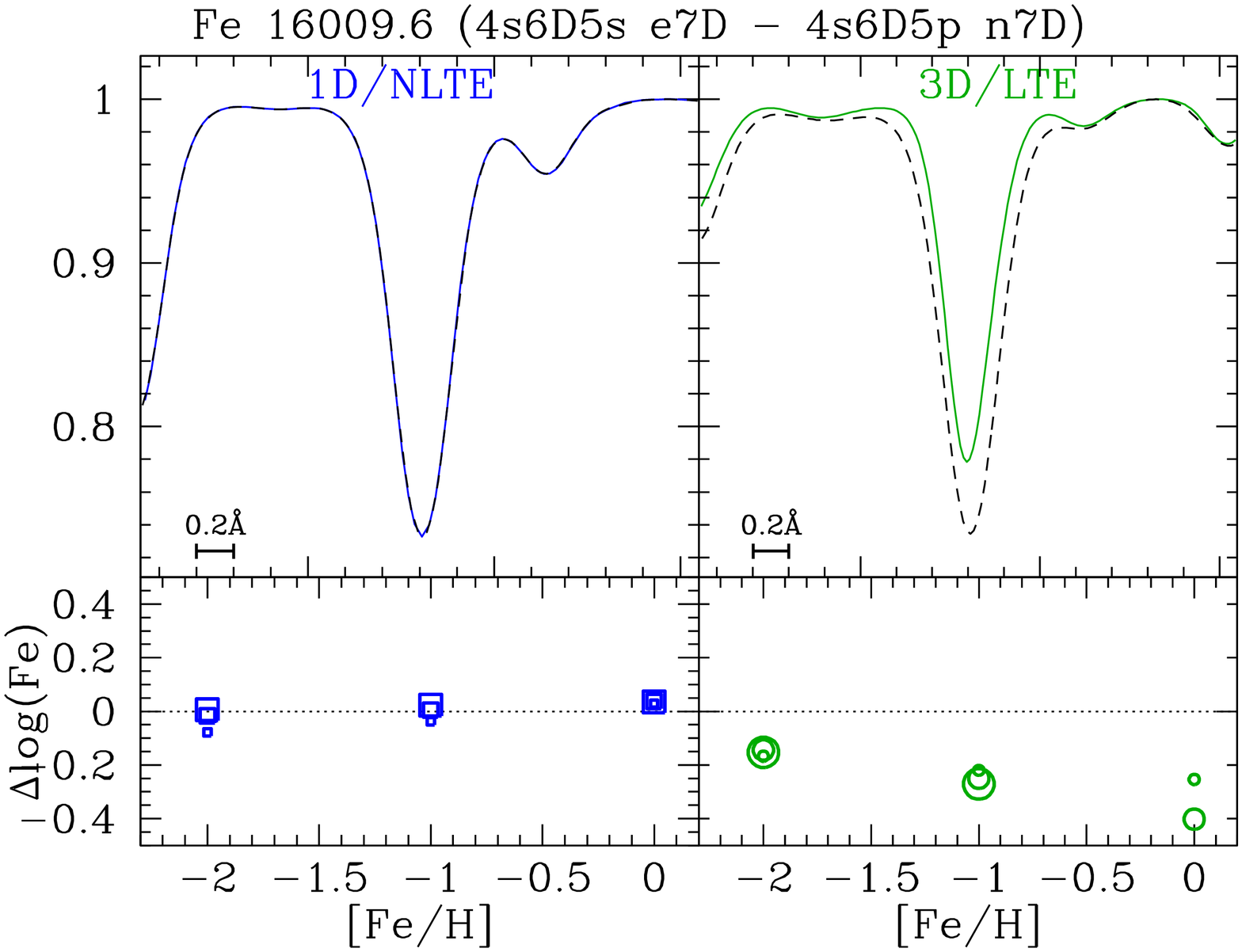}
        \includegraphics[angle=-90,width=0.33\textwidth]{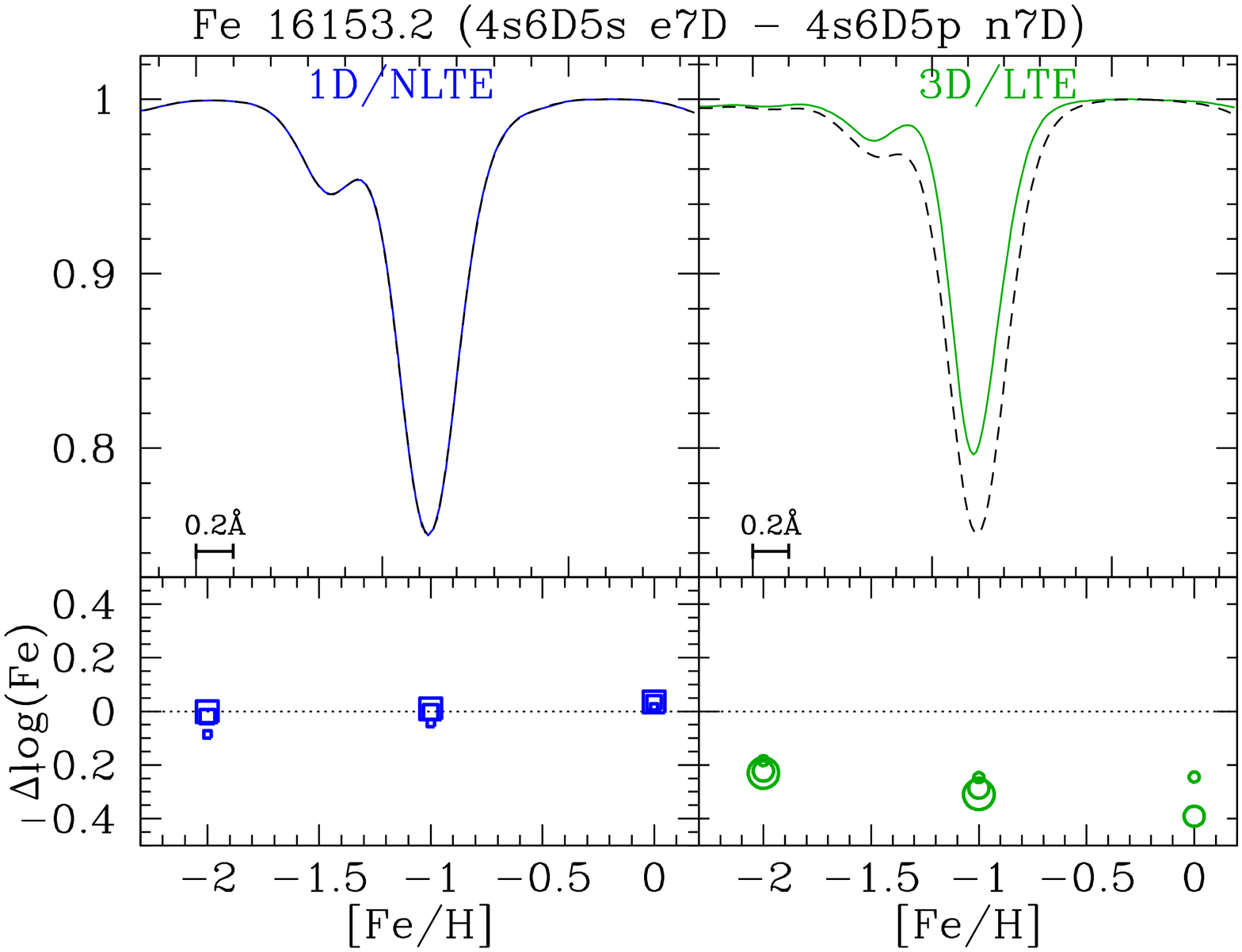}
    \end{subfigure}
\caption{\label{fig:Fe_lines} Synthesis of Fe lines used in this work. In each sub-panel the upper part shows the 1D--LTE synthesis (dashed black line), the 3D--LTE synthesis (green), and for some sub-panels the 1D--NLTE synthesis from \citet{Kovalev2018} (blue)  for a T$\rm{eff}$=4500K, $\rm \log$g=2.5, [M/H]=$-$1.0 model. The lower parts display the corresponding differences in abundances of the same line between the 1D--LTE model and the 3D model for various temperatures and metallicities (large symbols T$\rm{eff}$=4000K, $\rm \log$g=1.5; medium-size symbols  T$\rm{eff}$=4500K, $\rm \log$g=2.5; small symbols T$\rm{eff}$=5000K, $\rm \log$g=2.5).}
\end{figure*}

 \begin{figure*}[!ht]
\centering
        \begin{subfigure}[b]{\textwidth}
        \includegraphics[angle=-90,width=0.5\textwidth]{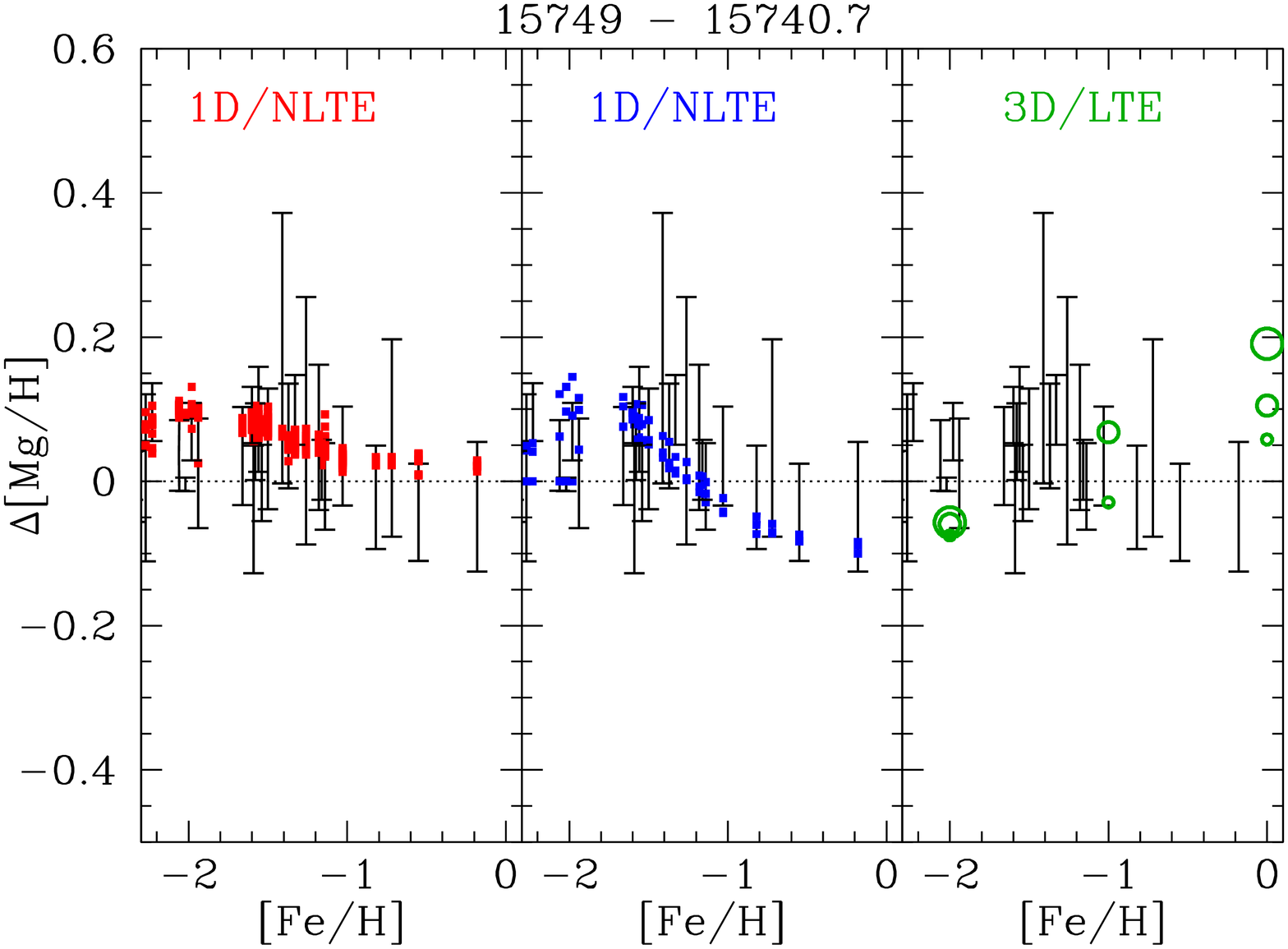}
        \includegraphics[angle=-90,width=0.5\textwidth]{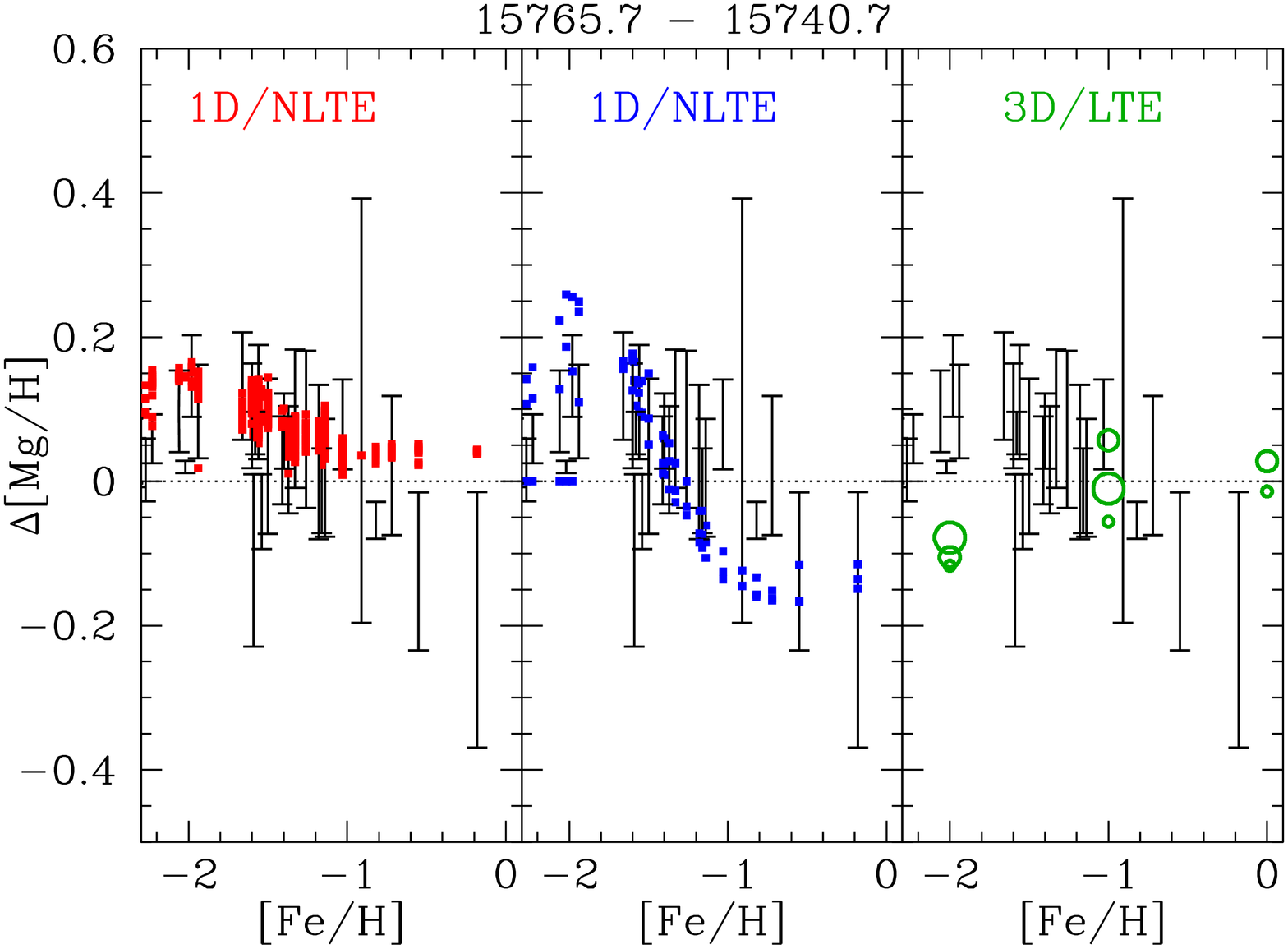}
    \end{subfigure}
        \begin{subfigure}[b]{\textwidth}
        \includegraphics[angle=-90,width=0.5\textwidth]{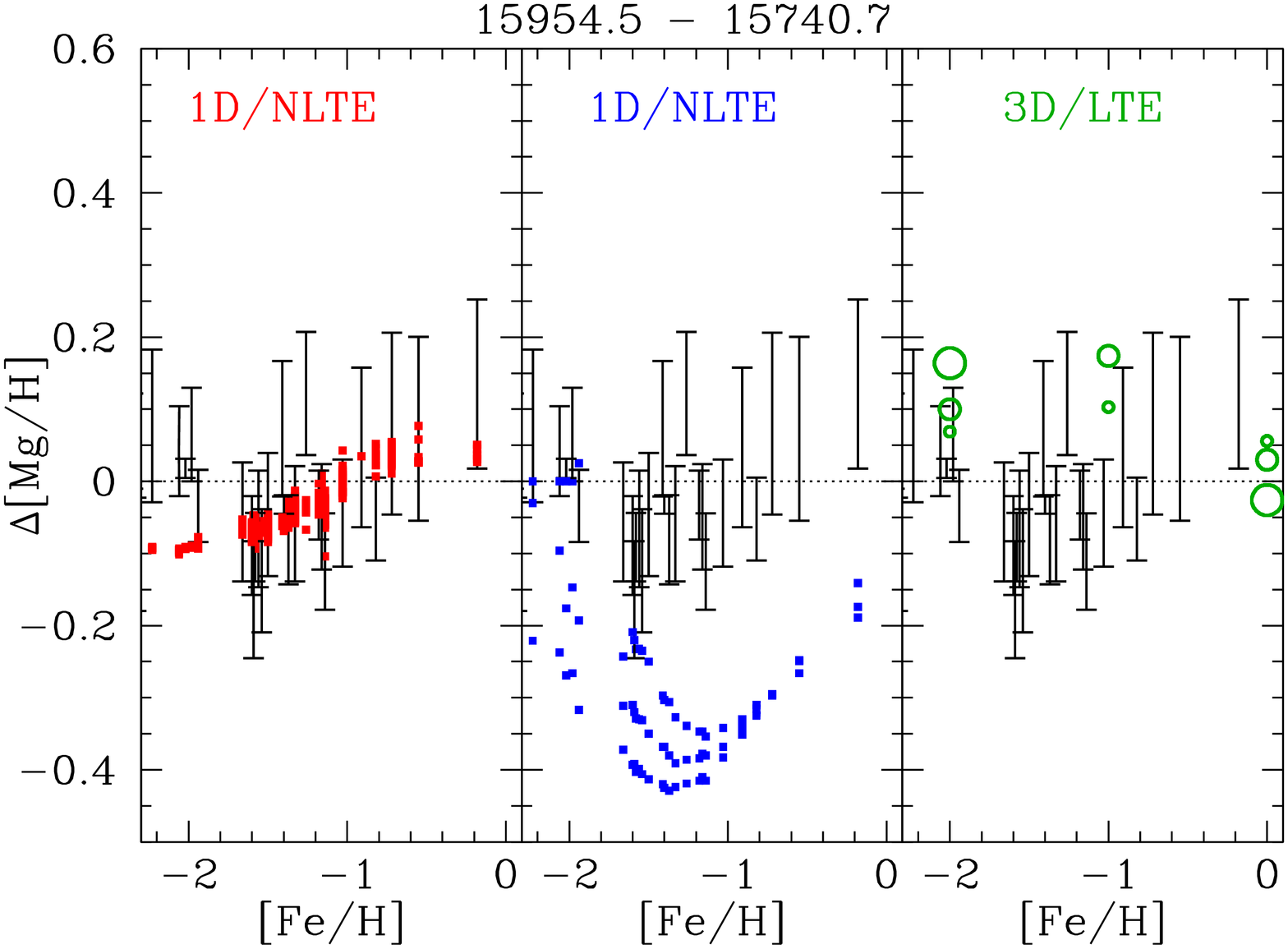}
        \includegraphics[angle=-90,width=0.5\textwidth]{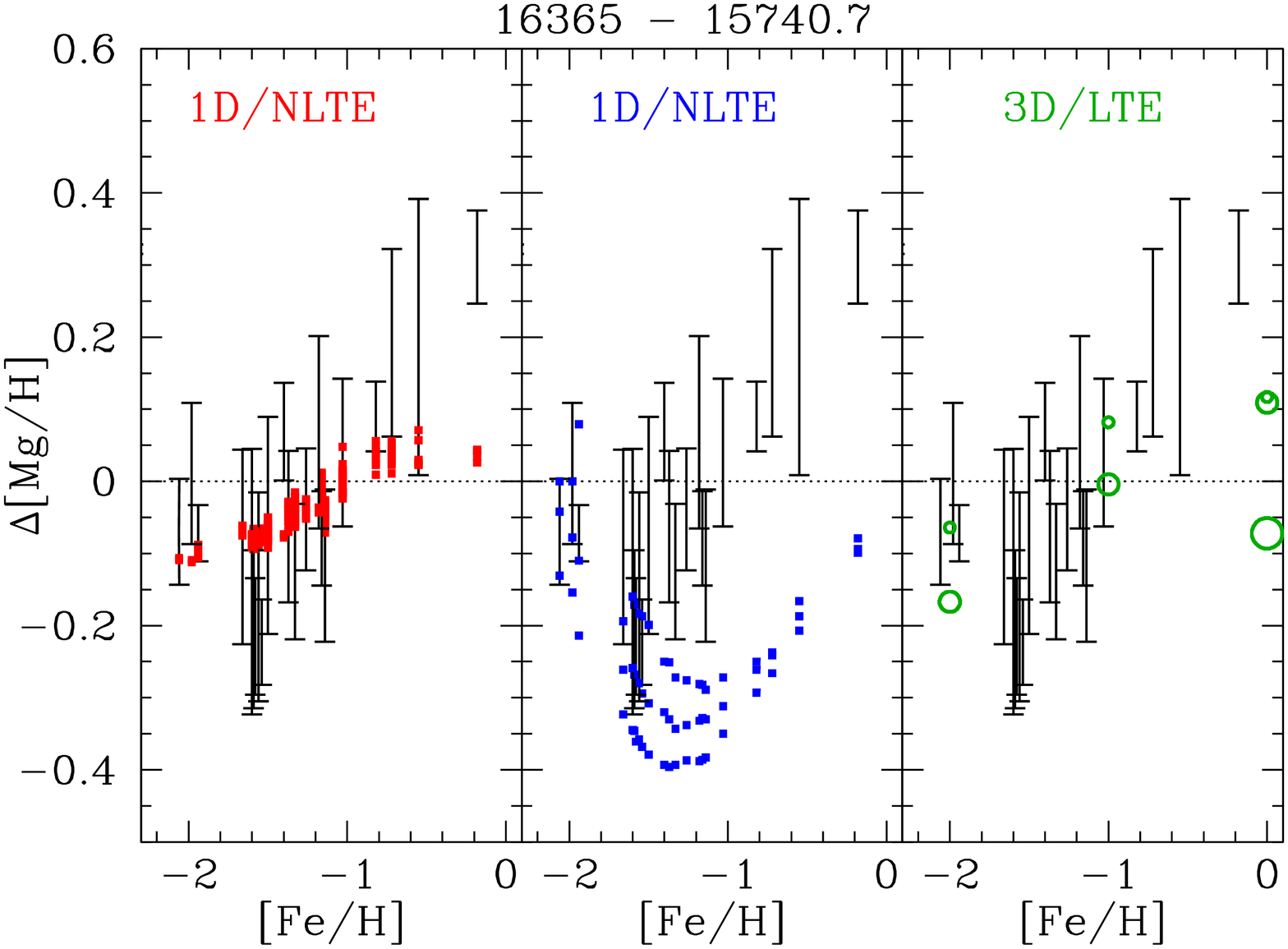}
    \end{subfigure}
\caption{\label{fig:Mg_diff} Differences in abundance obtained between two Mg lines. The black error bars show the star-to-star median of the 1D--LTE analysis of the APOGEE data. The red squares in the left panels show the prediction obtained with the 1D--NLTE models from this work. The blue squares in the middle panels show the prediction obtained with the 1D--NLTE models of \citet{Kovalev2018}. The green circles in the right panels show the prediction obtained by the 3D--LTE models of \citet{Ludwig2019}.}
\end{figure*}

 \begin{figure*}[!ht]
\centering
        \begin{subfigure}[b]{\textwidth}
        \includegraphics[angle=-90,width=0.5\textwidth]{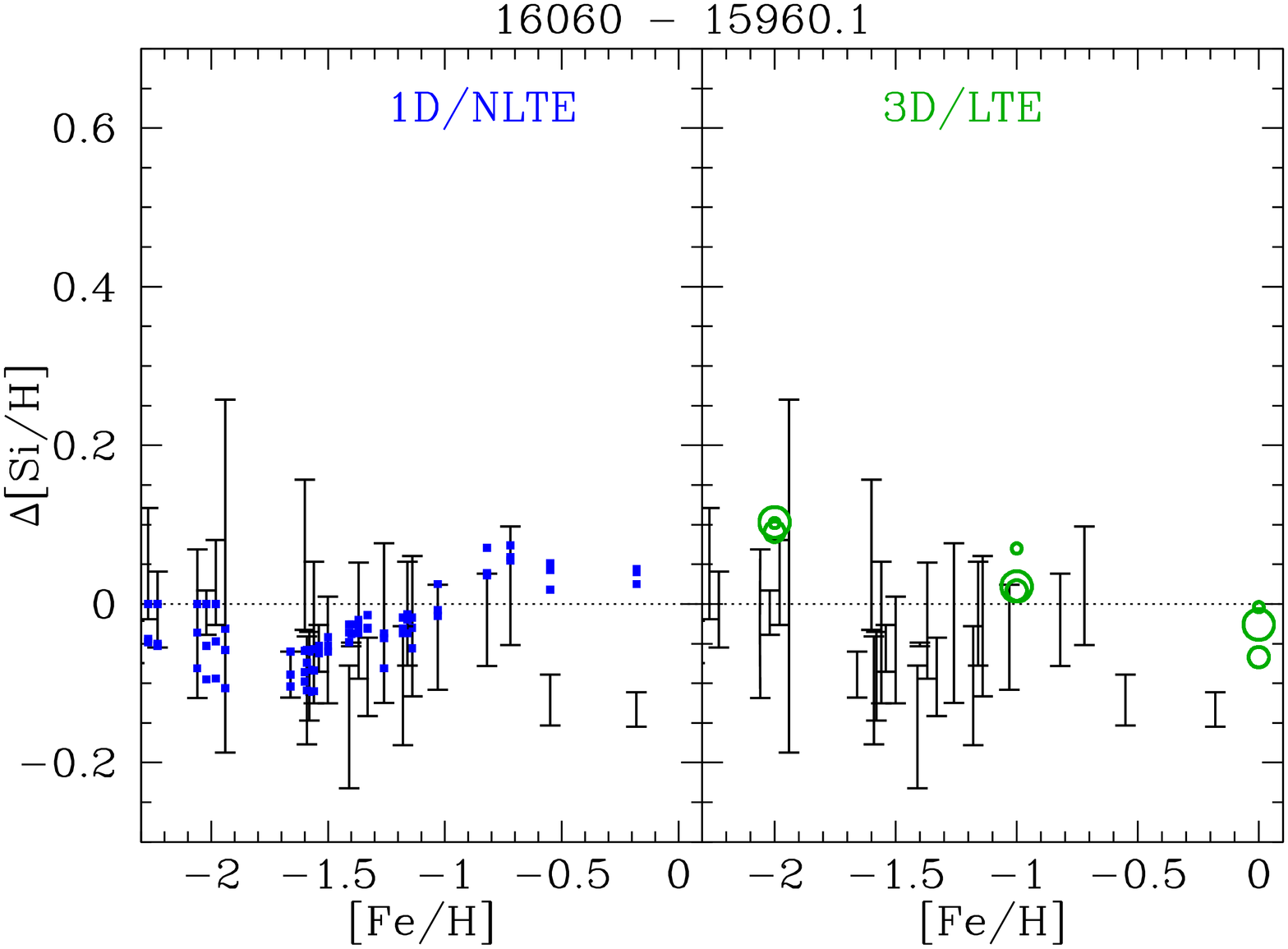}
        \includegraphics[angle=-90,width=0.5\textwidth]{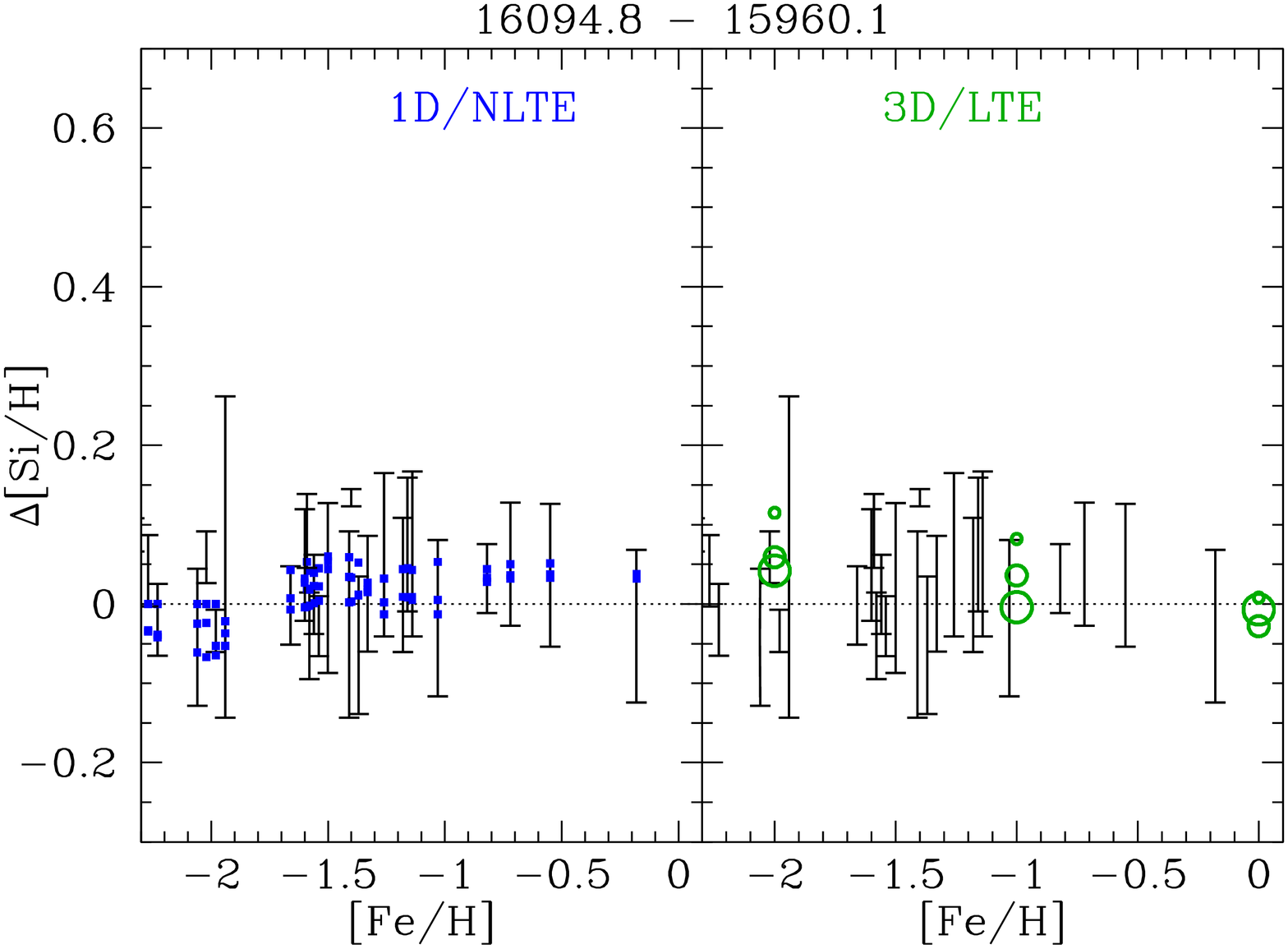}
    \end{subfigure}
        \begin{subfigure}[b]{\textwidth}
    \centering
        \includegraphics[angle=-90,width=0.33\textwidth]{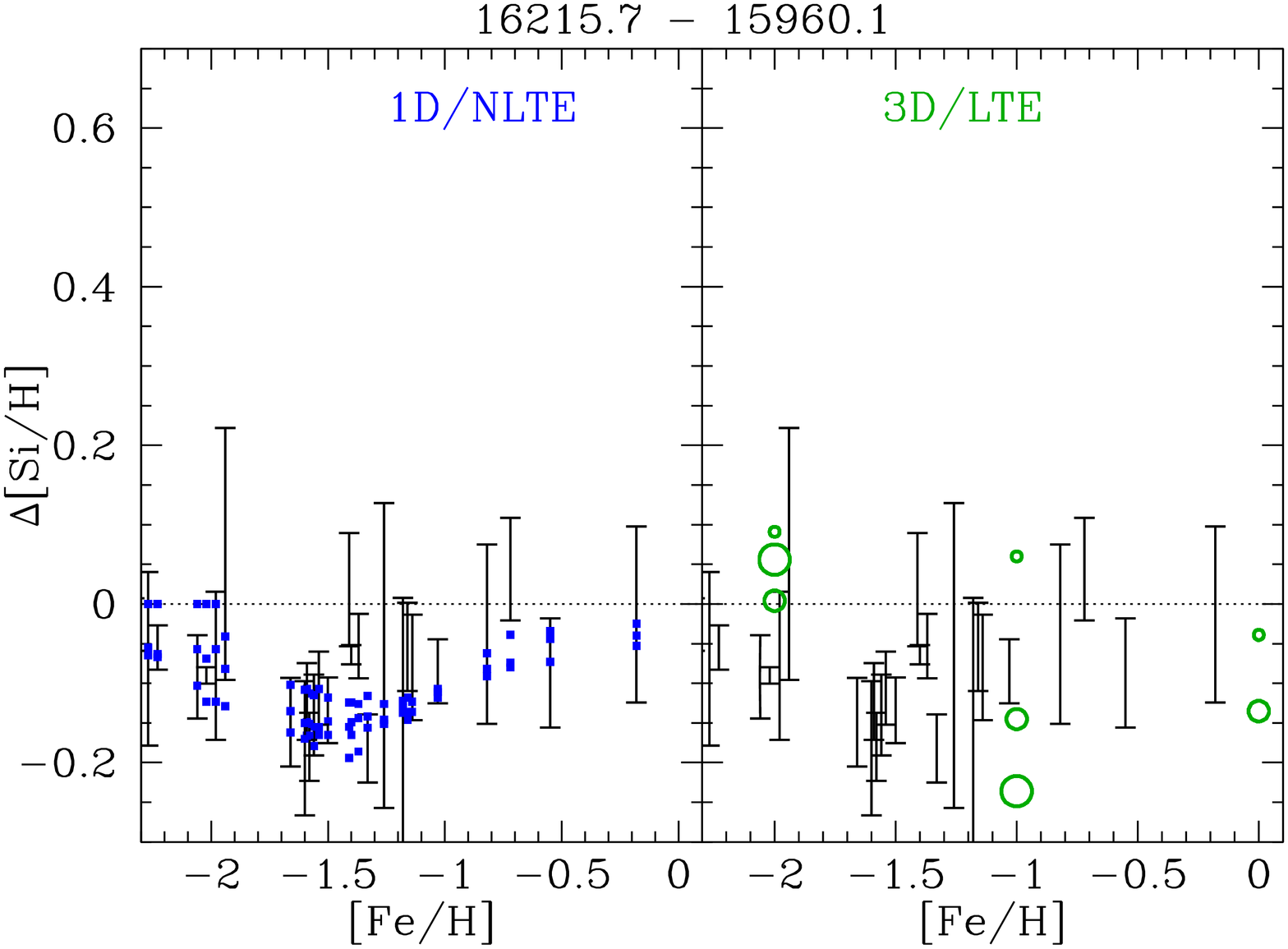}
        \includegraphics[angle=-90,width=0.33\textwidth]{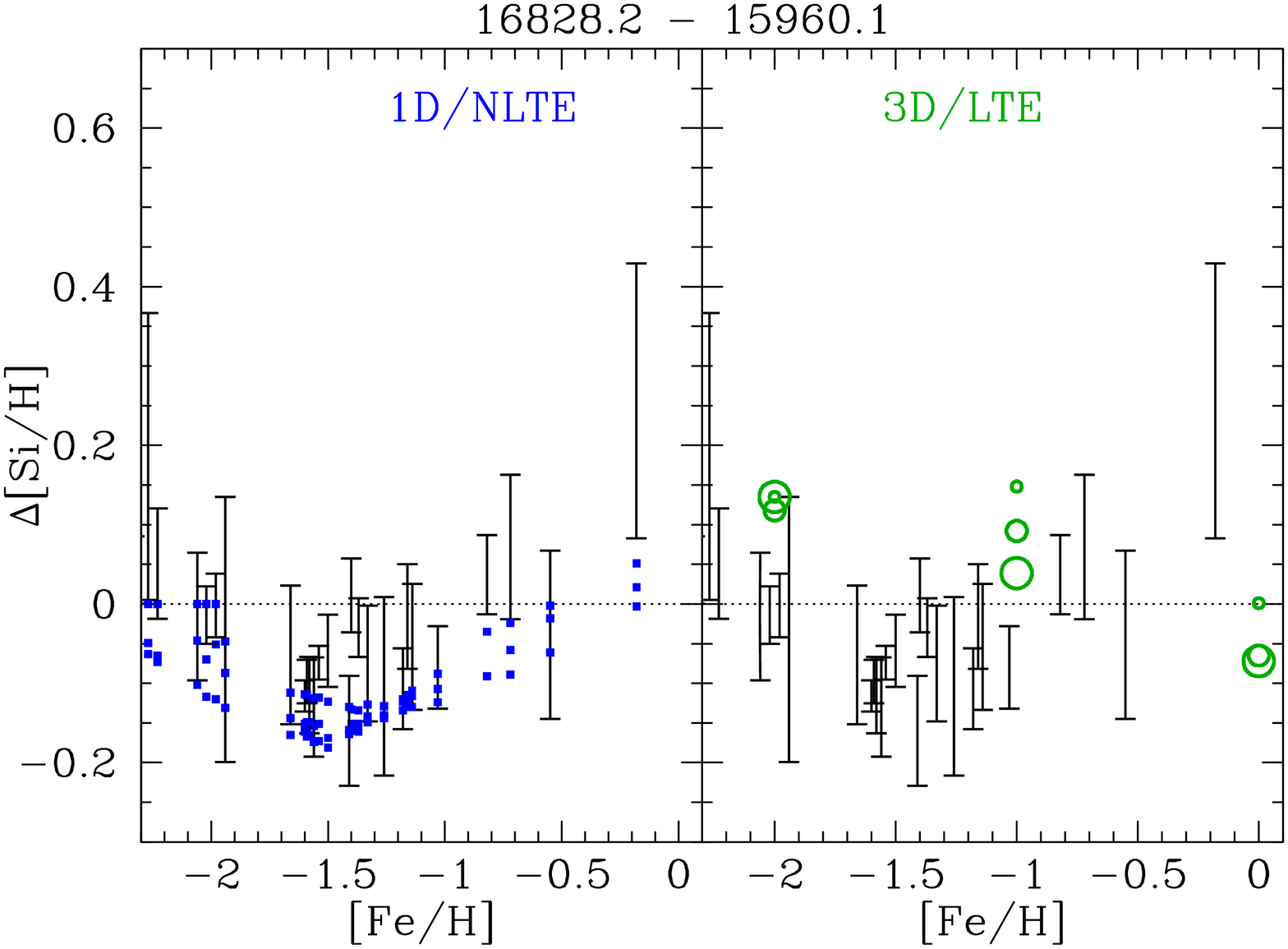}
        \includegraphics[angle=-90,width=0.33\textwidth]{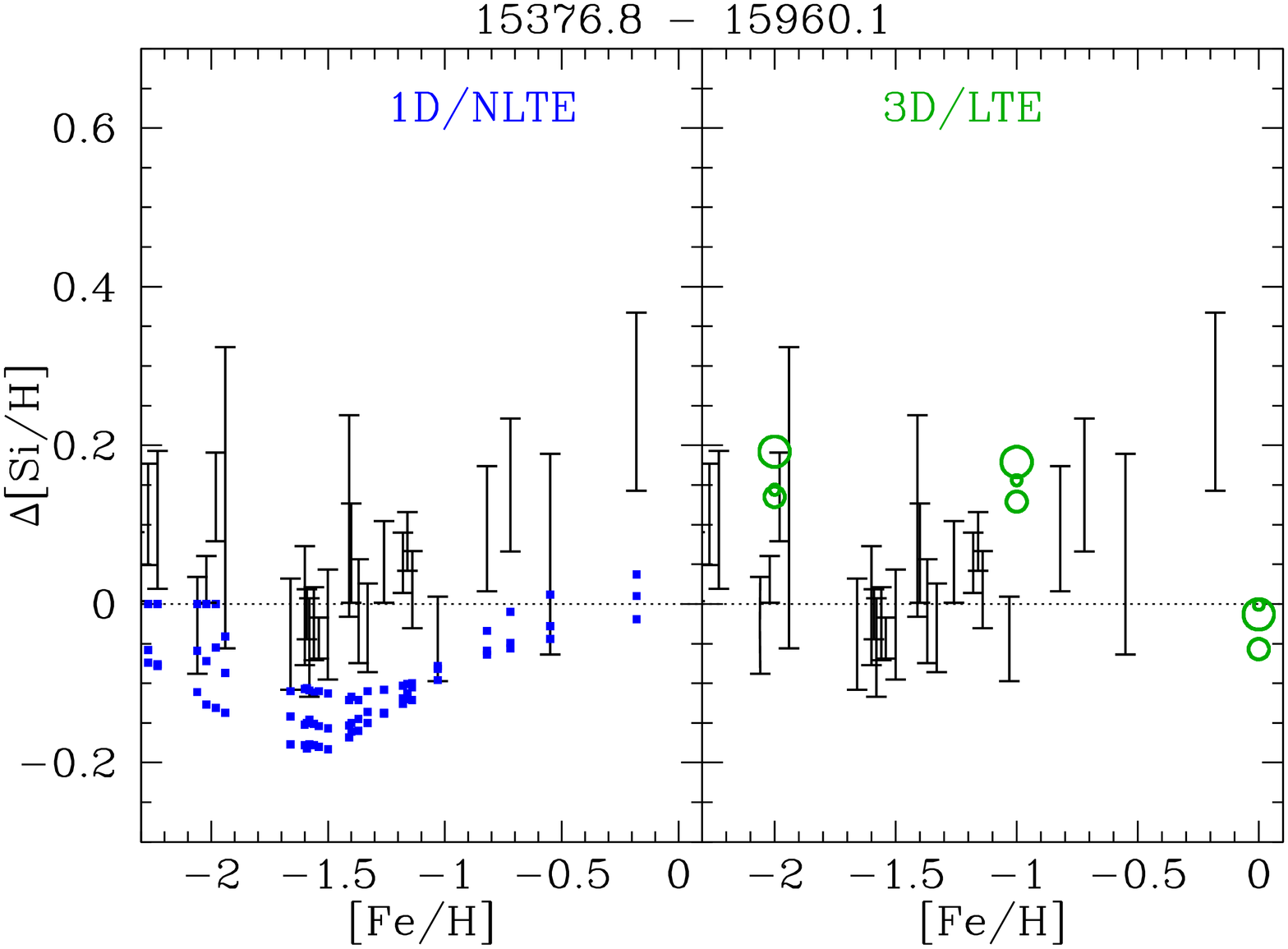}
    \end{subfigure}
\caption{\label{fig:Si_diff} Differences in abundance obtained between two Si lines. The black error bars show the star-to-star median of the 1D--LTE analysis of the APOGEE data. The blue squares in the left panels show the prediction obtained with the 1D--NLTE models of \citet{Kovalev2018}. The green circles in the right panels show the prediction obtained by the 3D--LTE models of \citet{Ludwig2019}.}
\end{figure*}


 \begin{figure*}[!ht]
\centering
        \begin{subfigure}[b]{\textwidth}
        \includegraphics[width=0.245\textwidth]{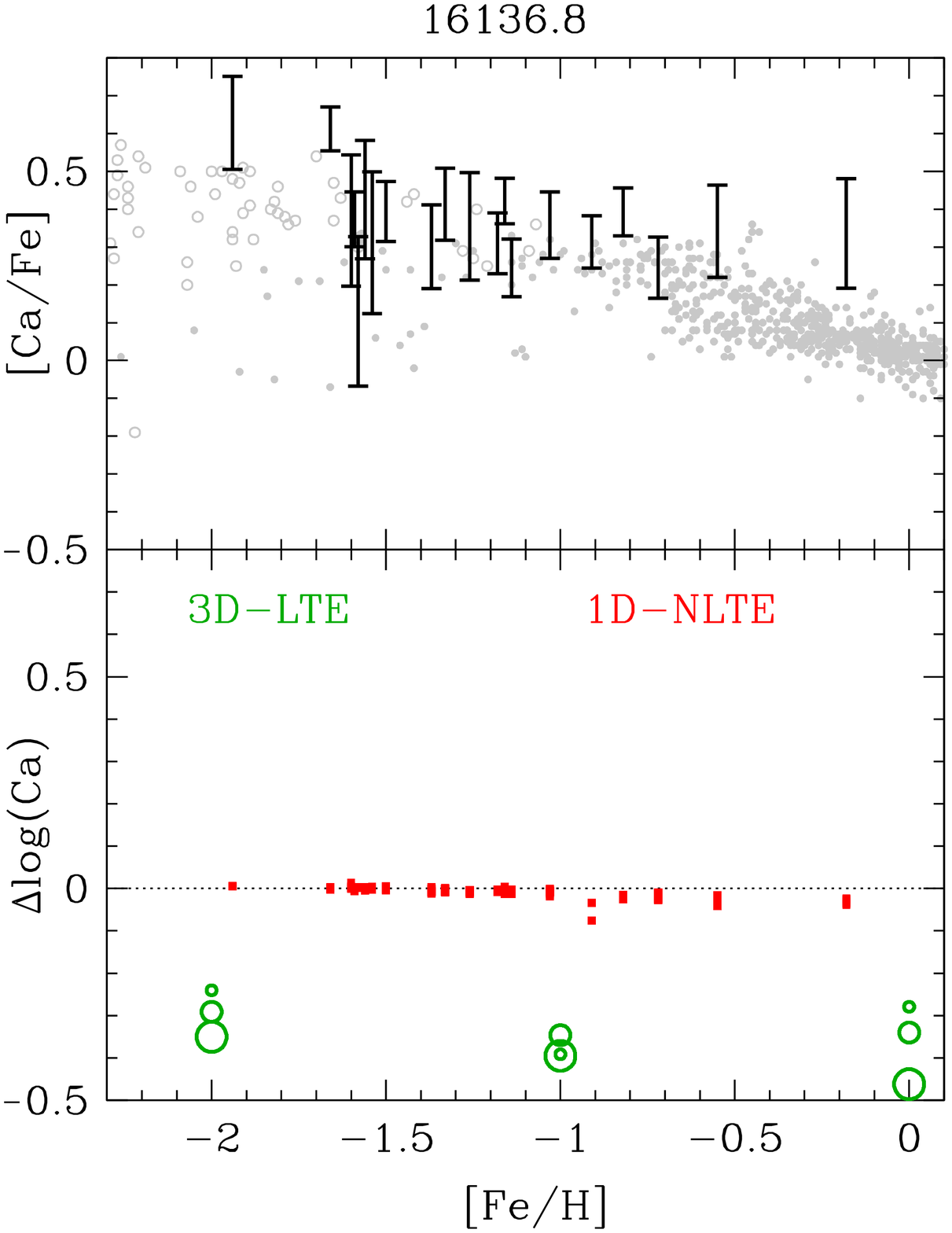}
        \includegraphics[width=0.245\textwidth]{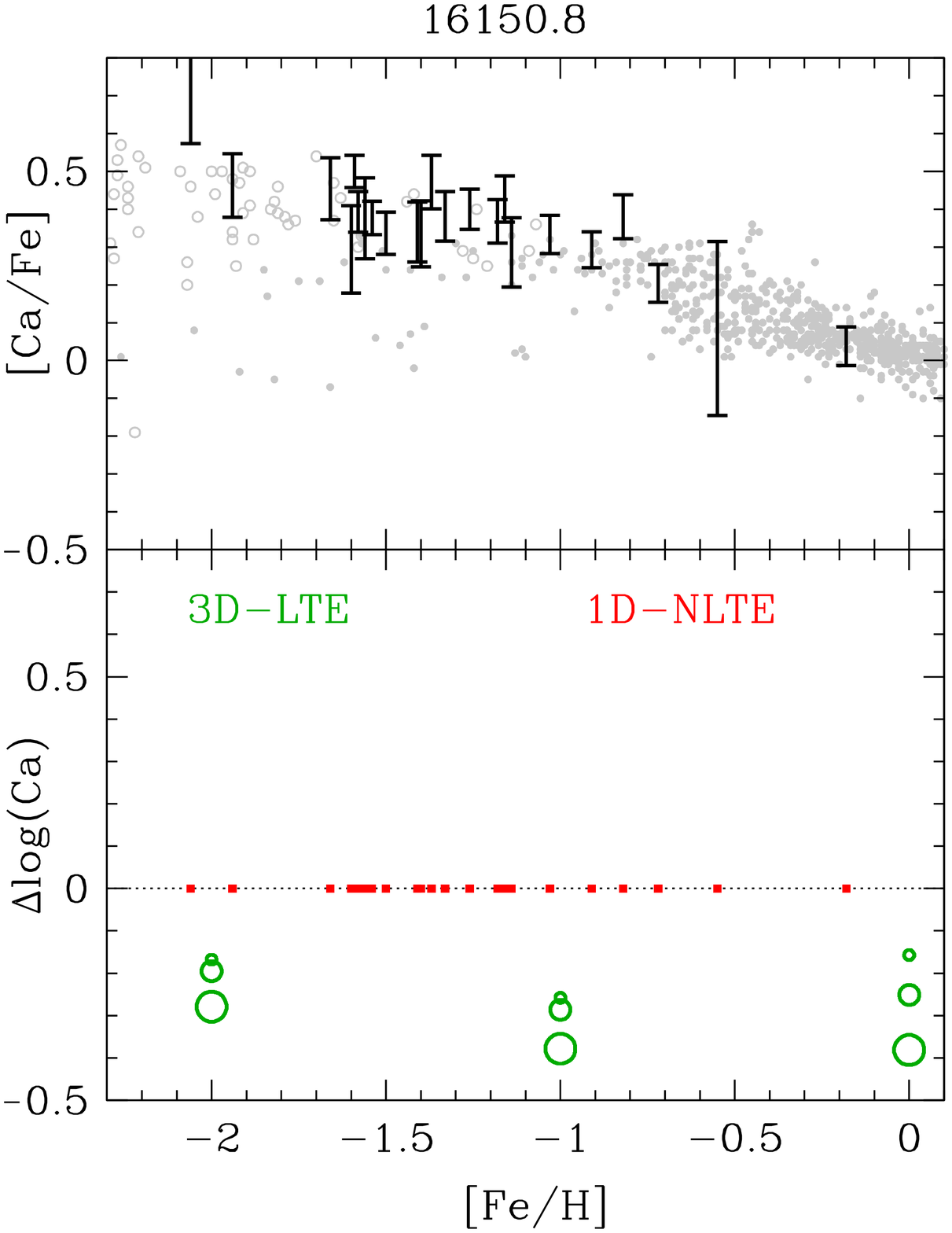}
        \includegraphics[width=0.245\textwidth]{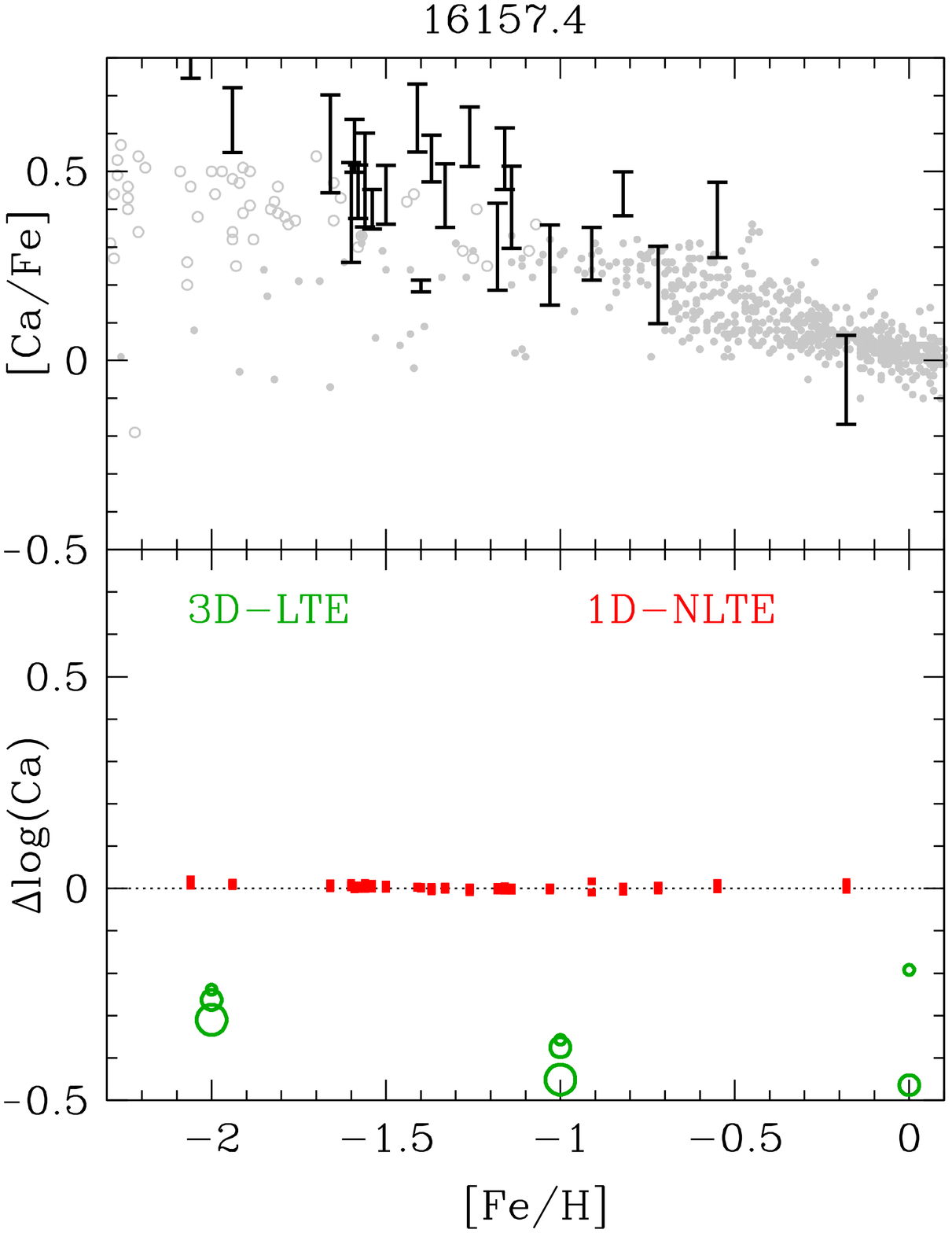}
        \includegraphics[width=0.245\textwidth]{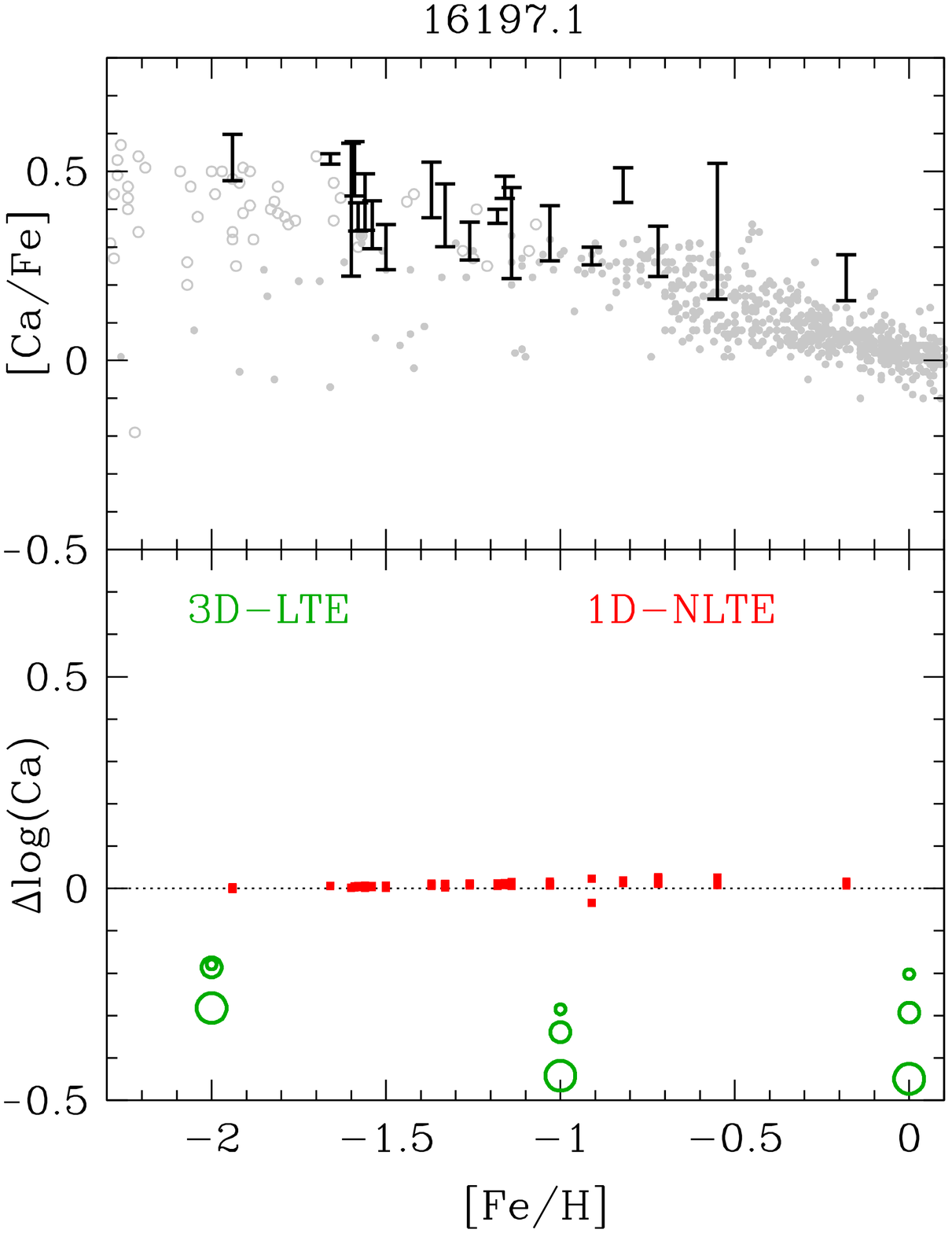}
    \end{subfigure}
\caption{\label{fig:CaFe} Upper panels: Abundances obtained for individual Ca lines. The black error bars show the star-to-star median of the 1D--LTE analysis of the APOGEE data. The grey open circles and grey dots are literature data from respectively \citet{Roederer2014} and \citet{Bensby2014} from optical dwarf spectra. Bottom panels: Corrections obtained with the  1D--NLTE models from this work (red squares) with the 3D--LTE models of \citet{Ludwig2019} (green circles). }
\end{figure*}

 \begin{figure*}[!ht]
\centering
        \begin{subfigure}[b]{\textwidth}
        \includegraphics[angle=-90,width=0.33\textwidth]{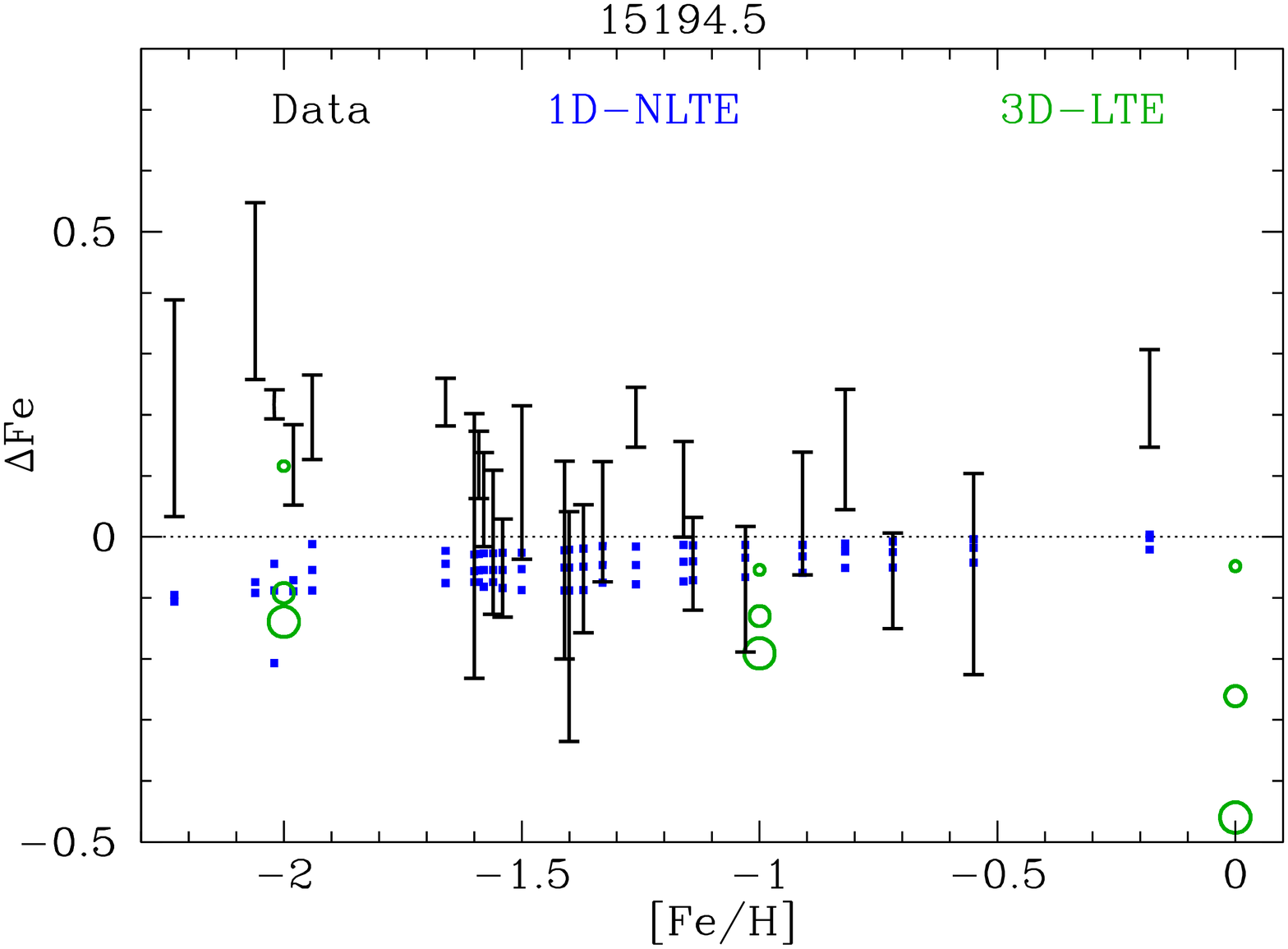}
        \includegraphics[angle=-90,width=0.33\textwidth]{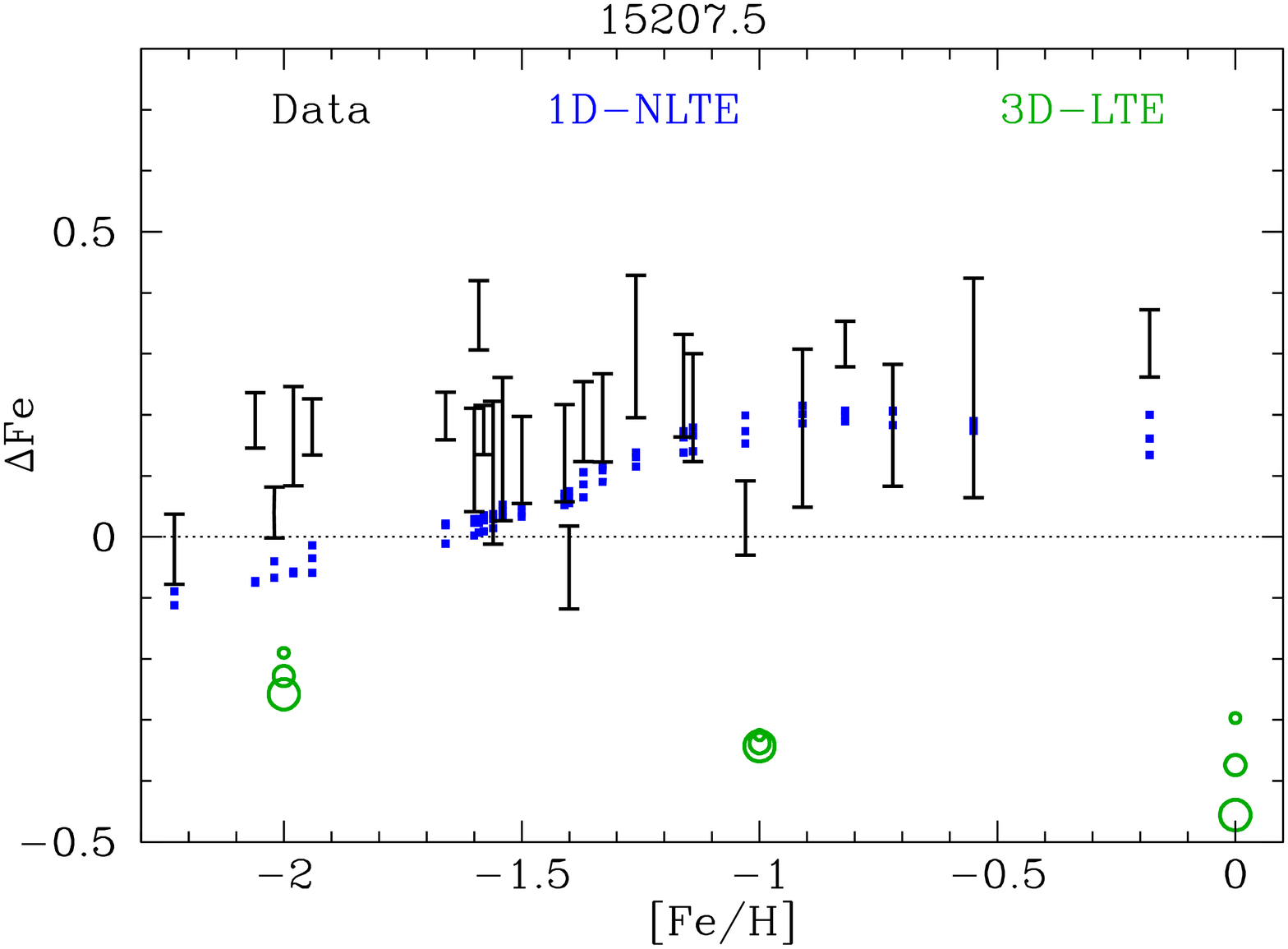}
        \includegraphics[angle=-90,width=0.33\textwidth]{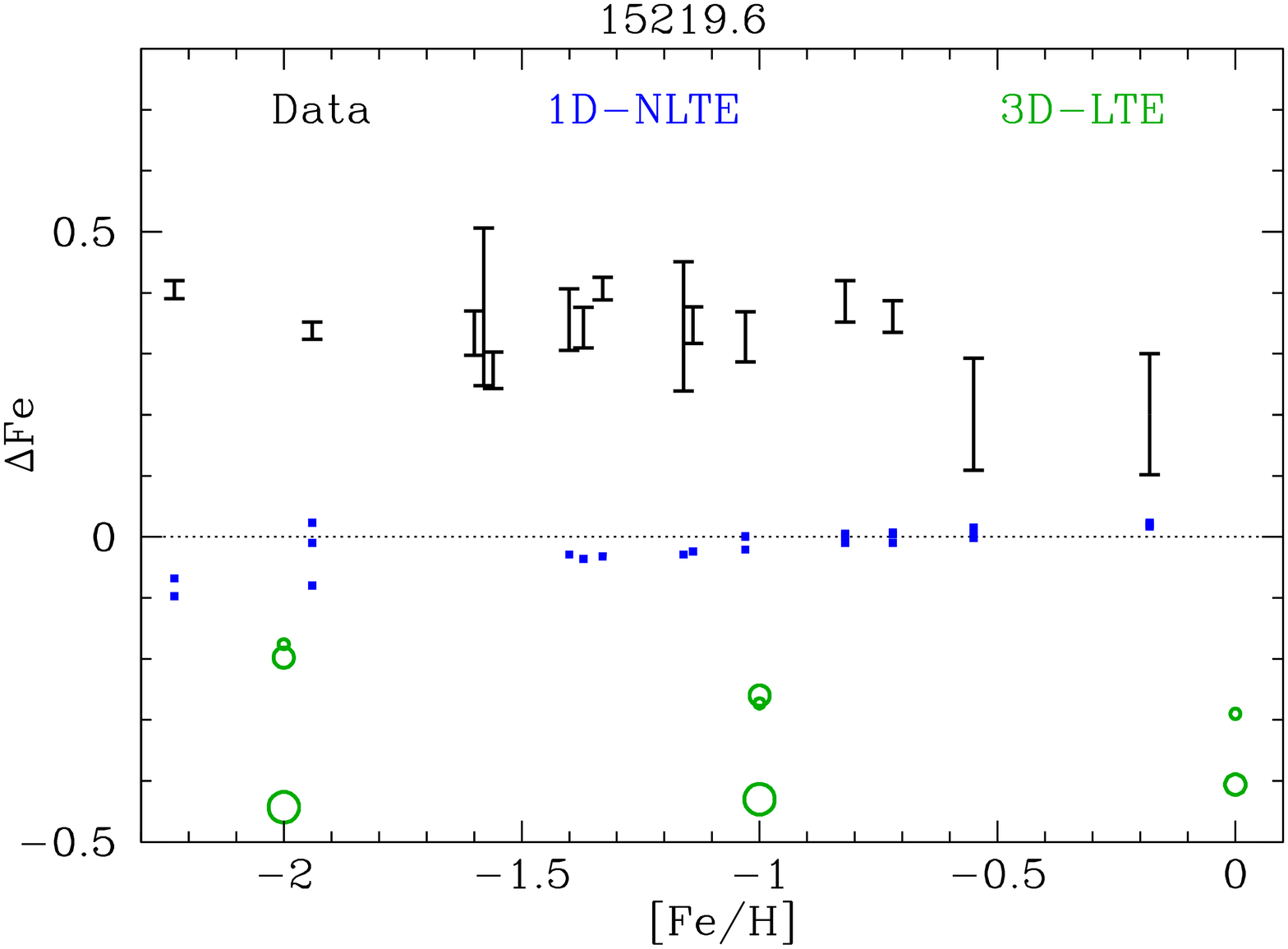}
    \end{subfigure}
        \begin{subfigure}[b]{\textwidth}
        \includegraphics[angle=-90,width=0.33\textwidth]{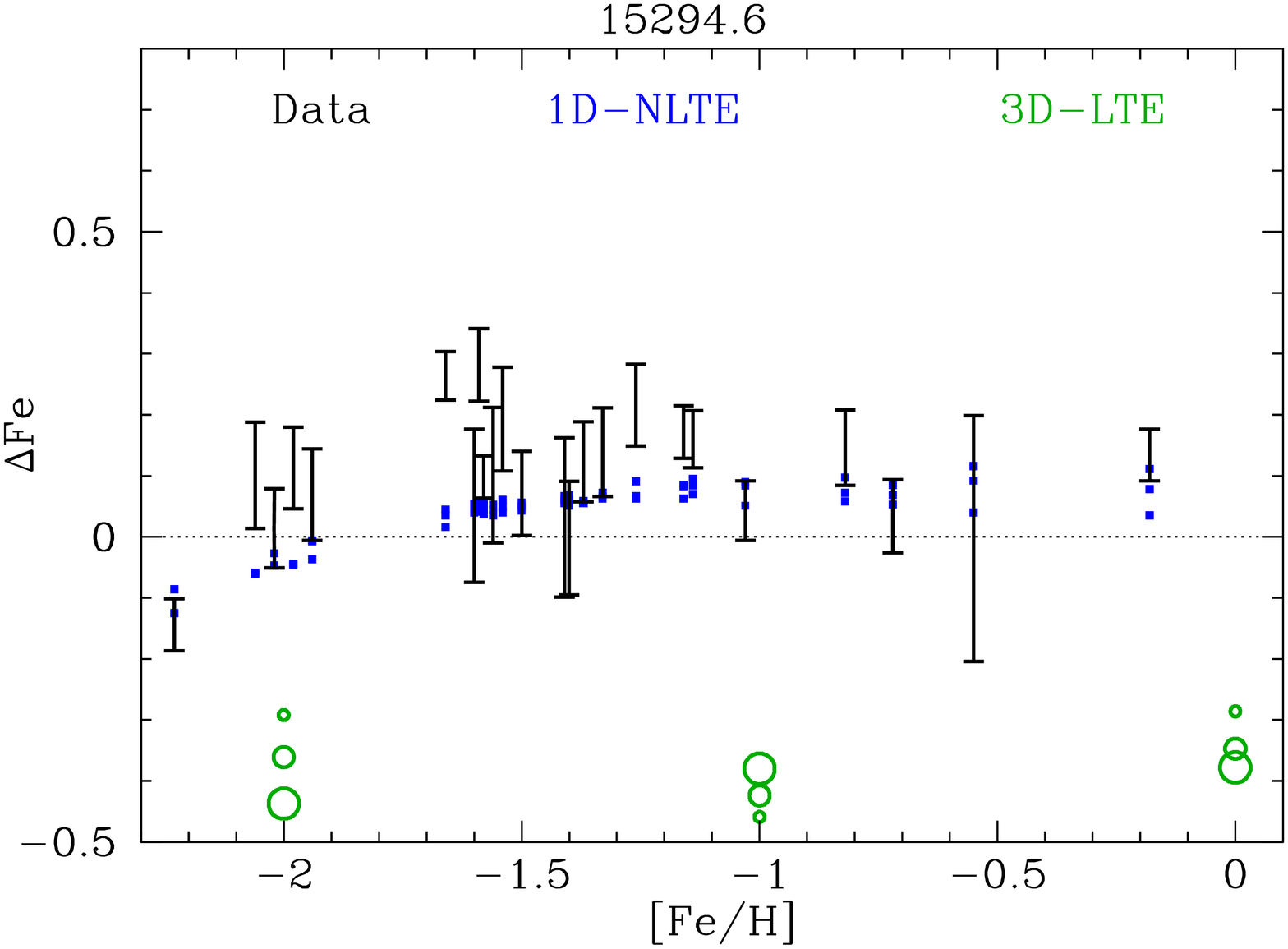}
        \includegraphics[angle=-90,width=0.33\textwidth]{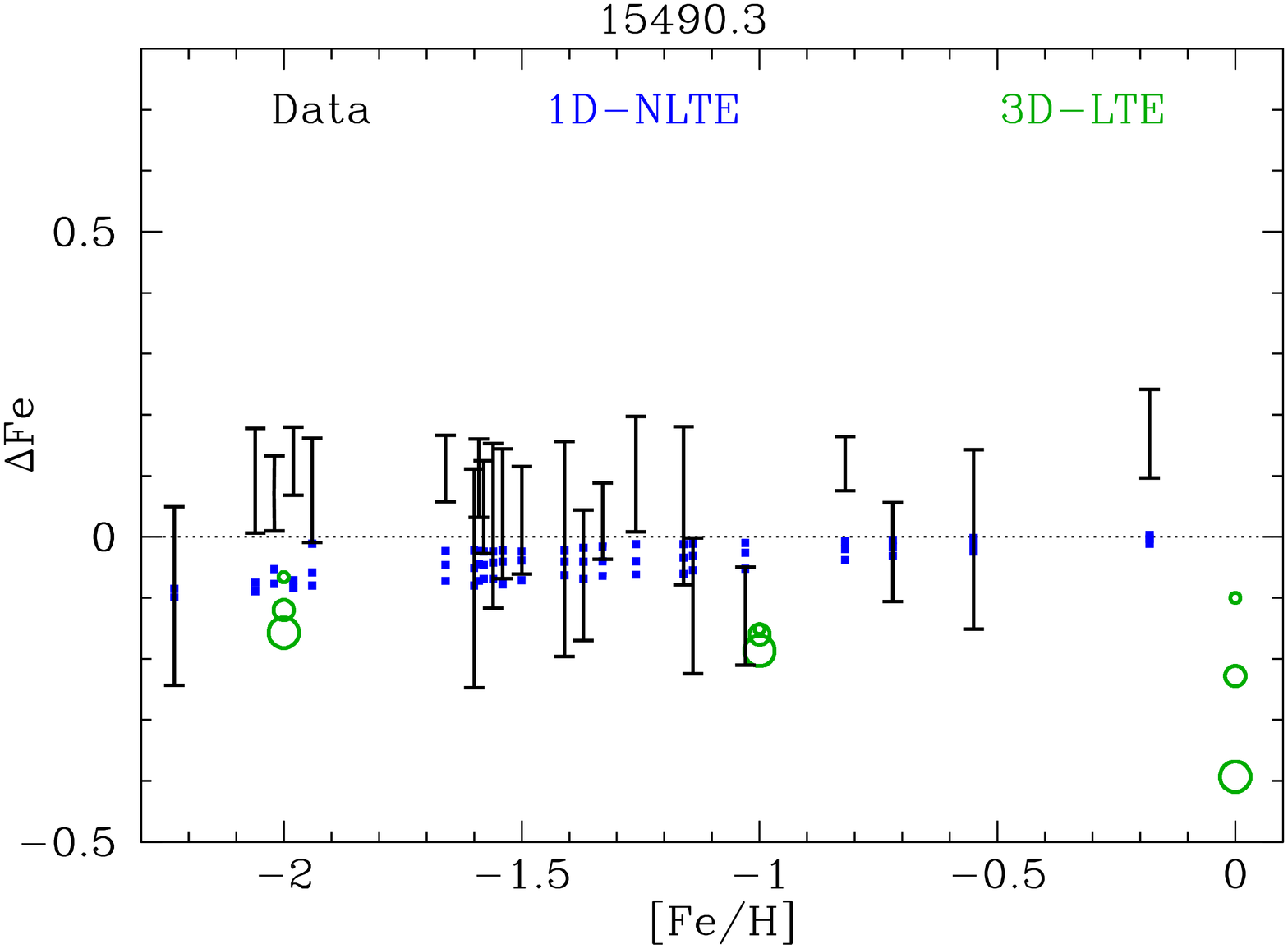}
        \includegraphics[angle=-90,width=0.33\textwidth]{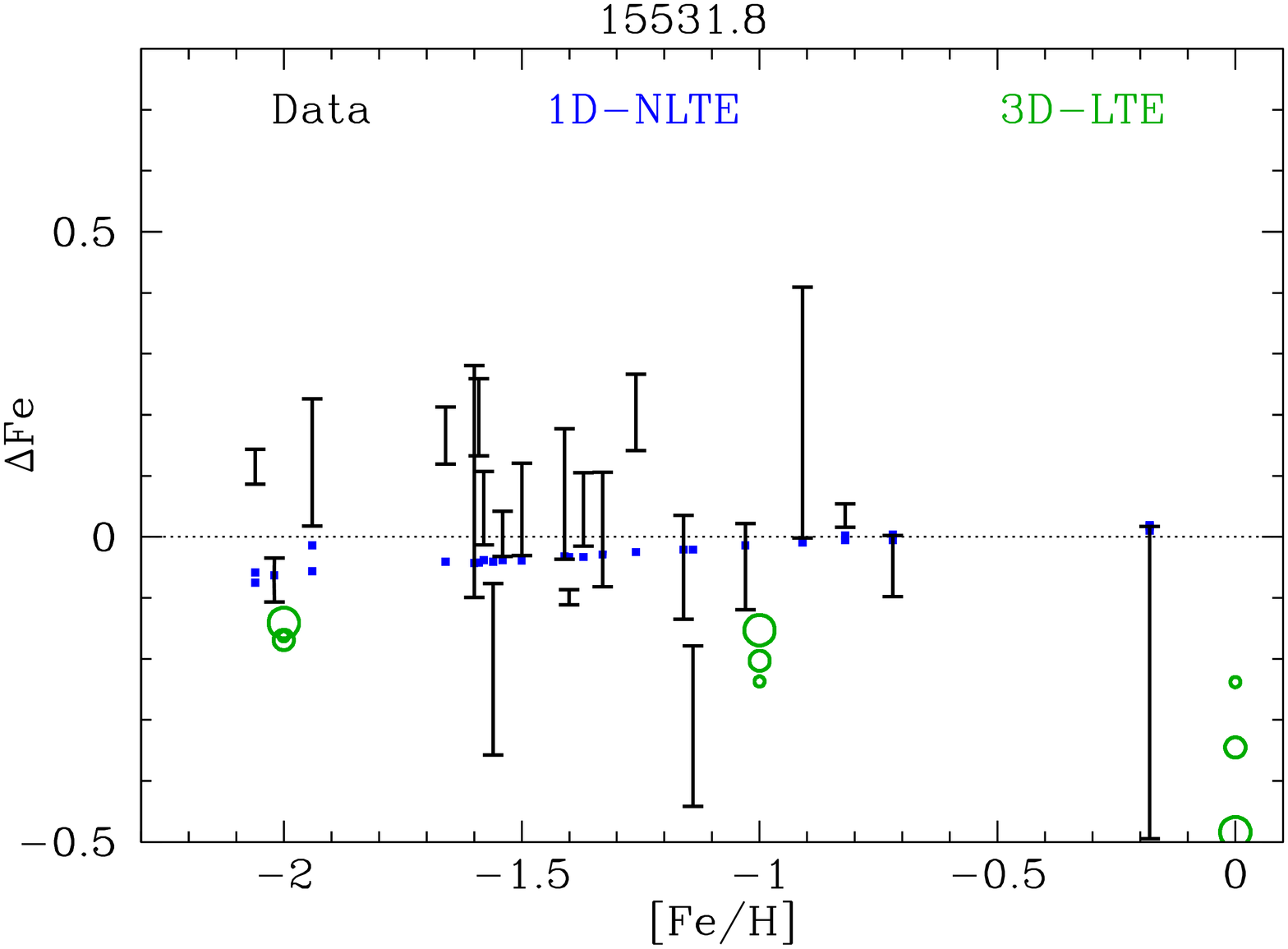}
    \end{subfigure}
        \begin{subfigure}[b]{\textwidth}
        \includegraphics[angle=-90,width=0.33\textwidth]{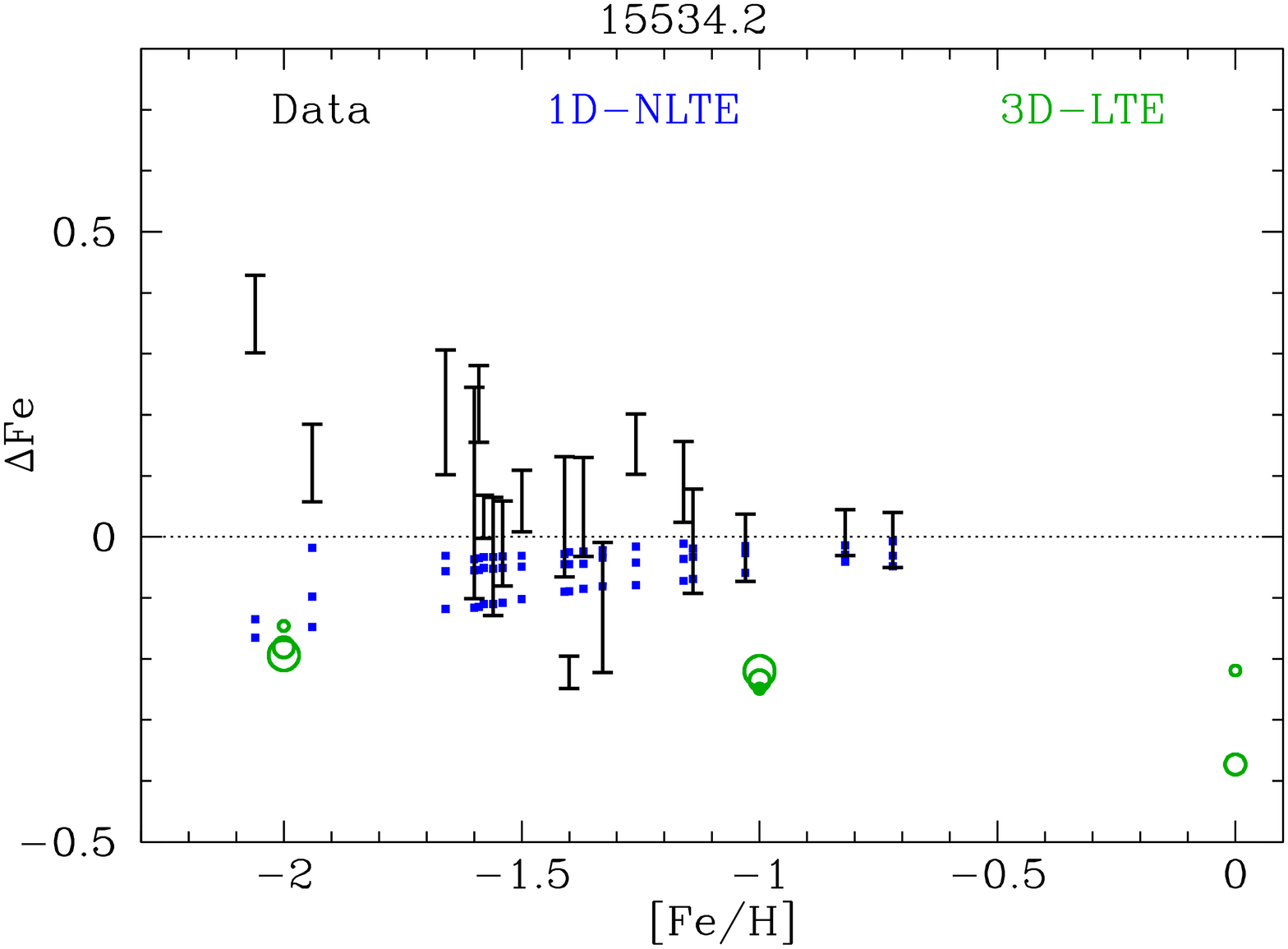}
        \includegraphics[angle=-90,width=0.33\textwidth]{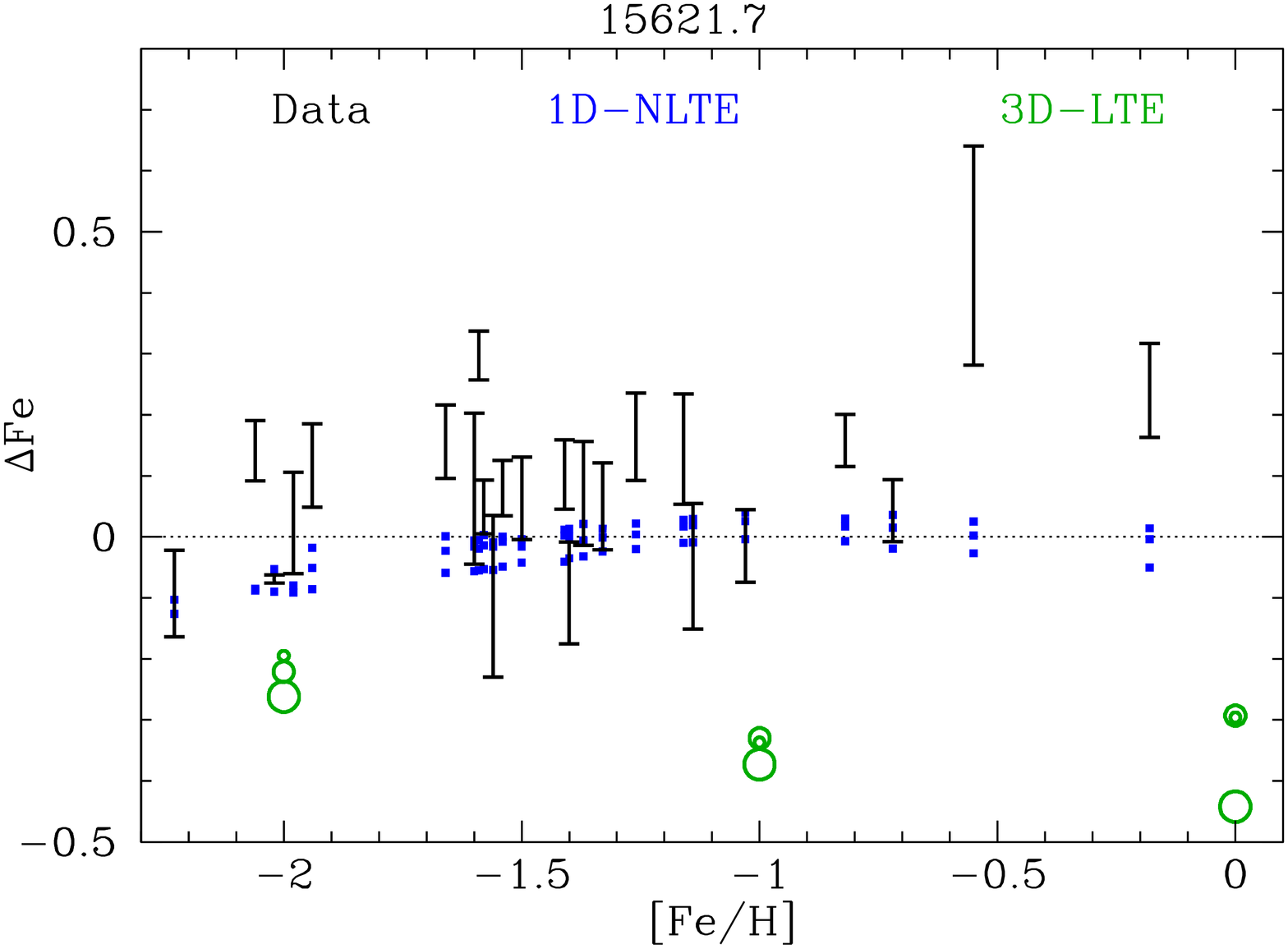}
        \includegraphics[angle=-90,width=0.33\textwidth]{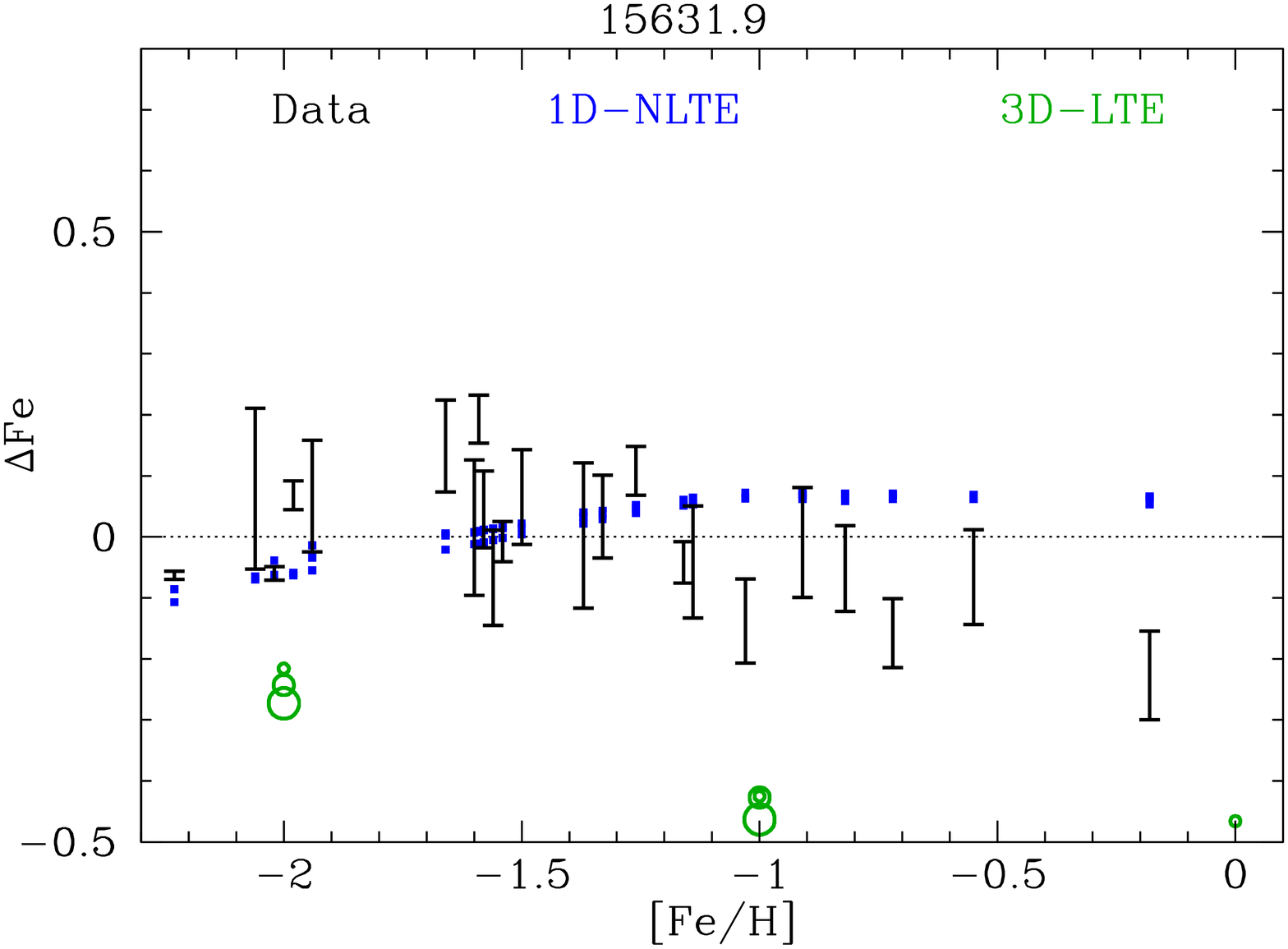}
    \end{subfigure}
        \begin{subfigure}[b]{\textwidth}
        \includegraphics[angle=-90,width=0.33\textwidth]{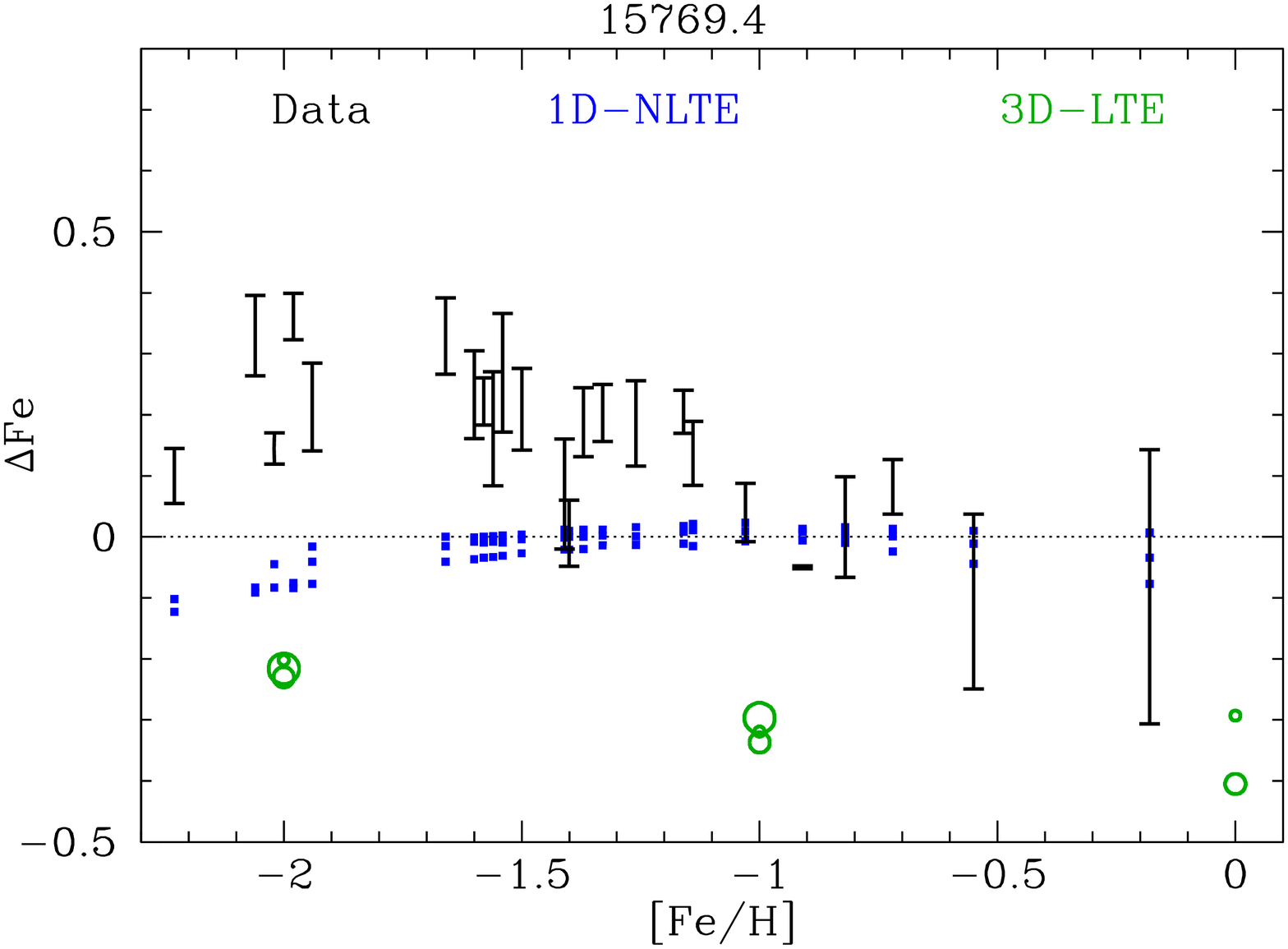}
        \includegraphics[angle=-90,width=0.33\textwidth]{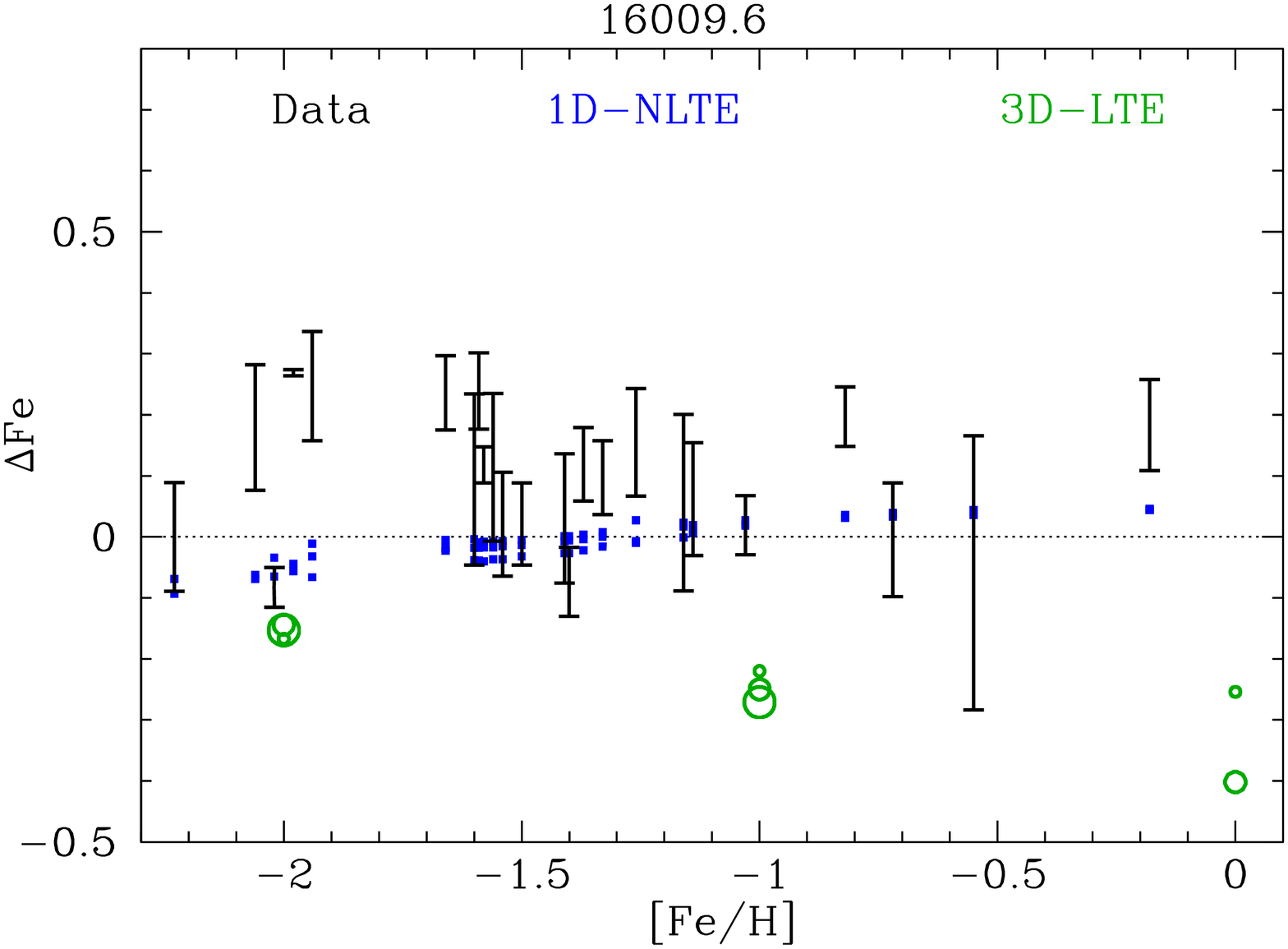}
        \includegraphics[angle=-90,width=0.33\textwidth]{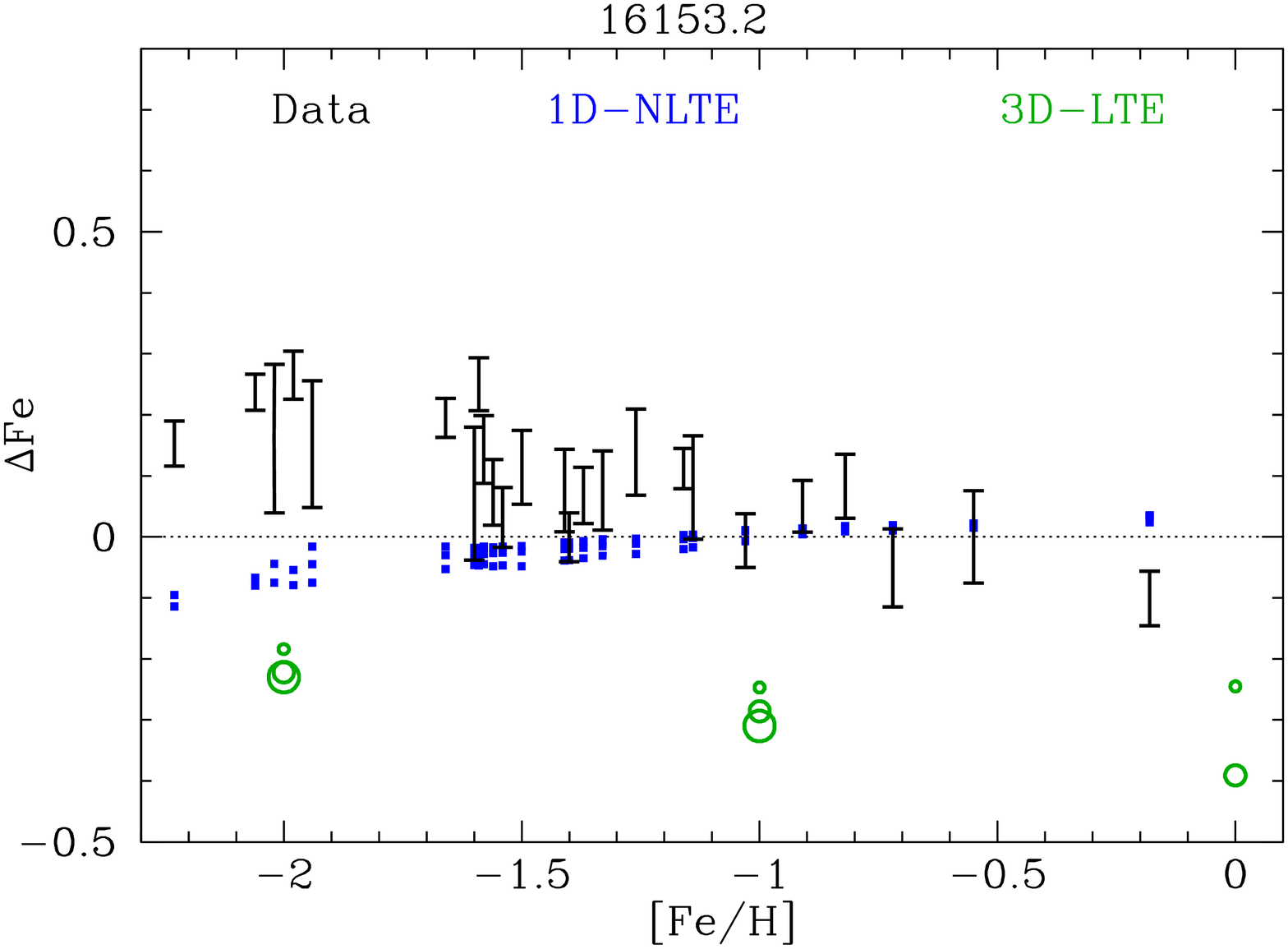}
    \end{subfigure}
\caption{\label{fig:DeltaFe} Difference of abundance of Fe lines obtained from individual lines with the literature metallicity of the cluster (black error bars).  The green circles show the prediction of the 3D--LTE models correction and blue squares indicate 1D--NLTE corrections  from \citet{Kovalev2018}. }
\end{figure*}

\section{Methodology}
All abundance results in this work were determined in a homogeneous way using the Brussels Automatic Code for Characterizing High accUracy Spectra (BACCHUS) \citep{Masseron2016}. In this version of the code, a pre-computed grid of 1D--LTE spectra were compared to an observed spectrum spectrum (Sec.~\ref{sec:data}), a 3D spectrum and a NLTE spectrum (Sec.~\ref{sec:models}). While all other parameters were fixed, each of the 1D--LTE spectra varies in one elemental abundance. The $\chi^2$ value of the 1D--LTE spectrum against the 3D models, the NLTE models or observed spectrum were determined within a  spectral range where the 1D--LTE flux changes significantly with the  abundance. Finally, the optimal abundance was determined by minimizing the $\chi^2$ value. In addition, the code  provides flags to reject any suspicious lines and upper limit values. \\
The studied lines were chosen according to two main criteria; they need to be strong enough to appear (and thus be measurable) in our sample and they need  to have the most reliable $\rm log$gf and collisional broadening parameters so that the most accurate abundance is obtained for the observations. 

\subsection{Observational data}\label{sec:data}
The observational data are from the SDSS~IV/APOGEE2 survey \citep{Gunn2006,Blanton2017,Majewski2017}. The data were reduced as described in \citet{Nidever2015}.
The selected sample corresponds to GCs members as selected by \citet{Masseron2019} and \citet{Meszaros2020}, with a minimum signal-to-noise ratio of 50   and with effective temperatures between 4000~K  and 4700~K. The final observational sample comprises a total of 673 spectra of red giants across 25 globular clusters. \\
The effective temperatures, surface gravities, microturbulence velocities, and abundances  were adopted from \citet{Masseron2019} and \citet{Meszaros2020}. However, both works noted some systematics in their Fe abundance scale compared to the literature values. Therefore, we   adopt the metallicities derived by \citet{Carretta2009_UVES,Carretta2009_GIRAFFE}, except for the $\omega$~Cen stars, which have been disregarded. We recall that the effective temperatures were determined from photometry instead of the APOGEE data releases values, thus independently of the observed spectra to reduce as much as possible covariances between metallicities and temperatures. Surface gravities were determined via isochrones, which  have been demonstrated to be more robust against NLTE effects \citep{Mucciarelli2015NLTE}.

\subsection{Model spectra} \label{sec:models}
For the comparison with the data we  used three sets of models: two sets of 1D--NLTE models from two distinct codes and one set of 3D--LTE models from one 3D code.

The first set of 1D--NLTE synthetic spectra was derived with the codes TLUSTY and SYNSPEC \citep{Hubeny2017}. The atom data and collisional cross-sections are the same as in \citet{Osorio2015_Mgatom} for Mg and in \citet{Osorio2019_Ca} for Ca. These atoms have been translated from MULTI to TLUSTY/SYNSPEC format \citep[details in][]{Osorio2019_inprep}. The atomic and molecular line lists used will be published in \citet{Smith2019}, except for Mg where we adopt the data from the model atom reference. For the  Ca lines we adopted the parameters from  \citet{YU2018263} for the model atom and the atomic line list. The stellar parameters and abundances used are  described in Sect.~\ref{sec:data}.

The second set of NLTE synthetic spectra was computed with the online synthesis tool from \citet{Kovalev2018}. The radiative transfer code of that tool uses the model atoms and collisional cross-sections  described in \citet{Bergemann2015_Mg} for Mg and in \citet{Bergemann2013_Si} for Si. For the stellar parameters the online tool could not provide as many spectra as we require in a reasonable time. Therefore, we   chose to compute only three representative effective temperatures for each cluster metallicity, such that T$\rm_{eff}=$ 4100K, 4500K, and 4700K. Corresponding values of $\log g$ and microturbulence velocities have been interpolated from the  observed data of the clusters (Sec.~\ref{sec:data}).\\
 It should be  noted that both 1D--NLTE sets use the same model atmospheres \citep[MARCS][]{Gustafsson2008}, but with differences in the line lists. Nevertheless,  we can assume  (for close enough $\rm \log gf$ values as  is true  for all the lines in this work) that  the differences in the line lists  cancel out when the corrections are computed consistently with the same code and the same line lists. In contrast, the two sets of 1D--NLTE models have noticeable differences regarding the implementation of the  NLTE correction. The NLTE calculations performed for this work is the first application of NLTE radiative transfer calculations for several atomic species simultaneously in cool stars \citep{Osorio2019_inprep}. The calculations from \citet{Kovalev2018} are traditional NLTE calculations (i.e. one atomic species is calculated at a time). 
 The two calculations also differ in the electron and hydrogen collisional data, in particular for the levels involved in the transitions in the H band, which directly affect the statistical balance of those levels. While for electron collisional excitation we used the CCC calculations from \cite{2017A&A...606A..11B} for Mg I, \citeauthor{Bergemann2015_Mg} used the formula from  \cite{1962PPS....79.1105S} taken from \cite{1998A&A...333..219Z}. For hydrogen collisional excitation between the low-lying levels of Mg I, the same data were used in \citeauthor{Bergemann2015_Mg} and our works \citep{2012A&A...541A..80B}; instead,  for the remaining transitions we used the free electron model from \cite{1991JPhB...24L.127K}, while in \citeauthor{Bergemann2015_Mg} no data are reported.

\citet{Zhang2016} and \citet{Zhang2017}  also made NLTE predictions for Si and Mg lines in the H band, but they did not provide the synthesis that would permit a direct comparison such as the one we discuss below. Nevertheless, their Mg model atom and collisional data were identical to the Kovalev  values;  their model atmospheres and statistical equilibrium code were also the same. Therefore, we expect nearly identical results to the Kovalev  results presented in this work. Similarly, \citet{Zhang2016} use comparable procedures and data to those used in \citet{Kovalev2018}, although they adopt a different scaling factor S$_h$=0.01 in the Drawin formula for H collisions while Kovalev   adopts 1. Nevertheless, as we show in the following, the Si lines we are discussing here have small predicted NLTE effects, leading to negligible differences between those two works.

The set of 3D spectral syntheses has been taken from \citet{Ludwig2019} based on the 3D radiative-hydrodynamical atmosphere grid of \citet{Ludwig2009}. Unfortunately, there are only nine models corresponding to giants ($\rm\{T_{eff},\log g\}=\{5000,2.5\},\{4500,2.5\},\{4000,1.0\}$ for four metallicities $\rm [M/H]=-3.0,-2.0,-1,0,0.0$).\\
Finally, we note that although the line lists between the 3D and NLTE computations may not always be rigorously identical, the 1D--LTE mini-grids  use the same corresponding line lists so that the computed abundance corrections are fully consistent.

\subsection{Comparison between theoretical models}

Figures~\ref{fig:Mg_lines}, \ref{fig:Ca_lines}, \ref{fig:Si_lines}, and \ref{fig:Fe_lines} illustrate the model outputs respectively for  the selected Mg, Ca, Si, and Fe lines in the H band. All of the figures compare the 3D or NLTE synthetic line profiles against 1D--LTE synthesis for a T$\rm{eff}$=4500K, $\rm \log$g=2.5, [M/H]=$-$1.0 model, typical of our sample stars.  In these figures we note that all 1D spectra have been convolved by a macroturbulence of 3km/s  with a rad-tan kernel to match the 3D simulation velocity field. \\
The figures also display in the lower panels the expected offset in abundance between the 3D or NLTE models and the 1D--LTE synthesis for different metallicities ($-$2.0, $-$1.0, 0.0) and different $\rm\{T_{eff},\log g\}$ pairs (\{5000,2.5\}, \{4500,2.5\}, \{4000,2.5\}). The abundance offsets in these figures is equivalent to $\rm \Delta\log(A) = log(A)_{\rm 1D/LTE} - log(A)_{\rm NLTE,3D}$. We also note that the models were also degraded to the APOGEE resolution (i.e. $R\approx22500$) to ensure that the $\chi^2$ minimisation process was homogeneous with the abundance analysis of the observations.

\subsubsection{Mg}\label{sec:Mgmodels}
From Fig.~\ref{fig:Mg_lines} we can observe two main expected behaviours regarding the  selected Mg lines. The first group (encompassing the 15954 and 16365\AA\ lines) displays very small NLTE effects (for  our computations and for  Kovalev's), but significant 3D effects strongly depending on metallicity. On the contrary, the second group of lines (15740, 15749, 15765\AA)  has   large negative 3D effects but larger NLTE effects which largely enhance the line core intensity, thus leading to an overestimation of a 1D--LTE analysis. The similarities in the second group is not surprising as the three lines belong to the same triplet. Therefore, their NLTE effects are expected to be very similar, while their relatively similar strength make them equally sensitive to 3D effects. \\
Furthermore, we can use Mg lines to compare NLTE predictions for two available codes and methodology. The signs of corrections generally agree between the two codes, but the amplitude may vary. The 15954\AA\ line and to a lesser extent the 16365\AA\ line are of particular interest because the \citet{Kovalev2018} models predict significant corrections while our models do not. We strongly suspect that the discrepancies are due to the difference in the collisional data prescriptions.
We also note that those of the 15954.5\AA\ line show a small blend of a Ti~I line on its left wing. In addition to  being weak, we note that this line is included in all our line-by-line syntheses, thus taken consistently into account when applying our differential approach. \\

\subsubsection{Ca}\label{sec:Camodels}
In contrast to Mg, none of the four Ca lines shows strong NLTE effects (Fig.~\ref{fig:Ca_lines}), well in line with those reported by \citet{Osorio2019_inprep}.  However, they display  large 3D effects, only slightly correlated with metallicity. Moreover, the sign  and amplitude of the abundance bias is nearly identical for all lines. This is easily explained  because three of them belong to the same triplet transition and the last one is the corresponding singlet transition.

\subsubsection{Si}\label{sec:Simodels}
The selected Si lines show moderate NLTE and large negative 3D effects (Fig.~\ref{fig:Si_lines}). Actually, the 15376\AA\ line does not show any NLTE effect, probably due to the fact that it is a forbidden transition. The 15960, 16060, and 16094\AA\ lines and the 16215 and 16828\AA\ are belonging to same triplets. Moreover, their respective strengths are similar, naturally leading to quite similar behaviour of the abundances offsets.

\subsubsection{Fe}\label{sec:Femodels}
 According to Kovalev's models, the NLTE corrections for the Fe lines in   H band are small (Fig.~\ref{fig:Fe_lines}).
 Despite the large number of lines under study, it appears that the 3D corrections all show negative offsets independently of temperature and metallicity, thus suggesting that the 3D effects lead to an overestimation of the mean Fe abundance.  \\

After examining the expected behaviours of the lines relative to the theoretical predictions of 3D and NLTE models, we can now use this information to isolate lines with a dominant NLTE or 3D effect.

\section{Discussion} 
After reviewing the 3D and NLTE model expectations for individual lines in Sect.~\ref{sec:models}, we now discuss them against the data. Although the resolution of the data does not allow a detailed line profile comparison, the effects   still bias the abundance determination. In this section we   use two ways to test the model predictions. The first   compares the abundances obtained between two different lines. This method provides only relative estimates of the 3D and NLTE effects, and we have applied it to the Mg and Si lines in our GCs stars sample. The second way consists in assuming that the absolute abundance of each star is known. In this regard, GCs stars offer an adequate sample as they are known to have generally homogeneous abundances of Ca and Fe, unlike Mg and Si. This method provides absolute estimates of the 3D and NLTE effects.

\subsection{Probing relative 3D and NLTE effects}\label{sec:relative}
In Fig.~\ref{fig:Mg_diff} we compare the difference in abundances between two Mg lines (Sect.~\ref{sec:Mgmodels}) in the data and for the 3D--LTE models and the 1D--NLTE models.
In this comparison we use the 15740\AA\ line as a common reference line between all the panels. In Sect.~\ref{sec:Mgmodels} we explain that similar 3D effects and similar NLTE effects are also expected between the reference line and the 15749 and 15765\AA\ lines.  Although the quantum properties of those lines are identical,  we observe in Fig.~\ref{fig:Mg_diff} that the 3D and NLTE models still predict some residual behaviour probably due to the slight difference in their strengths, suggesting slightly different formation depths in the atmospheres. Both NLTE models predict a decreasing trend with increasing metallicity, which appears compatible with the data within the error bars. However, the 3D--LTE models predict a trend in the opposite direction which does not seem compatible with the data. \\
Concerning the two other Mg lines (15954 and 16364.8\AA) the interpretation of their abundance against the reference line is more complex because these lines are expected to be 3D sensitive but not NLTE sensitive, whereas the reference line is sensitive to both. The data of 15954$-$15740 differences and the 16365$-$15740 differences  do not show any offset, but do show a weak positive trend that is reproduced by our 1D--NLTE predictions but not by the Kovalev predictions. This implies either that   Kovalev's NLTE effects in a 3D hydrodynamical atmosphere environment are drastically different, or that the NLTE effects of this work cancel out between the two lines, but that the 3D effects are smaller than predicted. In this case, only self-consistent 3D--NLTE simulations could sort out the correct interpretation.\\
It is also interesting to note that in all panels of this figure that the predicted dispersion around the trends is relatively small even though all the stars have various stellar parameters and Mg abundances, and that we consistently took this into account   in our calculations. Therefore, we conclude that the dependence of NLTE effect for those Mg lines is dominated by metallicity.       

Regarding the Si lines, we observe a similar situation to that for the Mg lines (Fig.~\ref{fig:Si_diff}). There are three sets of lines: two sets of triplet transitions (15960, 16060, 16094 \AA) and (16215, 16828\AA), and a forbidden transition (15376 \AA). We chose again one of the triplet lines as a reference line (15960\AA). As for the Mg triplet lines, the observed 16060$-$15960 and the 16094-15960 differential abundances are nearly null with a slight residual trend. The predicted NLTE corrections seem compatible with the data. The agreement between the data and the NLTE models appears particularly good for the  16094$-$15960, and a bit less in the 16060$-$15960 case.  The differential abundance against the other lines of the other triplet (16215, 16828\AA) not surprisingly gives very similar results between the two lines, and is still in very close  agreement with   Kovalev's NLTE simulations. \\
The Kovalev predictions are in good agreement with the Si lines abundance data while they are not in good agreement with the Mg lines abundance data. Beyond the fact that they are two distinct atoms, this could also be explained by differences in the quality and completeness of the collisional data adopted: for Mg, the \citet{Kovalev2018} code uses the \citet{Bergemann2015_Mg} model where hydrogen collisions between high excitation levels were neglected; for Si, the \citet{Kovalev2018} code uses the model atom of \citet{Bergemann2013_Si}. This latter work used one scaling relation of the Drawin formula from \cite{Lambert_1993} to assign cross-section values for all energy levels and check it against some lines of the solar spectrum. Given that the test lines of \citet{Bergemann2013_Si} and our lines have similar energy levels, this may explain why  Kovalev's NLTE predictions provide good agreement with the data. However, our study also involves the  forbidden transition at 15376 \AA. For that line, the data seem to  disagree  with   Kovalev's predictions, suggesting that the scaling relation they used to compute the Si cross-sections is not  valid for all transitions. This supports the idea that the scaling relation in NLTE calculations should be avoided and, as quoted from \citet{Bergemann2013_Si},   that `more accurate estimates of the collision rates are desirable and, once available, will significantly improve the accuracy of the calculations'. 

Regarding the 3D predictions, they show moderate amplitude with a negative dependence with metallicity,  a trend which seems hardly compatible with the data. The case of the 15376\AA\ line is of particular interest because it is a forbidden transition whose NLTE effects can be considered  negligible. Therefore, the NLTE effect should only be due to the 16960\AA\ line alone. The data suggest a v-shaped dependence against metallicity, and so do the NLTE simulations. However, there is an offset between the models and the data that might be explained by a combination of the NLTE effects with the 3D effects. 
Therefore, the 3D--LTE spectra for the weak lines do not provide a good prediction of reality, and 3D--NLTE should be more appropriate. To verify this hypothesis a full 3D--NLTE calculation would be necessary.
We finally note that the microturbulence and macroturbulence velocities in the 1D--LTE calculations could be tuned to match the 3D computations. However, while in principle these empirical parameters can be freely varied, the values we use in this work are compatible with  optical studies and with APOGEE spectra standard analysis \citep{Holtzman2015}. Moreover, we argue that if this value had to be changed, it would have a drastic impact on other strong lines such as the one in Figs.~\ref{fig:Mg_lines} and \ref{fig:Si_lines}.

\subsection{Probing absolute 3D/NLTE effects}
Figure~\ref{fig:CaFe} shows the line-by-line abundances of Ca obtained for the whole sample and is compared to the analysis of optical dwarf spectra of \citet{Roederer2014} and \citet{Bensby2014}. All data sets show a global decreasing trend with increasing metallicity as expected by chemical evolution. The 3D and NLTE expected corrections are also displayed. As explained in Sect.~\ref{sec:Camodels}, the models predictions behave very similarly between the three Ca lines, with nearly null NLTE corrections and large 3D corrections. The agreement between the optical data and the IR data in Fig.~\ref{fig:CaFe} suggest that the 3D corrections should not be larger than 0.1 dex overall. 
Another possibility would imply that optical data are affected by NLTE while IR data are affected by 3D of the same amplitude. However, 3D abundance corrections are known to be larger in dwarfs than in giants \citep{Ludwig2008} in contradiction with the fact that the optical data are composed of dwarfs while the IR are composed of giants. Moreover, we argue that if the 3D corrections as estimated in this work were to be applied (and so for NLTE in the optical data), the chemical evolution of Ca would be affected in an unrealistic way that is incompatible with chemical evolution expectations. The line-differential work we  did for Mg and Si (see Sect.~\ref{sec:relative}) certainly supports to some extent that 3D effects on abundances are not accurate enough, or at least require  that NLTE effects   be simultaneously taken into account. This conclusion also corroborates theoretical predictions where 3D--LTE abundances are often worse than 1D--LTE values \citep{Nordlander2017} and is related to the NLTE-masking effect described by \citep{Rutten1982}. 

Regarding Fe, it is striking in Fig.~\ref{fig:DeltaFe} that the data systematically show a positive offset of $\sim$0.1 dex, implying that the overall Fe abundance is overestimated compared to the literature. First of all, this bias in metallicity is not likely due to our specific procedure because a similar metallicity offset is observed in other works with distinct temperature scales dealing with the APOGEE spectra \citep{Meszaros2013,Holtzman2015,Meszaros2015,Masseron2019}. Figure~\ref{fig:DeltaFe} shows that NLTE effects do not seem to  significantly affect the Fe lines in the H band. The literature reports that  optical Fe lines also do not seem to be affected by NLTE.  \citet{Carretta2009_UVES,Carretta2009_GIRAFFE} checked that their metallicity determined with optical Fe~I and with Fe~II lines are in agreement, suggesting that NLTE effects are negligible in the optical. In contrast, recent self-consistent 3D--NLTE models of one very metal-poor giant \citep{Amarsi2016} estimate that metallicities determined from optical Fe~I and Fe~II lines are respectively underestimated by 0.17 and 0.08 dex. In a similar way, \citet{Mashonkina2016_CaFe} compute 1D--NLTE corrections for many optical Fe lines in giants and provide corrections of $\sim$0.1 dex at low metallicities. Such corrections onto the GCs metallicities of Carretta's work may suggest that the APOGEE data provide quite accurate GC metallicities. But before drawing any firm conclusion, these are quite coarse estimates because the giants used in the Amarsi and Monshonkina works are not identical to our GCs stars, and it is not clear whether the Fe lines between the studies are fully compatible. In addition, as we demonstrate in the case of Mg and Si in Sect.~\ref{sec:relative}, all these results are strongly dependent on the current knowledge of the Fe atom and its cross-section values, which are known to be still quite poor.

Moreover, the unrealistically large predicted 3D corrections also highlight the problem of applying corrections self-consistently. An illustration of this argument can be seen when looking at the Fe lines in Fig.~\ref{fig:DeltaFe}, which show the difference   between the Fe abundances and the literature cluster metallicity \citep{Carretta2009_UVES,Carretta2009_GIRAFFE}. In this figure the 3D--LTE estimated corrections are overplotted and predict a negative offset, which is  the opposite of the data. 
If we had to derive the metallicity of the 3D models with 1D models, we would certainly  find a lower metallicity. By extension, we can also   derive different T$\rm _{eff} and \log g$ \citep[see also][where different 1D effective temperatures are derived in 3D supergiant model spectra   depending on whether you use optical molecular band strength or infrared flux]{Chiavassa2011}. In other words, the derived stellar parameters derived with 3D models may be different to the one derived in 1D, implying that 3D abundance corrections applied on 1D abundances with the same stellar parameters between 1D and 3D are not consistent; thus, a direct comparison (as  in Fig.~\ref{fig:CaFe} and Fig.~\ref{fig:DeltaFe}) may not be adequate.
Actually,  if we had to apply the average $\sim$-0.3 metallicity correction from Fig.~\ref{fig:DeltaFe} in addition to  the $\sim$-0.3 Ca corrections in Fig.~\ref{fig:CaFe}, the net resulting corrections would be at more realistic levels.
Nevertheless, we note that self-consistency is certainly an important argument for our discussion for probing absolute corrections as we do for Ca and Fe, but that does not affect our method for testing relative corrections (Sect.~\ref{sec:relative}) where the line-differential approach ensures that the differences in stellar parameters between 1D and 3D cancel out.

\section{Conclusions}
In this paper we have tested the theoretical predictions of some 3D--LTE and 1D--NLTE models against a homogeneous analysis of a large sample of GC spectra from the APOGEE large spectroscopic survey. In particular, we have discussed the expected effects on the individual lines abundances of four elements (Mg, Si, Ca, and Fe). 
Using Mg lines, we could compare two distinct sets of models for 1D--NLTE and show that the predictions are in the same direction. Nevertheless, we observe some quantitative differences that are mainly due to the inclusion or not of good and complete collisional cross-sections, and not by the differences in the codes themselves.\\
The analysis of the abundance differences between several Mg lines (and similarly  Si lines) allows us to also verify the impact of the 3D effects. However, some 3D--LTE predictions seem to provide a satisfactory explanation for some of the lines abundance behaviour, but most do not. Nevertheless, while waiting for more extensive 3D model grids, our work supports and generalises the fact that 1D--LTE provides a better approximation than 3D--LTE, and thus that 1D--LTE calculation partially cancel out the 3D--NLTE effects \citep{Asplund2003}. We reach similar conclusions regarding Ca abundances where the 3D--LTE estimates on the Ca abundances seem too large. However, we recommend strong caution when applying 3D corrections directly to 1D abundances without assuming self-consistent 3D and 1D stellar parameters.   \\
Finally, we have checked the Fe abundances for 12 Fe lines and show a systematic offset against the literature metallicity values. Although more NLTE models with accurate atom data for Fe lines are crucially needed, our results may suggest that the globular clusters metallicity maybe underestimated in the optical-based literature, whereas the Fe lines of the H band offer more accurate values. \\
In the end our study suggests that the largest effect on abundances for late-type giants is NLTE, and that 3D is only a second-order effect that cannot be treated separately. This idea finds theoretical support in the detailed study \citet{Amarsi2016} where 1D--LTE, 1D--NLTE, 3D--LTE, and 3D--NLTE abundances of Fe lines in a few giants are compared.

The line-differential technique we develop here to probe NLTE and/or 3D models can be applied to any other elements and other data sets, and offers the great advantage of being free from knowing the abundance a priori. Nevertheless, to be valid this method still requires three things,   that the stellar parameters are well constrained, that  the line's oscillator strength and collisional broadening coefficient are accurately known,  and that the selected lines are relatively clean of blends.

\begin{acknowledgements}
We deeply thank H.G. Ludwig for his careful checks of the 3D spectra and thoughtful comments. 

We thank M. Kovalev and M. Bergemann for their great support of their NLTE online synthesis tool. We also thank A. Amarsi for fruitful discussion about 3D/NLTE effects.

 T.M. acknowledges support from Spanish  Ministry  of  Economy and Competitiveness (MINECO) under the 2015 Severo Ochoa Program SEV-2015-0548. T.M., D.A.G.H. and O.Z. also acknowledge support from the State Research Agency (AEI) of the
Spanish Ministry of Science, Innovation and Universities (MCIU) and the 
European Regional Development Fund (FEDER) under grant AYA2017-88254-P. Y.O, and C.A.P. acknowledge support from the same agency under grant AYA2017-86389-P.
  
SzM has been supported by the J{\'a}nos Bolyai Research Scholarship of 
the Hungarian Academy of
Sciences, by the Hungarian NKFI Grants K-119517 and 
GINOP-2.3.2-15-2016-00003 of the Hungarian National
Research, Development and Innovation Office, and by the {\'U}NKP-19-4 
New National Excellence Program of the
Ministry for Innovation and Technology.

This paper made use of the IAC Supercomputing facility HTCondor (http://research.cs.wisc.edu/htcondor/), partly financed by the Ministry of Economy and Competitiveness with FEDER funds, code IACA13-3E-2493. The authors also acknowledge the technical expertise and assistance provided by the 
  Spanish Supercomputing Network (Red Española de Supercomputación), as well as the computer 
  resources used: the LaPalma Supercomputer, located at the Instituto de Astrofísica de Canarias.

Funding for the Sloan Digital Sky Survey IV has been provided by the Alfred P. Sloan Foundation, the U.S. Department of Energy Office of Science, and the Participating Institutions. SDSS acknowledges support and resources from the Center for High-Performance Computing at the University of Utah. The SDSS web site is www.sdss.org.

SDSS is managed by the Astrophysical Research Consortium for the Participating Institutions of the SDSS Collaboration including the Brazilian Participation Group, the Carnegie Institution for Science, Carnegie Mellon University, the Chilean Participation Group, the French Participation Group, Harvard-Smithsonian Center for Astrophysics, Instituto de Astrofísica de Canarias, The Johns Hopkins University, Kavli Institute for the Physics and Mathematics of the Universe (IPMU) / University of Tokyo, the Korean Participation Group, Lawrence Berkeley National Laboratory, Leibniz Institut für Astrophysik Potsdam (AIP), Max-Planck-Institut für Astronomie (MPIA Heidelberg), Max-Planck-Institut für Astrophysik (MPA Garching), Max-Planck-Institut für Extraterrestrische Physik (MPE), National Astronomical Observatories of China, New Mexico State University, New York University, University of Notre Dame, Observatório Nacional / MCTI, The Ohio State University, Pennsylvania State University, Shanghai Astronomical Observatory, United Kingdom Participation Group, Universidad Nacional Autónoma de México, University of Arizona, University of Colorado Boulder, University of Oxford, University of Portsmouth, University of Utah, University of Virginia, University of Washington, University of Wisconsin, Vanderbilt University, and Yale University.

\end{acknowledgements}


\bibliographystyle{aa}
\bibliography{NLTEAPOGEE}

\end{document}